# The Moving Born-Oppenheimer Approximation

**Bernardo Barrera**[a], **Daniel P. Arovas**[b], **Anushya Chandran**[a,c], **and Anatoli Polkovnikov**[a,1]

**We develop a mixed quantum-classical framework, dubbed the Moving Born-Oppenheimer Approximation (MBOA), to describe the dynamics of slow degrees of freedom (DOFs) coupled to fast ones. As in the Born-Oppenheimer Approximation (BOA), the fast degrees of freedom adiabatically follow a state that depends on the slow ones. Unlike the BOA, this state depends on both the positions *and the momenta* of the slow DOFs. We study several model systems: a spin-1/2 particle and a spinful molecule moving in a spatially inhomogeneous magnetic field, and a gas of fast particles coupled to a piston. The MBOA reveals rich dynamics for the slow degree of freedom, including reflection, dynamical trapping, and mass renormalization. It also significantly modifies the state of the fast DOFs. For example, the spins in the molecule are entangled and squeezed, while the gas of fast particles develops gradients that are synchronized with the motion of the piston for a long time. The MBOA can be used to describe both classical and quantum systems and has potential applications in quantum chemistry, correlated materials, atomic physics, molecular dynamics, and quantum sensing.**

The full quantum-mechanical simulation of complex systems is often intractable. It is thus important to develop tools to study their dynamics, even if only approximately. The Born-Oppenheimer Approximation (BOA) is the starting point to describe systems with a clear time-scale separation between a set of slow and fast degrees of freedom (DOFs) (1, 2). See, for example, the left panels of Fig. 1, which show: a) a swinging bucket of water, and b) a molecule with three spins that moves in a spatially varying magnetic field $\vec{\mathcal{B}}(x)$. The BOA assumes that the fast DOFs (the water/spins) adiabatically follow the instantaneous equilibrium state determined by the position of the slow DOF (the bucket/molecule) at every instant of time. Thus, for example, the water surface is always horizontal and the spins are fully polarized along the local field direction as the bucket/molecule moves in space. The power of the BOA is that it greatly reduces the dimensionality of the problem at hand. It is widely used in computational chemistry (3) to simulate the dynamics of molecules composed of slow nuclei coupled to fast electrons, and in condensed matter systems (4) to approximately describe quasiparticles in the background of slow degrees of freedom like phonons.

The BOA fails when the fast DOFs are not in instantaneous equilibrium, either because the nuclear motion is fast, and/or because the energy gap between electronic levels is small. A description of the system in these settings is however important, e.g. to compute reaction rates and molecular spectra (5–8). There is significant interest in developing computational methods that can capture non-adiabatic processes beyond the BOA (see Ref. (9) for an overview of modern techniques).

In particular, the BOA does not account for pseudo forces which can non-perturbatively modify the instantaneous equilibrium state of the fast DOFs. As a familiar example, in Fig. 1(a), the centrifugal force tilts the water surface away from the horizontal and prevents it from falling to the ground even when the bucket is upside down. While in motion, the water reaches a new state of "moving" (or MBO) equilibrium, analogous to mechanical equilibrium in an accelerated or rotating frame. Likewise, in solid state systems, the BOA does not account for (momentum-dependent) pseudo forces on the fast electrons due to the motion of the slow nuclei. These corrections are necessary to recover e.g. accurate electronic momenta and current densities (10, 11).

The subject of this work is a systematic and non-perturbative framework to compute the moving equilibrium state in generic quantum and classical systems. The framework is called "The Moving Born-Oppenheimer Approximation" (MBOA), as it is the BOA applied to the effective Hamiltonian of the fast DOFs in frames of reference co-moving with the slow ones.

## Significance Statement

Complex systems often exhibit phenomena on widely differing timescales, which simplifies their description. An everyday example is a swinging bucket of water. When the bucket moves slowly, the water surface remains nearly horizontal, an approximation known as the Born-Oppenheimer Approximation (BOA). This method is fundamental in condensed matter physics, where similar timescale separations arise due to e.g. the mass difference between nuclei and electrons. However, as the bucket rotation rate increases, the BOA breaks down, and pseudo forces cause the water to reach a state of new "moving" equilibrium. We develop a systematic framework to compute this state in generic quantum and classical systems, revealing rich dynamical effects. Our approach generates spin entanglement, squeezing, and synchronized time-dependent states through motion.

Author affiliations: [a]Department of Physics, Boston University, Boston, Massachusetts 02215, USA; [b]Department of Physics, University of California at San Diego, La Jolla, California 92093, USA; [c]Max-Planck-Institut für Physik komplexer Systeme, 01187 Dresden, Germany

Author contributions: B.B., D.P.A., A.C., and A.P. designed research, performed research, contributed new reagents/analytic tools, analyzed data, and wrote the paper.

The authors declare no conflict of interest.

[1]To whom correspondence should be addressed. E-mail: asp28@bu.edu

At a technical level, the MBOA is implemented as follows. First, we perform a transformation to the basis co-moving with the slow DOFs, dubbed the "moving frame". We postulate that the fast DOFs adiabatically follow the eigenstates of the Hamiltonian in the moving frame $\mathbf{H}^{\mathrm{M}}(x,p)$, which are qualitatively different from the BO states. In particular, they depend on both the slow coordinates and momenta, involve *coherent* superpositions of the instantaneous BO states, and are generically entangled. In the example shown in Fig. 1(b), instead of remaining in a fully polarized state, the spins lag behind the instantaneous direction of the magnetic field and are entangled and squeezed.

Second, the expectation value $\langle \mathbf{H}^{\mathrm{M}} \rangle(x,p)$ evaluated in the moving equilibrium state is used as an effective Hamiltonian for the slow DOFs. To compute the time-evolution of observables, trajectories are propagated using a generalized version of the Truncated Wigner Approximation (TWA). The resulting dynamics can be dramatically different from that in the BOA. For example, the slow DOFs can be dynamically trapped or reflected from momentum-dependent barriers.

There is a large class of existing computational methods in which non-adiabatic processes are treated incoherently. For example, in surface-hopping (12, 13) and multiple spawning (14), the fast DOFs are allowed to transition between different BO surfaces through self-consistently determined rates or spawning events. As such, the fast DOFs are in a classical mixture of BO states. In contrast, the MBOA retains coherence between BO states, allowing it to capture phenomena like entanglement and spin squeezing.

The closest approach to the MBOA is that intuited by Shenvi (15), who obtained the moving Hamiltonian in phase space for model systems of nuclei coupled to electronic DOFs, and showed that surface-hopping between MBO surfaces is quantitatively more accurate than that between BO surfaces. Recent works in the chemistry literature have developed Shenvi's formalism further and applied it to a larger class of systems (16, 17). Other phase space approaches are perturbative and perform expansions in an adiabatic parameter (18–21) (commonly a dimensionless parameter involving the ratio of the electronic and nuclear masses) or truncations of the Moyal product (22–25).

Our work significantly generalizes Ref. (15) and discovers new phenomena in coupled systems of slow and fast degrees of freedom. First, we derive the MBOA from first principles using the path integral, which allows for connections with semi-classical methods like the TWA, and applications to spin systems. Second, we derive the MBOA in both classical and quantum many-particle systems. We show how the Moyal product expansion has a well-defined classical limit that can be understood through the adiabatic gauge potential (AGP). While the exact AGP does not usually exist in complex chaotic systems, one can develop accurate local approximations for it using variational methods (26). Third, we demonstrate rich MBOA dynamics in model classical and quantum systems. In particular, for a spinful molecule moving in a spatially inhomogeneous magnetic field, the fast degrees of freedom (the spins) adiabatically follow entangled/squeezed states. In the textbook classical system of a piston coupled to a classical gas of particles, the fast gas particles settle into a long-lived state with gradients that are synchronized with the motion of the piston. These phenomena are non-perturbatively beyond the BOA.

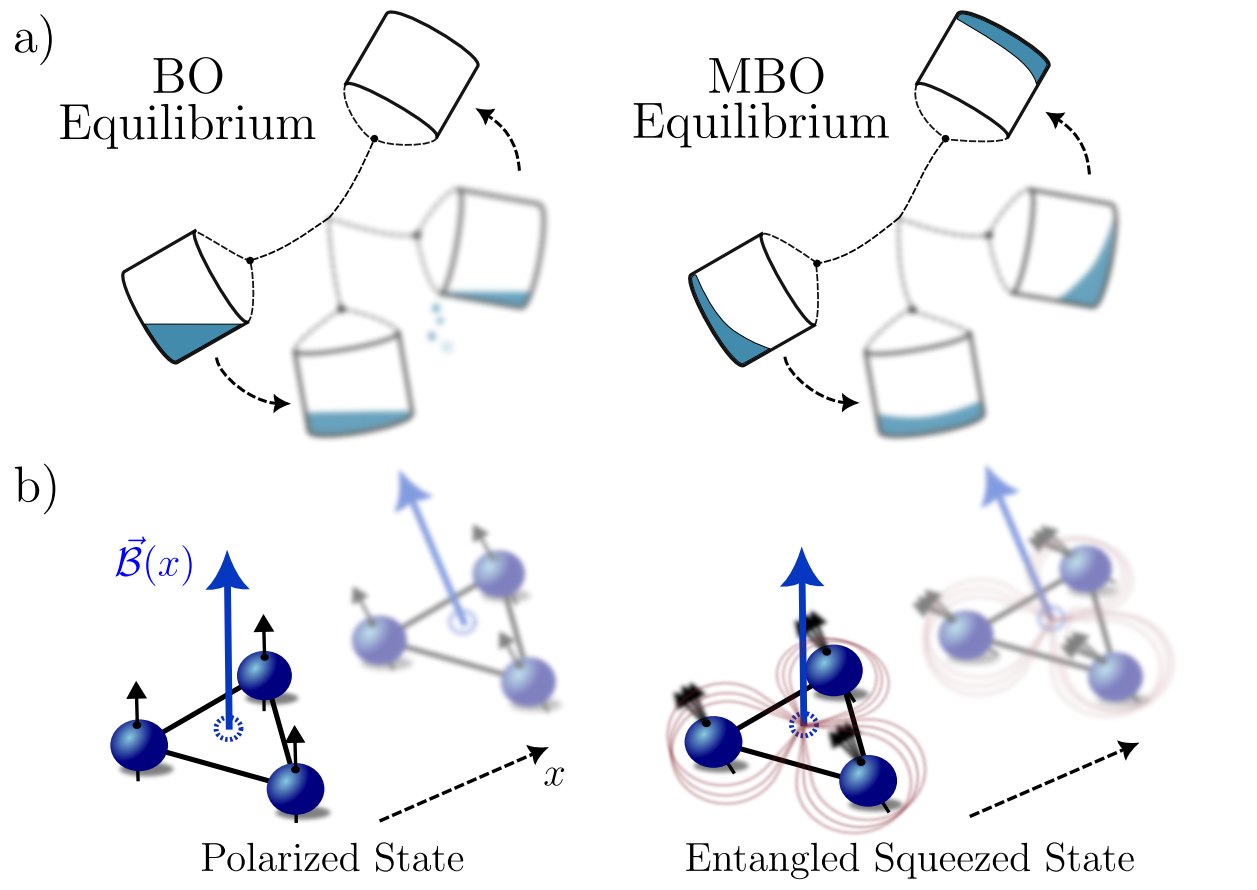

**Fig. 1.** Schematic depiction of BO vs. MBO equilibrium for two examples: a) a swinging bucket of water, and b) a molecule with three spins moving in a spatially varying magnetic field $\vec{\mathcal{B}}(x)$. In the BOA, the fast DOFs (the water/spins) adiabatically follow their instantaneous ground state as the slow DOF (the position of the bucket/molecule) evolves in time. Thus, the water surface is horizontal, and the spins are fully polarized along the local field direction. In the MBOA, the fast DOFs adiabatically follow the ground state of an effective Hamiltonian which contains analogs of pseudo forces. The centrifugal pseudo force tilts the water surface away from the horizontal and prevents it from falling out of the bucket even when it is upside down. Analogous forces cause the spins to lag behind the local magnetic field and follow a many-body state that is entangled and squeezed.

The paper is organized as follows. We begin by working through a familiar example of a rotating bucket of water to intuitively motivate the common failures of the BOA and their connection to our formalism. Sec. 1 then lays out the general MBOA formalism. Sec. 2 studies several model examples demonstrating numerical agreement of the formalism with exact dynamics. In particular, we analyze a quantum spin-1/2 particle and a spinful molecule moving in a spatially inhomogeneous magnetic field, as well as a classical gas of fast particles coupled to a slow piston. Finally, Sec. 3 details the technical steps to implement the MBOA. The derivation of our results is given in Sec. 2 of Supplementary Information.

**Example: A Rotating Water Bucket.** Consider a bucket ($b$) of water ($w$) swinging on a rigid rod of length $R$, as shown in Fig. 1(a). The angular coordinate of the bucket is denoted by $\hat{\theta}$, while the masses, angular momenta and moments of inertia of the bucket/water are denoted by $M_j$, $\hat{L}_j$, and $I_j$ with $j = w, b$.

The BOA assumes that the water is instantaneously in equilibrium at each $\hat{\theta}$. Formally, one first converts $\hat{\theta}$ and $\hat{L}_b$ to classical numbers in the Hamiltonian: $\hat{H}(\hat{\theta}, \hat{L}_b) \to \hat{\mathbf{H}}(\theta, L_b)$ *, and assumes that the water molecules adiabatically follow eigenstates of $\hat{\mathbf{H}}$. As the water particles only couple to the bucket coordinate $\theta$, these states are independent of $L_b$. The expectation value $\langle \hat{\mathbf{H}} \rangle_{\mathrm{BO}}$ then serves as an effective Hamiltonian for the evolution of the bucket.

Physically, the BOA predicts that the water surface is horizontal at all times. If the bucket motion is slow, this prediction is accurate. The BOA also predicts that the

*We use bold font throughout the paper to denote operators that act on the fast DOFs, where the slow ones are fixed as external parameters. This operation corresponds to the Wigner-Weyl transform, which we review in Sec. 1.

angular momentum of the water is zero, and thus that the inertial mass appearing in the equations of motion for the bucket is the bare bucket mass $M_b$. However, this is obviously incorrect even when the bucket is moving slowly; indeed, it violates the equivalence principle. The correct mass is the combined mass $M_b + M_w$.

The MBOA identifies a different slow variable: a *collective* canonical momentum $\hat{\mathcal{L}}$ of the composite bucket+water system,

$$\hat{\mathcal{L}} \equiv \hat{L}_b + \hat{A}_\theta. \quad [1]$$

The operator $\hat{A}_\theta$ is the adiabatic gauge potential (AGP). In the absence of gravity, the AGP is equal to the angular momentum operator of the water molecules $\hat{L}_w$, such that $\hat{\mathcal{L}}$ corresponds to the *total* angular momentum of the system. In the presence of gravity, the AGP acquires additional corrections (see Sec. 12 of S.I). Whereas $\hat{L}_b$ is a noisy fluctuating quantity, the collective $\hat{\mathcal{L}}$ decouples from the quantum noise of the fast DOFs (see Fig. 6) and behaves as a classical degree of freedom that is smoothly evolving in time.

The MBOA assumes that the water molecules are in equilibrium at each $\hat{\theta}$ *and* $\hat{\mathcal{L}}$. It then proceeds exactly as the BOA.

Physically, the MBOA predicts that the water molecules i) equilibrate under the centrifugal force and ii) acquire a mean non-zero angular momentum $\langle \hat{L}_w \rangle \neq 0$ (Sec. 12 of S.I.). Both follow from the effective action of the bucket's kinetic energy operator on the fast space $\hat{L}_b^2/2I_b \to (\mathcal{L} - \hat{\mathbf{A}}_\theta)^2/2I_b$. Analogously, in molecular dynamics, corrections from the nuclear motion are needed in order to recover correct electronic momenta and current densities (10, 11).

In turn, the modified equilibrium of the water provides new equations of motion for the bucket. For example, the expectation value of the angular kinetic energy of the water in the MBO state,

$$\frac{\langle \hat{\mathbf{L}}_w^2 \rangle_{\mathrm{MBO}}}{2I_w} \approx \frac{\langle \hat{\mathbf{L}}_w^2 \rangle_{\mathrm{BO}}}{2I_w} + \frac{M_w R^2 \dot{\theta}^2}{2}, \quad [2]$$

has an extra term as compared to the BOA, which adds to the kinetic energy of the collective degree of freedom. This fixes the inertial mass appearing in the equations of motion for $(\theta, \mathcal{L})$ to be the total mass, $M_b + M_w$.

This intuitive example illustrates how the MBOA restores the correct inertial mass for the motion of the combined water+bucket system. In more complex settings, the corrections are non-trivial and non-perturbative, as we illustrate in the rest of this work.

## 1. Formalism

Consider a system composed of a slow DOF with coordinate $\hat{x}$ and conjugate momentum $\hat{q}$, as well as a fast quantum DOF described by a set of spin operators $\hat{\vec{\sigma}}$. Both the slow and fast DOFs can also be multi-dimensional. For concreteness, we consider the following class of Hamiltonians,

$$H(\hat{\vec{\sigma}}, \hat{x}, \hat{q}) = \frac{\hat{q}^2}{2M} + V(\hat{x}) + H_{\mathrm{int}}(\hat{\vec{\sigma}}, \hat{x}), \quad [3]$$

where $V(\hat{x})$ is an external potential and $H_{\mathrm{int}}(\hat{\vec{\sigma}}, \hat{x})$ describes the interaction between the fast DOF and the slow coordinate. We illustrate all general expressions with a specific example of a spin-1/2 particle moving in a spatially varying magnetic field. The fast DOF (the spin) couples to the slow one (the position of the particle) through,

$$H_{\mathrm{int}}(\hat{x}, \hat{\vec{\sigma}}) = -\mu \vec{\mathcal{B}}(\hat{x}) \cdot \hat{\vec{\sigma}}. \quad [4]$$

In this work we are interested in carrying out so-called mixed quantum-classical dynamics (9), where the slow DOFs are treated classically while keeping the fast DOFs within quantum mechanics. To do so, it is convenient to adopt the Wigner-Weyl formalism, which we review below.

### A. Background.

**The Wigner-Weyl Formalism.** The Wigner-Weyl formalism is an exact formulation of quantum mechanics using phase space variables (27, 28). Quantum-mechanical operators are transformed to functions of the phase space variables $(x, p)$ via the one-to-one map, $\hat{\Omega} \leftrightarrow \Omega(x, p)$, where $\Omega(x, p)$ is known as the Weyl symbol of operator $\hat{\Omega}$. Formally, the Weyl symbol is obtained through the Wigner-Weyl transform,

$$\Omega(x, p) = \int_{-\infty}^{\infty} \mathrm{d}\xi \, \langle x - \xi/2 | \, \hat{\Omega} \, | x + \xi/2 \rangle \, e^{i\frac{p\xi}{\hbar}}. \quad [5]$$

A few examples of Weyl symbols include $f(\hat{x}) \leftrightarrow f(x)$ and $g(\hat{q}) \leftrightarrow g(p)$. For operators that contain both $\hat{x}$ and $\hat{q}$, the Wigner-Weyl mapping is particularly simple for the so-called Weyl-ordered operators; for example, $(\hat{x}\hat{q} + \hat{q}\hat{x})/2 \leftrightarrow xp$.

The Wigner-Weyl mapping can be extended to mixed quantum-classical systems through the use of the partial Wigner transform (29). For a system described by Eq. (3), the Hilbert space is given by a tensor product of the slow (classical) and fast (quantum) Hilbert spaces, $\mathbb{H} = \mathbb{H}^C \otimes \mathbb{H}^Q$. The Weyl symbol is obtained using Eq. (5) in the position eigenbasis of $\mathbb{H}^C$ and remains a quantum mechanical operator acting in $\mathbb{H}^Q$. Throughout the text, we use bold font to denote such mixed quantum-classical operators. For example, the Weyl symbol of Eq. (3) is $\hat{\mathbf{H}}(x, p) = (p^2/(2M) + V(x))\, \hat{\mathbf{I}} + \hat{\mathbf{H}}_{\mathrm{int}}(x) = (p^2/(2M) + V(x))\, \hat{\mathbf{I}} - \mu \vec{\mathcal{B}}(x) \cdot \hat{\vec{\sigma}}$, where $\hat{\mathbf{I}}$ denotes the identity operator in $\mathbb{H}^Q$.

**The Born-Oppenheimer Approximation.** The BOA assumes that the fast DOFs adiabatically follow the instantaneous eigenstates of $\hat{\mathbf{H}}(x, p)$ as $x$ and $p$ slowly change in time. We denote the $n$-th eigenstate at a given $x$ by $\left|\psi_n^{\mathrm{BO}}(x)\right\rangle$, which we refer to as the "BO state". When the fast variables only couple to the coordinate as in Eq. (3), $\left|\psi_n^{\mathrm{BO}}(x)\right\rangle$ is obtained from the unitary transformation $\hat{\mathbf{U}}_{\mathrm{BO}}(x)$ which diagonalizes $\hat{\mathbf{H}}_{\mathrm{int}}(x)$. For the Hamiltonian given by Eq. (4), the two BO states correspond to spin states that are aligned or anti-aligned with the local magnetic field $\vec{\mathcal{B}}(x)$.

The energy of the $n$-th BO state then provides an effective Hamiltonian for the evolution of the slow coordinate and momentum,

$$\mathcal{H}_n^{\mathrm{BO}}(x, p) = \langle \psi_n^{\mathrm{BO}}(x) | \hat{\mathbf{H}}(x, p) | \psi_n^{\mathrm{BO}}(x) \rangle. \quad [6]$$

The time-evolved position $x(t)$ and momentum $p(t)$ are obtained through the standard equations of motion $\dot{x} = \partial_p \mathcal{H}_n^{\mathrm{BO}}$, $\dot{p} = -\partial_x \mathcal{H}_n^{\mathrm{BO}}$. For the spin example, the two BO Hamiltonians corresponding to $n = +, -$ are $\mathcal{H}_\pm^{\mathrm{BO}}(x, p) = p^2/2M + V(x) \pm \mu \mathcal{B}(x)$. The functions $V_n^{\mathrm{BO}}(x) \equiv \langle \psi_n^{\mathrm{BO}}(x) | \hat{\mathbf{H}}_{\mathrm{int}}(x) | \psi_n^{\mathrm{BO}}(x) \rangle$ are called the Born-Oppenheimer potentials or surfaces.

**The Moving Frame.** In the MBOA, one first performs a *time-independent* unitary transformation $\hat{U}_{\rm BO} = U_{\rm BO}(\hat{x})$ and only afterwards computes the Wigner–Weyl transform of the Hamiltonian. This choice of basis is termed the moving frame (30), or equivalently the adiabatic representation in the chemistry literature (31).

Specifically, an arbitrary lab frame operator $\hat{O}$ is mapped to $\hat{U}^\dagger_{\rm BO}\hat{O}\hat{U}_{\rm BO}$ in the moving frame. Note that the unitary rotation and the Wigner-Weyl transform do not commute, and as a result we obtain a different representation of Weyl symbols in the moving frame (see Sec. 3 for details). In particular, the Weyl symbols of position and momentum operators transform non-trivially as,

$$\hat{\mathbf{x}} \to \hat{\mathbf{x}}^{\rm M} \equiv x\,\hat{\mathbf{I}}, \quad \hat{\mathbf{q}} \to \hat{\mathbf{q}}^{\rm M} \equiv p\,\hat{\mathbf{I}} - \hat{\mathbf{A}}_x(x), \qquad [7]$$

where the superscript M is added to refer to the Weyl symbols of various operators in the moving frame. We term such Weyl symbols as the "dressed operators", such that Eq. (7) represents the dressed coordinate and momentum of the particle.

We emphasize that $p$ is the canonical momentum conjugate to $x$ which, like in electromagnetism, should be distinguished from the physical momentum operator $\hat{\mathbf{q}}^{\rm M}$. Only the expectation value of the latter can be identified as the mass times the mean velocity of the particle. The canonical momentum $p$ is a function of both slow and fast DOFs, and as such represents a *collective* variable of the system.

The operator $\hat{\mathbf{A}}_x(x) = i\hbar(\partial_x\hat{\mathbf{U}}_{\rm BO})\hat{\mathbf{U}}^\dagger_{\rm BO}$ is known as the adiabatic gauge potential (AGP) (30), or equivalently the non-adiabatic derivative coupling in the chemistry literature (32). Its expectation values $\langle\hat{\mathbf{A}}_x\rangle_n$ in the $n$-th BO state are gauge-dependent quantities known as Berry connections. If the eigenstate manifold is degenerate, the corresponding projection is called a non-Abelian Berry connection (33, 34). Although we consider examples where the AGP can be found exactly, there exist efficient variational approaches that do not require exactly diagonalizing the interaction Hamiltonian to compute $\hat{\mathbf{A}}_x$ (26, 35–37).

In this work, we consider the *Kato AGP*, whose BO expectation values $\langle\hat{\mathbf{A}}_x\rangle_n$ are zero, and which can be uniquely written in terms of eigenstate projectors (38),

$$\hat{\mathbf{A}}_x(x) = -\frac{1}{2}\sum_n \left[\hat{\mathbf{\Pi}}_n(x), i\hbar\partial_x\hat{\mathbf{\Pi}}_n(x)\right], \qquad [8]$$

with $\hat{\mathbf{\Pi}}_n(x) \equiv \left|\psi_n^{\rm BO}(x)\right\rangle\left\langle\psi_n^{\rm BO}(x)\right|$. Since we only work with examples where the slow coordinate $x$ is one-dimensional and non-compact, Eq. (8) is a valid gauge choice for the BO unitary †. The Kato AGP for the spin-1/2 example has a simple expression in terms of the magnetic field unit vector $\vec{n}(x) \equiv \vec{\mathcal{B}}(x)/\mathcal{B}(x)$,

$$\hat{\mathbf{A}}_x(x) = \frac{\hbar}{2}(\vec{n}(x)\times\partial_x\vec{n}(x))\cdot\hat{\vec{\sigma}}. \qquad [9]$$

It is proportional to the product of the angular momentum operator orthogonal to the plane of rotation of the field and the in-plane rotation rate $\theta'(x) \equiv |\vec{n}(x)\times\partial_x\vec{n}(x)|$. For concreteness, if the field rotates in the $y$-$z$ plane, with $\theta(x)$ denoting the angle with the $z$-axis, then $\hat{\mathbf{A}}_x(x) = -\hbar\theta'(x)\hat{\sigma}_x/2$.

The structure of the dressed operators has important consequences on both equilibrium and dynamical properties of the system. For example, Eq. (7) and Eq. (9) imply that $\hat{\mathbf{q}}^{\rm M}$ is spin-dependent. As a result, if the particle's canonical momentum $p$ has small fluctuations, then measurements of the particle's physical momentum $q$ at position $x$ will produce bimodal distributions peaked around the values $p\pm\hbar\theta'(x)/2$ ‡. Another effect of $\hat{\mathbf{q}}^{\rm M}$ being an operator in $\mathbb{H}^{\rm Q}$ is the emergence of quantum interference even in spin-independent observables, as we illustrate in Sec. 2.

Most importantly, the Hamiltonian acquires new terms in the moving frame that are momentum-dependent and modify both the dynamics and the adiabatic state followed by the fast DOFs. By performing the substitution in Eq. (7),

$$\begin{aligned}\hat{\mathbf{H}}^{\rm M}(x,p) &= \frac{\left(p-\hat{\mathbf{A}}_x(x)\right)^2}{2M} + V(x) + \hat{\mathbf{H}}_{\rm int}(x),\\ &= \hat{\mathbf{H}}(x,p) - \frac{p\hat{\mathbf{A}}_x(x)}{M} + \frac{\hat{\mathbf{A}}_x^2(x)}{2M}.\end{aligned} \qquad [10]$$

The new terms have well-defined classical analogues. As motivated in the introduction, they give rise to pseudo forces due to the moving frame transformation. For example, if the slow variable represents the center of mass coordinate of a system of particles, then the AGP is the total momentum operator of these particles, and the Hamiltonian in Eq. (10) is a standard Galilean transformation. Likewise, if $x$ represents the angular variable of fast particles, then the AGP is the angular momentum operator, and Eq. (10) is the transformation to the rotating frame. For the former, the emergent force is the inertial force, while for the latter they are the centrifugal and the Coriolis forces. Whereas for these simple examples the corrections from the MBOA reduce to well-known expressions, our formalism generalizes this procedure to more general situations.

Several important corrections to the BOA were found early on by M. Berry (33). In Eq. (6), if the average is taken not with respect to the bare Hamiltonian $\hat{\mathbf{H}}$, but instead the moving one, $\left\langle\psi_n^{\rm BO}(x)\right|\hat{\mathbf{H}}^{\rm M}(x,p)\left|\psi_n^{\rm BO}(x)\right\rangle$, one obtains important corrections to $\mathcal{H}_n^{\rm BO}$ that enter in the form of vector and scalar potentials. In particular, the Berry connection plays the role of an emergent vector potential for the slow coordinate, and the metric tensor, given by the variance of the AGP, appears as a correction to the BO scalar potential. The key difference between these standard approaches and the one developed here are the corrections to the dynamics of both slow *and* fast DOFs. For example, Ref. (33) assumes that the fast DOFs follow the BO equilibrium and neglects the off-diagonal terms in $\hat{\mathbf{A}}_x$. However, these off-diagonal corrections are, for example, responsible for the emergence of the inertial force, which can significantly modify the BO state (30, 41).

The transformation to the moving frame is also possible when both slow and fast variables are classical. In this case, it corresponds to a canonical transformation of the fast variables which keeps their trajectories invariant under adiabatic changes in $x$. Then, Eq. (7) enforces canonical relations between fast and slow variables. We illustrate this transformation in Sec. 2 using a toy model of a piston coupled to a gas of fast particles.

† The Kato AGP corresponds to Kato's generator of adiabatic parallel transport (39, 40). In higher-dimensional parameter spaces with nonzero Berry curvature, parallel transport around a closed loop produces a geometric phase. As a consequence, one cannot in general choose a gauge in which the Berry connection vanishes everywhere. Extending the MBOA to such higher-dimensional settings is the subject of future work.

‡ These fluctuations, in particular, ensure the position-momentum uncertainty relation, which can be violated within the BOA. For example, a measurement of the direction of the spin when $\theta'(x)$ is sufficiently large localizes the particle's position in space. The AGP term ensures that this localization is accompanied by an increased uncertainty in the physical momentum $\hat{q}$ such that $\Delta q\Delta x \geq \hbar/2$ irrespective of the measurement protocol.

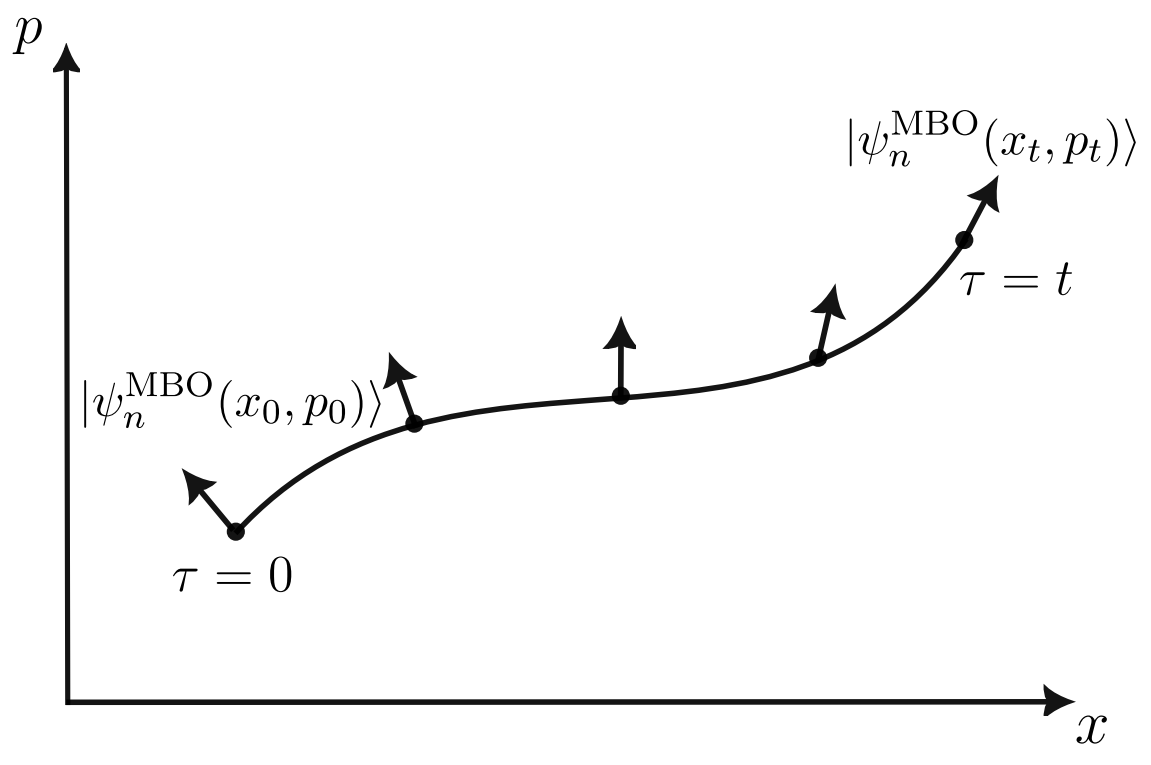

**Fig. 2.** Schematic of the MBO framework. The fast DOF, for simplicity shown as a spin, adiabatically follows an MBO eigenstate $\left|\psi_n^{\rm MBO}(x,p)\right\rangle$ of the moving Hamiltonian $\hat{\mathbf{H}}^{\rm M}(x,p)$ as the slow DOF traces a path through its phase space. MBO states are determined by both the position $x$ and collective momentum $p$.

### B. Moving Born-Oppenheimer Approximation.

**Dynamics of Individual MBO States.** In the MBOA, the fast DOFs are assumed to adiabatically follow the eigenstates of the moving Hamiltonian $\hat{\mathbf{H}}^{\rm M}(x,p)$,

$$\hat{\mathbf{H}}^{\rm M}(x,p)\left|\psi_n^{\rm MBO}(x,p)\right\rangle = \mathcal{H}_n^{\rm MBO}(x,p)\left|\psi_n^{\rm MBO}(x,p)\right\rangle. \quad [11]$$

We refer to these states as the MBO states, and remark that they will be in general both $x$ and $p$ dependent, in contrast to the usual BO states. Thus, the picture to have in mind is sketched in Fig. 2. The fast degree of freedom, for simplicity shown as a spin, adiabatically follows a particular eigenstate $\left|\psi_n^{\rm MBO}(x,p)\right\rangle$ as the slow DOF traces a path through its phase space.

The energy of each MBO state then serves as an effective Hamiltonian governing the dynamics of the slow variables,

$$\mathcal{H}_n^{\rm MBO}(x,p) = \left\langle\psi_n^{\rm MBO}(x,p)\right|\hat{\mathbf{H}}^{\rm M}(x,p)\left|\psi_n^{\rm MBO}(x,p)\right\rangle. \quad [12]$$

Note the resemblance between Eq. (6) and Eq. (12). However, $\mathcal{H}_n^{\rm MBO}$ has non-trivial dependence on both $x$ and $p$ both explicitly through the moving Hamiltonian $\hat{\mathbf{H}}^{\rm M}(x,p)$ and implicitly through the state $|\psi_n^{\rm MBO}(x,p)\rangle$.

In order to compute the time evolution of some observable $\hat{\Omega} = \Omega(\hat{x},\hat{p},t)$ within the MBOA, we solve the non-linear equations of motion [§],

$$\frac{dx}{dt} = \frac{\partial\mathcal{H}_n^{\rm MBO}(x,p)}{\partial p} = \left\langle\psi_n^{\rm MBO}\right|\partial_p\hat{\mathbf{H}}^{\rm M}\left|\psi_n^{\rm MBO}\right\rangle, \quad [13]$$

$$\frac{dp}{dt} = -\frac{\partial\mathcal{H}_n^{\rm MBO}(x,p)}{\partial x} = -\left\langle\psi_n^{\rm MBO}\right|\partial_x\hat{\mathbf{H}}^{\rm M}\left|\psi_n^{\rm MBO}\right\rangle [14]$$

The last equalities follow from the Hellmann-Feynman theorem. Denote the solutions of these equations as $(x_n(t), p_n(t))$, where the subscript $n$ stands for the index of the eigenstate of $\hat{\mathbf{H}}^{\rm M}$. Then, we evaluate the expectation value of the (dressed) Weyl symbol of the observable $\Omega_n^{\rm M}(t) \equiv \langle\psi_n^{\rm MBO}(x,p)|\hat{\mathbf{\Omega}}^{\rm M}(x,p)|\psi_n^{\rm MBO}(x,p)\rangle$ at $x = x_n(t)$ and $p = p_n(t)$. For example, if the fast degrees of freedom follow the MBO ground state $n = 0$, the expectation value of the momentum operator is given by $\langle\hat{\mathbf{q}}^{\rm M}(t)\rangle = p_0(t) - A_x(t)$, where $A_x(t) \equiv \langle\psi_0(t)|\,\hat{\mathbf{A}}_x(x_0(t))\,|\psi_0(t)\rangle$ is the AGP expectation value in the MBO ground state. Here, we use the short hand notation $|\psi_0(t)\rangle \equiv \left|\psi_0^{\rm MBO}(x_0(t),p_0(t))\right\rangle$. Likewise $\langle(\hat{\mathbf{q}}^{\rm M})^2(t)\rangle = [p_0(t)-A_x(t)]^2 + g_0^{\rm M}(t)$, where $g_0^{\rm M}(t) = \langle\psi_0(t)|\,\hat{\mathbf{A}}_x^2(x_0(t))\,|\psi_0(t)\rangle - A_x^2(t)$ is the fidelity susceptibility or equivalently the Fubini-Study metric tensor also evaluated in the moving ground state [¶]. Finally, $\Omega_n^{\rm M}(t)$ is averaged over a distribution of initial conditions $\{x,p\} \equiv \{x_0,p_0\}$ given by the (dressed) Wigner function of the initial density matrix, as detailed in Sec. 3.

When the fast DOFs are prepared not in a single MBO state, but in a coherent superposition of them, there can be important phenomena resulting from their interference. For example, the phase information between MBO states is important to describe the precession of the spin about the magnetic field (see Sec. 6 of S.I.), or transient oscillations in the momenta of the slow DOF, as examined in Fig. 3. To recover these effects within the MBOA, one must consider the time evolution of the entire matrix elements of $\hat{\mathbf{\Omega}}^{\rm M}(x,p)$ in the MBO basis. The procedure for doing this is more involved and is described in detail in Sec. 3.

## 2. Examples

We now apply the MBOA to select examples. We demonstrate numerical agreement with exact quantum dynamics and highlight new physics which lie beyond the standard BOA.

### A. Spin-1/2 Particle in a Spatially Varying Magnetic Field.

Consider the textbook example of the spin-1/2 particle in 1d interacting with a position-dependent magnetic field (33, 42, 43), with the Hamiltonian in Eq. (3), Eq. (4), and $V(\hat{x}) = 0$. For concreteness, we take the magnetic field to rotate in the plane perpendicular to the motion, $\vec{\mathcal{B}}(x) = \mathcal{B}(x)(\cos\theta(x)\hat{z} + \sin\theta(x)\hat{y})$. General magnetic field profiles are discussed in Sec. 6 of S.I.

Within the Born-Oppenheimer approximation, the spin follows the instantaneous eigenstates of the magnetic potential energy as it moves in space. That is, the spin remains aligned with the local magnetic field, and the BO potential only depends on the magnitude of the field $\mathcal{B}(x)$.

However, if the motion is sufficiently fast, corrections to the BOA become important. Using Eq. (10) and the explicit form of the Kato AGP in Eq. (9), $\hat{\mathbf{H}}^{\rm M}$ reads,

$$\hat{\mathbf{H}}^{\rm M} = \frac{p^2}{2M} + \frac{\hbar^2(\theta'(x))^2}{8M} - \mu\vec{\mathcal{B}}(x)\cdot\hat{\vec{\sigma}} + \frac{\hbar}{2M}p\,\theta'(x)\hat{\sigma}_\perp, \quad [15]$$

where $\hat{\sigma}_\perp \equiv \hat{\sigma}_x$ is the spin component in the direction orthogonal to the plane of the magnetic field. The first two terms clearly do not affect the spin eigenstates. The last two terms represent the Hamiltonian for a spin in an effective

[§] The evolution of the slow coordinate can also be treated quantum mechanically by performing the inverse Wigner-Weyl transform of $\mathcal{H}_n^{\rm MBO}(x,p)$. In this work, we focus on semiclassical dynamics.

[¶] Note that the Berry connection is often defined with an extra $1/\hbar$ prefactor, while the Fubini-Study metric tensor has an extra $1/\hbar^2$ prefactor. It is more convenient, however, to work with units where $A_x$ has dimensions of a momentum and a well-defined classical limit.

$p$-dependent magnetic field $\vec{\mathcal{B}}_{\rm eff}(x,p)$ with magnitude,

$$\mathcal{B}_{\rm eff}(x,p)=\sqrt{\mathcal{B}(x)^2+\left(\frac{p\hbar\theta'(x)}{2M\mu}\right)^2}, \qquad [16]$$

which is tilted in the $x$-direction, i.e. the direction of motion. In particular, it makes an angle $\phi$ with the real magnetic field given by,

$$\phi(x,p)=\tan^{-1}\left(\frac{\hbar\theta'(x)p}{2M\mu\mathcal{B}(x)}\right). \qquad [17]$$

Interestingly, $\mathcal{B}_{\rm eff}$ does not vanish even if the real magnetic field is zero. This result might look surprising at first sight, since if the magnetic field is zero then the two BO states are exactly degenerate. However, hiding behind is an interesting physical result that the states polarized along the $x$-axis, i.e. perpendicular to the magnetic field rotation plane, are gapped in the moving frame and hence robust to non-adiabatic effects.

**Reflection and Dynamical Trapping.** Consider a magnetic field with constant magnitude $\mathcal{B}(x)=\mathcal{B}$, but with a region of non-zero rotation at the origin. The rotation of the field is set by $\theta(x)=\theta_0(1+\mathrm{erf}(x/d))$, such that it rotates by a total angle of $2\theta_0$ over a distance of the order of $d$. The particle is prepared in a wavepacket at some position $x_0$ that is far away from any field rotation, $|x_0|\gg d$, and moving with mean velocity $p_0/m$ towards the origin. The spin is initially polarized in the direction of the local magnetic field.

Within the BOA, the predicted dynamics is trivial. The spin remains aligned with the field and the Hamiltonian of the motional DOF is modified by a constant potential energy of $-\mu\mathcal{B}$. The particle consequently moves through the origin with constant velocity.

Within the MBOA, non-adiabatic corrections significantly modify the dynamics of the particle. As it enters the region of non-zero $\theta'$, the spin follows the MBO ground state, $\left|\psi_-^{\rm MBO}\right\rangle=\cos(\phi/2)\left|\psi_-^{\rm BO}\right\rangle-\sin(\phi/2)\left|\psi_+^{\rm BO}\right\rangle$ of $\hat{\mathbf{H}}^{\rm M}(x,p)$, with $\phi$ given by Eq. (17). Then, according to Eq. (12), the two classical Hamiltonians for the ground and excited states read,

$$\mathcal{H}_{\mp}^{\rm MBO}(x,p)=\frac{p^2}{2M}+\frac{\hbar^2(\theta'(x))^2}{8M}\mp\mu\mathcal{B}_{\rm eff}(x,p). \qquad [18]$$

If non-adiabatic effects are small, virtual excitations of the spin effectively renormalize the mass of the particle (30). For concreteness, consider a spin in the MBO ground state with Hamiltonian $\mathcal{H}_-^{\rm MBO}(x,p)$. In the perturbative regime with $|\tan(\phi(x,p))|\ll 1$, one can Taylor expand $\mathcal{B}_{\rm eff}(x,p)$. The leading non-adiabatic term proportional to $p^2$ produces a position-dependent correction to the particle mass as $M\to M+\kappa(x)$, with $\kappa(x)=\frac{(\hbar\theta'(x))^2}{4\mu\mathcal{B}(x)}$. This same term was discussed in Ref. (43) and constitutes only the leading order correction within the MBOA. In higher spatial dimensions there is an additional Lorentz-type force (42), which is also recovered in the perturbative regime, as shown in Sec. 6 of S.I.

The MBOA dynamics are however most interesting when non-adiabatic effects are large, i.e. in the non-perturbative $\kappa/M>1$ regime. The top two panels of Fig. 3 show the energy contours in phase space of $\mathcal{H}_\pm^{\rm MBO}$ for $\kappa(x=0)=2M$. In the panel corresponding to $\mathcal{H}_-^{\rm MBO}$, we highlight in blue the appearance of orbits in which the particle reflects from the origin. Indeed, the field rotation creates an effective energetic barrier of height $V_{\rm eff}$ at $x=0$ (see Sec. 6 of S.I.). Within the MBOA, a spin with initial kinetic energy below the barrier height cannot cross the region of field rotation and is reflected back. The red trajectory highlights another interesting effect. If the spin is initially prepared in the MBO ground state within some phase space region bounded by a separatrix (shown as a dashed line), it remains dynamically trapped around the origin. In both cases, the emergent dynamics is qualitatively different from that predicted by the BOA. Similar results for a related model were reported earlier in Ref. (15).

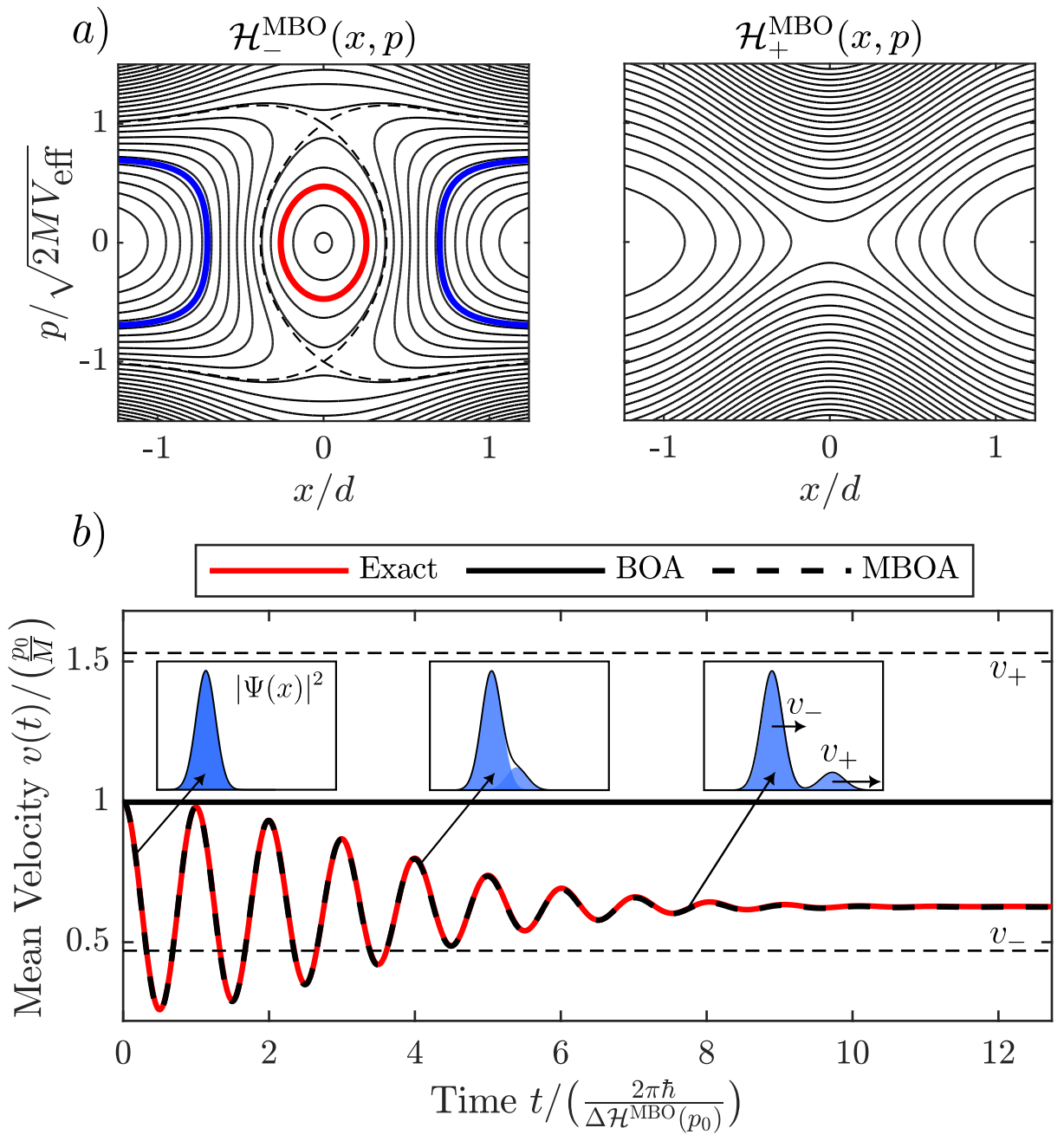

**Fig. 3.** Dynamical effects predicted by the MBOA for a spin-1/2 particle moving in a rotating magnetic field in the strongly non-adiabatic regime $\kappa(x=0)=2M$. a) Energy contours of the two eigenvalues $\mathcal{H}_\pm^{\rm MBO}(x,p)$ of the moving Hamiltonian $\hat{\mathbf{H}}^{\rm M}(x,p)$. The highlighted orbits show two dynamical effects absent within the standard BOA: reflection (blue) and dynamical trapping (red). b) Time evolution of a wavepacket with spin initially prepared in a superposition of the two MBO states. At long times, the initial packet splits into two moving at different velocities $v_\pm$, each corresponding to one of the two MBO states (see insets). However, while the two packets have spatial overlap, the mean velocity of the wavepacket oscillates as a consequence of phase coherence between the MBO states. The initial state is given by Eq. 63 of S.I., with initial mean momentum $p_0=(2/3)\hbar\theta'$, non-adiabatic tilt angle $\tan(\phi(p_0))=1$, and wavepacket width $\sigma_x=7.5/\theta'$.

We stress again that the momentum $p$ does not correspond to the mechanical momentum of the particle, which is rather obtained as $M\dot{x}=M\langle\partial_p\hat{\mathbf{H}}^{\rm M}\rangle=p-\langle\hat{\mathbf{A}}_x(x)\rangle=\langle\hat{\mathbf{q}}^{\rm M}\rangle$. Here, the average is taken with respect to the moving state and hence the Berry connection term $A(x,p)\equiv\langle\hat{\mathbf{A}}_x\rangle$, which plays a similar role as the vector potential in electromagnetism, does not coincide with the one usually appearing in the literature (33), which would be identically zero using the Kato AGP. For the same reason, the non-equilibrium correction to the force on the particle, which is obtained from $\dot{p}=-\langle\partial_x\hat{\mathbf{H}}^{\rm M}\rangle$, generally differs from the diagonal Born-Huang (metric) correction to the BO surfaces (see Fig. S3 for more details).

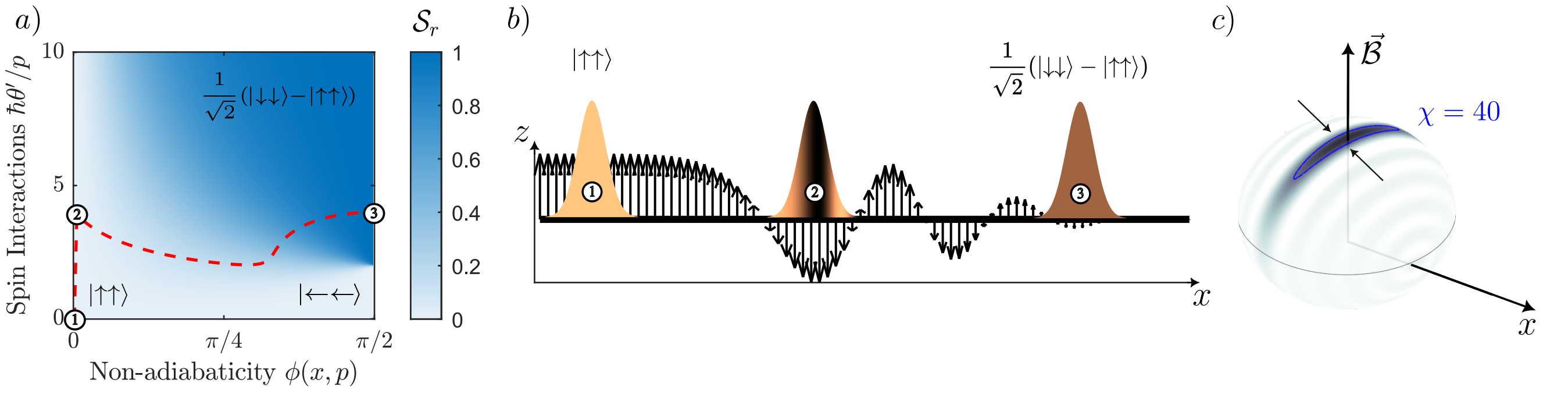

**Fig. 4.** Motion generated entanglement and spin squeezing. a) Phase diagram of the ground state of $\hat{\mathbf{H}}^{\mathrm{M}}$ for $N = 2$ spins. Shown in blue is the entanglement entropy $\mathcal{S}_r = -\mathrm{Tr}(\rho_r \log_2 \rho_r)$ of the single-spin reduced density matrix $\rho_r$ of the MBO ground state. b) Schematic protocol to generate a Bell pair through motion. The spins are initially prepared in a product state, polarized in the direction of the magnetic field, and are sent through a region of high field rotation. After the rate of rotation increases past the threshold value $\hbar\theta'/p > 2$, the field magnitude is adiabatically turned off. The corresponding path through parameter space is shown in red in panel (a). The outgoing state is a Bell pair. c) Wigner function of the collective spin of the ground state of $\hat{\mathbf{H}}^{\mathrm{M}}$, for $N = 20$ and $\chi = 40$. With increasing $\chi$, the spin is squeezed along the direction of motion.

**Coherent Dynamics and Interference.** We now apply the MBOA scheme to study the dynamics of initial states composed of superpositions of the MBO states. Consider a magnetic field profile with both constant magnitude $\mathcal{B}(x) = \mathcal{B}$ and rate of rotation $\theta'(x) = \theta'$. The spin is initially prepared in a BO ground state, i.e. in a wavepacket polarized along the instantaneous field direction. The wavepacket is given a mean (canonical) momentum $p_0$. This setup can be realized by e.g. sending the packet into a region of strong rotating field and then suddenly reducing the field magnitude. Note that a spin aligned with the field corresponds to a superposition of the two MBO states, which are tilted towards the $x$-axis.

Figure 3(b) shows the exact time evolution, along with the BOA and MBOA predictions. Naively, we expect the two MBO states to propagate independently, each with a corresponding velocity given by $\dot{x}_\pm = \langle \partial_p \hat{\mathbf{H}}^{\mathrm{M}} \rangle_\pm \big|_{p=p_0} = \frac{p_0}{M} \pm \frac{\hbar\theta'}{2M} \sin\phi(p_0)$. This is indeed observed at long times, where the initial packet splits into two, each with an internal spin structure corresponding to one of the two eigenstates of $\hat{\mathbf{H}}^{\mathrm{M}}$ (see insets in Fig. 3). The mean speed at long times is a weighted average of these two contributions. However, there is a transient period of oscillations that persists while the two packets have spatial overlap. These are due to the interference between the two MBO states, which, within our formalism, are captured by the time-dependent phases in the off-diagonal entries of $\hat{\mathbf{q}}^{\mathrm{M}}_{mn}(t)$ (see Eq. (38) and Sec. 6 of SI). The frequency of oscillations is set by the MBO gap, $\hbar\omega = \Delta\mathcal{H}^{\mathrm{MBO}}(p_0) = \mathcal{H}^{\mathrm{MBO}}_+(p_0) - \mathcal{H}^{\mathrm{MBO}}_-(p_0)$.

**B. Entanglement and Spin Squeezing.** The MBOA predicts new physics for the fast DOFs that are not captured by conventional frameworks. In particular, the $\hat{\mathbf{A}}_x^2$ term in Eq. (10) mediates effective interactions between otherwise non-interacting degrees of freedom, leading to non-trivial correlations in the MBO states.

For example, consider a moving molecule containing several non-interacting spin degrees of freedom coupled to the same local magnetic field, as shown in Fig. 1(b). The Hamiltonian is given by the multi-spin version of Eq. (4),

$$\hat{H} = \frac{\hat{p}^2}{2M} - \sum_{i=1}^{N} \mu \vec{\mathcal{B}}(\hat{x}) \cdot \hat{\vec{\sigma}}_i. \qquad [19]$$

The adiabatic gauge potential is simply the sum of the single-spin AGPs, $\hat{\mathbf{A}}_x(x) = -\hbar\theta'(x)\hat{\mathbf{S}}_x$, with $\hat{\mathbf{S}}_x = \sum_{i=1}^{N} \hat{\sigma}_x^i/2$, the total spin along the $x$-direction.

The $\hat{\mathbf{A}}_x^2$ term thus naturally induces all-to-all spin interactions in the moving frame. Indeed, the moving Hamiltonian is of a form commonly used for spin-squeezing applications,

$$\hat{\mathbf{H}}^{\mathrm{M}} = \frac{p^2}{2M} - 2\mu\vec{\mathcal{B}}(x) \cdot \hat{\vec{\mathbf{S}}} + \frac{p\hbar\theta'(x)}{M}\hat{\mathbf{S}}_\perp + \frac{(\hbar\theta'(x))^2}{2M}\hat{\mathbf{S}}_\perp^2, \quad [20]$$

where $\hat{\mathbf{S}}_\perp \equiv \hat{\mathbf{S}}_x$. Eq. (20) corresponds to a one-axis twisting Hamiltonian with a control field (44), whose eigenstates are optimally spin-squeezed states (45). In the following sections, we study properties of the eigenstates of $\hat{\mathbf{H}}^{\mathrm{M}}$ and their potential use for state preparation.

**Bell State Generation.** The MBO states are entangled states of the fast spins. The left panel of Fig. 4 shows the phase diagram for the ground state of $\hat{\mathbf{H}}^{\mathrm{M}}$ for $N = 2$ spins.

The ground state is determined by two dimensionless parameters, $\phi = \tan^{-1}(\hbar\theta' p/2M\mu\mathcal{B})$ and $\hbar\theta'/p$. The angle $\phi$ compares the magnitude of the BO energy splitting $\mu\mathcal{B}$ to the linear in AGP coupling, and is thus a measure of non-adiabaticity. The ratio $|\hbar\theta'/p|$ compares the de Broglie wavelength of the particle to the characteristic length scale over which the angle $\theta$ locally varies. For large $|\hbar\theta'/p|$, the quadratic in AGP term dominates over the linear one, and $\hat{\mathbf{H}}^{\mathrm{M}}$ can therefore mediate strong spin-spin interactions. In the following analysis, we take for simplicity $p\,, \hbar\theta' > 0$.

In the regime $\hbar\theta'/p \ll 1$ (lower edge of the phase diagram), the spins are non-interacting. We may thus use the same intuition as developed for the single spin model. The dominant term in the moving Hamiltonian $\hat{\mathbf{H}}^{\mathrm{M}}$ is determined by an effective momentum-dependent magnetic field $\vec{\mathcal{B}}_{\mathrm{eff}}$ which is tilted away from the real magnetic field by an angle $\phi(x,p)$ given in Eq. (17). Each spin independently follows the effective field. For $\phi = 0$, the ground state $|\uparrow\uparrow\rangle$ is polarized along the direction of the real magnetic field, while for $\phi = \pi/2$, it is maximally tilted in the $x$-direction, $|\leftarrow\leftarrow\rangle$.

For non-zero $\hbar\theta'/p$, the MBO ground state is entangled. In Fig. 4(b), the color indicates the magnitude of the entanglement entropy $\mathcal{S}_r = -\mathrm{Tr}(\rho_r \log_2 \rho_r)$ of the single-spin reduced density matrix $\rho_r$ in the MBO ground state. In particular, for $\hbar\theta'/p > 2$ and $\phi \to \pi/2$ (upper right quadrant), the ground state is close to the $S_x = 0$ triplet $\frac{1}{\sqrt{2}}\left(|\downarrow\downarrow\rangle - |\uparrow\uparrow\rangle\right)$ ‖.

These considerations motivate the idea of generating spin entanglement through motion. Starting from a product state, it is possible to design adiabatic protocols that realize a path from the lower left to the upper right quadrant of the diagram, producing a Bell state of the spins. The subtlety is that the path through parameter space must be self-consistently generated by the MBOA equations of motion as given by Eq. (13) and Eq. (14). As a proof of concept, we provide one such realization of a protocol, producing the red path shown. The full details are given in Sec. 7 of S.I.

In Fig. 4(b), we describe the setup schematically. The process consists of three parts. First, both spins are initially prepared in a product state $|\uparrow\uparrow\rangle$, polarized in the direction of the magnetic field. The wavepacket is given some mean momentum $p$. Second, the packet moves through a region where the field rotation rate is larger than a threshold value: $\hbar\theta'/p > 2$. The field magnitude is kept large enough such that $\phi$ remains small - thus, the spins remain locally in a BO product state. However, since the de Broglie wavelength of the wavepacket becomes comparable to $1/\theta'$, the local spin direction varies significantly along the width of the wavepacket. This is shown in the middle of the right panel, where the (dark)bright colors denote local spin (anti)alignment with the $z$-axis. At this moment, the spin state and the position of the particle are entangled. Lastly, the field magnitude is slowly decreased while keeping the rate of rotation constant. By doing this, $\hat{\mathbf{S}}_x^2$ becomes the dominant term in $\hat{\mathbf{H}}^{\mathrm{M}}$, and the spins are adiabatically driven towards the $S_x = 0$ triplet. Due to the invariance of the triplet under rotations about the $x$-axis, spin and position become unentangled once again, and the outgoing packet has a constant spin texture everywhere in space. The resulting state is the desired Bell state. We remark that to design the trajectory of the particle through phase space, an additional external potential was added to Eq. (20) (see Sec. 7 of S.I. for details). This extra potential, however, does not modify the eigenstates of $\hat{\mathbf{H}}^{\mathrm{M}}$.

**Spin Squeezing.** Another interesting observation is that MBO states can be spin-squeezed. In particular, in the large $N$ and strongly interacting $\hbar\theta'/p \gg 1$ limit of $\hat{\mathbf{H}}^{\mathrm{M}}$, one can engineer ground states where the collective spin points along the direction of the local magnetic field, but is squeezed in the direction of motion.

Taking $\vec{\mathcal{B}}$ to be along the $z$-axis, the ratio between the spin uncertainties along the $x$ and $y$ directions in the MBO ground state is set by the dimensionless parameter $\chi \equiv \frac{N(\hbar\theta')^2}{4M\mu\mathcal{B}}$ as,

$$\frac{\Delta\hat{\mathbf{S}}_x}{\Delta\hat{\mathbf{S}}_y} = \frac{1}{\sqrt{1+\chi}}, \qquad [21]$$

where $\Delta\hat{\mathbf{S}}_i = \sqrt{\langle\hat{\mathbf{S}}_i^2\rangle - \langle\hat{\mathbf{S}}_i\rangle^2}$. Figure 4(c) shows the Wigner function for the collective spin of the ground state of $\hat{\mathbf{H}}^{\mathrm{M}}$, for a total number of spins $N = 20$, with $\chi = 40$. The blue contour encloses an uncertainty region of $\Delta\hat{\mathbf{S}}_x\Delta\hat{\mathbf{S}}_y = N/4$.

Eq. (20) can be used to produce optimally squeezed states with scaling $\Delta\hat{\mathbf{S}}_x/N \sim 1/N$ (45). To obtain the quadrature ratio, we have performed a Holstein-Primakoff transformation (see Sec. 8 of S.I.), which holds for $\chi \ll N^2$. Therefore, the minimum quadrature ratio one can achieve, while remaining in the regime where the ground state still resembles a squeezed coherent state as in Fig. 4(c), is of the order of $\sim 1/N$. In the limit $\chi \to \infty$, the ground state reduces to the $S_x = 0$ eigenstate with total spin $S = N/2$.

As the magnetic field $\vec{\mathcal{B}}(x)$ rotates in space, the collective spin orientation also changes such that it becomes entangled with the position similarly to the wavepacket labelled by (2) in Fig. 4(b). In order to utilize spin squeezing in the lab frame, a step function can be introduced into the spatial profile of the magnetic field, such that once the particle leaves the field region, the collective spin points in the direction of the field at the step. Alternatively, one can extract the spin state using time-dependent protocols. For instance, by applying an $x$-dependent magnetic field pulse that rotates the local spin about the $x$-axis.

**C. The MBOA in the Classical Limit.** The methodology of the MBOA can be extended to study systems where both fast and slow DOFs are classical. Interestingly, while the derivation of the MBOA seems intrinsically quantum, based on expanding the Moyal product (see Sec. 3), the emergent adiabatic gauge potential has a well-defined classical interpretation as a generator of trajectory preserving adiabatic transformations (30, 46). As we discuss below, the transformation Eq. (7) then ensures proper canonical relations between the fast and slow DOFs.

As an illustration, consider a system with one slow DOF described by a coordinate and its conjugate momentum $(x, q)$, coupled to $N$ fast DOFs described by a vector of coordinates $\vec{\xi} = (\xi_1, \ldots, \xi_N)$ and their conjugate momenta $\vec{\eta} = (\eta_1, \ldots, \eta_N)$. The Hamiltonian of the system is of the form,

$$H(\vec{\xi}, \vec{\eta}, x, q) = \frac{q^2}{2M} + V(x) + H_{\mathrm{int}}(\vec{\xi}, \vec{\eta}, x), \qquad [22]$$

in analogy to Eq. (3). For concreteness, consider $V(x) = kx^2/2$ and,

$$H_{\mathrm{int}}(\vec{\xi}, \vec{\eta}, x) = \frac{|\vec{\eta}|^2}{2m} + U_0\sum_{i=1}^{N}(\Theta(-\xi_i) + \Theta(\xi_i - x)), \qquad [23]$$

with $U_0 \to \infty$ and $\Theta$ the Heaviside step function. This corresponds to the setup shown in Fig. 5. The slow DOF is a heavy piston of mass $M$ that is bound to a spring which has a resting equilibrium position at $x = 0$. A non-interacting gas of $N$ fast particles of mass $m$ lie between the piston and the left wall at $x = 0$. The mass ratio $m/M$ is taken to be small. If the initial kinetic energy of the fast particles is given by $\frac{|\vec{\eta}|^2}{2m} = E_0^{\mathrm{fast}}$, then the position of mechanical equilibrium, where the mean pressure from the particles on the piston is equal to the restoring force from the spring, is found at $x_* = \sqrt{2E_0^{\mathrm{fast}}/k}$. In what follows, we consider perturbing the piston away from equilibrium and studying the resulting dynamics.

In analogy to the Wigner-Weyl transform Eq. (5), we introduce bold font to highlight functions of the fast phase space coordinates $\vec{\xi}$ and $\vec{\eta}$, where the slow ones are treated as fixed external parameters. For example, $\mathbf{H}_{\mathrm{int}}(x) \equiv$

‖In the exact $\hbar\theta'/p > 2$ and $\phi = \pi/2$ limit, the $S_x = 0$ triplet and singlet become degenerate. However, non-adiabatic transitions between them are forbidden by total spin conservation.

$H_{\rm int}(\vec\xi,\vec\eta,x)$. As in the quantum case, we reserve the symbol $p$ for the canonical momentum, while using $q$ for the physical one. In the lab frame, the two are equivalent and we have $\mathbf{q} = p\,\mathbf{I}$, where $\mathbf{I}$ is the identity function of $\vec\xi,\vec\eta$. This is the analogous classical statement for the Wigner-Weyl mapping $\hat{\mathbf{q}} = p\,\hat{\mathbf{I}}$. In this way, the quantum and classical language can be used interchangeably by simply adding or removing hats over the corresponding bold functions.

We first study Eq. (23) within the BOA. In quantum systems, the BOA assumes that the fast DOFs adiabatically follow the eigenstates of the Hamiltonian at the instantaneous value of $x$. For classical systems, the analogous statement is that fast degrees of freedom adiabatically follow a micro-canonical distribution $\boldsymbol{\rho}(x) = \delta(E^{\rm fast}(x) - \mathbf{H}_{\rm int}(x))$ [**], where $E^{\rm fast}(x)$ is fixed such that the entropy of fast particles,

$$S(x,E^{\rm fast}) = \ln\Omega(x,E^{\rm fast}),$$
$$\Omega(x,E^{\rm fast}) = \int {\rm d}\vec\xi{\rm d}\vec\eta\,\delta(E^{\rm fast} - \mathbf{H}_{\rm int}(x)), \qquad [24]$$

is conserved.

A straightforward calculation yields the entropy $S = N/2\ln\left(x^2E^{\rm fast}\right) + {\rm const}$ (see Sec. 9 of S.I.), which determines the energy of the fast particles as a function of the piston position: $x^2E^{\rm fast}(x) = {\rm const}$. As expected, the gas cools as the piston expands, and it heats up as it contracts.

Then, averaging the Hamiltonian over the fast particles at the instantaneous energy shell $E^{\rm fast}(x)$ yields,

$$\mathcal{H}^{\rm BO}(x,p) = \frac{p^2}{2M} + \frac{1}{2}kx^2 + E_0^{\rm fast}\left(\frac{x_0}{x}\right)^2, \qquad [25]$$

which serves as the emergent Hamiltonian for the motion of piston. Here, $E_0^{\rm fast}$ denotes the initial kinetic energy of the fast particles, and $x_0$ the initial position of the piston. Note that $V^{\rm BO}(x) = E_0^{\rm fast}(x_0/x)^2$ is precisely the Born-Oppenheimer potential. The same result is obtained if the particles are treated quantum-mechanically. There, the single-particle BO energies are given by $\hbar^2\pi^2n^2/(2mx^2)$. Thus, for adiabatic transformations preserving $n$, the Born-Oppenheimer potentials are inversely proportional to $x^2$.

We now derive the MBOA Hamiltonian. In classical mechanics, the role of eigenstates is played by time-averaged trajectories. The classical analogue of the moving frame transformation is a canonical transformation of the fast variables such that the time-averaged trajectories generated by $\mathbf{H}_{\rm int}(x)$ are independent of $x$. For this example, the required transformation is the phase space dilation,

$$\vec\xi\,'(x) = \vec\xi/x, \qquad \vec\eta\,'(x) = x\vec\eta. \qquad [26]$$

Indeed, particle collisions in the dilated coordinates always occur at either $\xi'_i(x) = 0$ or $\xi'_i(x) = 1$, independent of the piston position.

The canonical transformation also affects the momentum of the slow DOF. As it stands, Eq. (26) yields non-zero Poisson brackets $\{\xi'_i(x), q\}$ and $\{\eta'_i(x), q\}$. To remedy this, we define a new canonical momentum $p$, such that,

$$\mathbf{q} = p - \mathbf{A}_x(x) = p - \sum_{i=1}^{N} A_x^{(i)}(\xi_i,\eta_i,x). \qquad [27]$$

[**] Alternatively, one can work in the canonical ensemble, assuming that the fast DOFs follow a Gibbs state $\boldsymbol{\rho}(x) \propto e^{-\beta(x)\mathbf{H}_{\rm int}(x)}$, with inverse temperature $\beta(x)$ fixed such that entropy is conserved. See Sec 9 of S.I. for details.

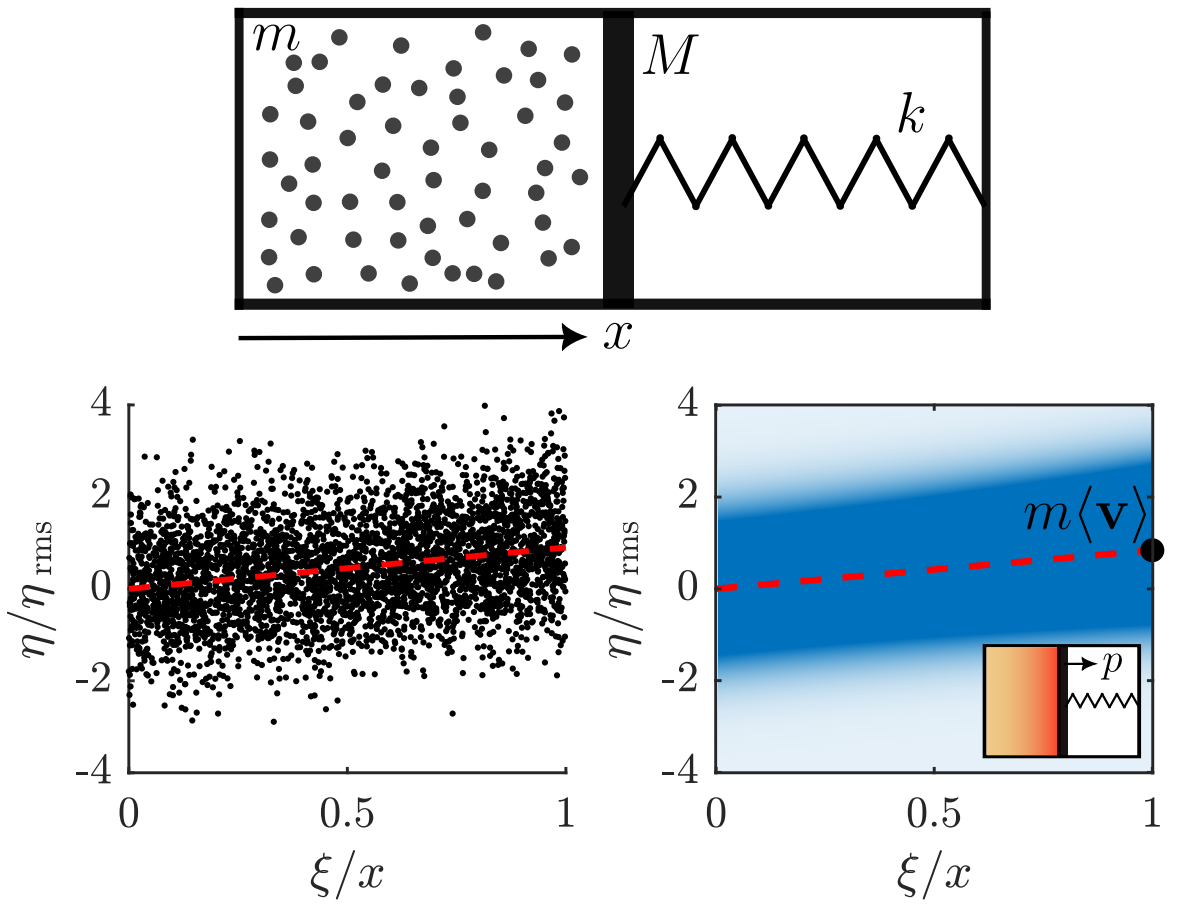

**Fig. 5.** The MBOA applied to a classical system. The slow DOF is a heavy piston of mass $M$, which interacts with a gas of light particles each of mass $m$, shown in the top panel. When the piston is in motion, the fast particles reach a state of moving equilibrium with position-momentum correlations. In particular, particles closer to the piston have a higher average kinetic energy than those near the static wall. The bottom panels show a snapshot of the distribution of fast particles in phase space from exact numerical simulation (left), as compared to the MBOA prediction (right).

Note the similarity of this expression to Eq. (7). Here, the $A_x^{(i)}(\xi_i,\eta_i,x)$ are the (classical) adiabatic gauge potentials, which are generators of canonical transformations satisfying ${\rm d}\xi'_i/{\rm d}x = \{A_x^{(i)}(\xi'_i,\eta'_i,x),\xi'_i\}$ and ${\rm d}\eta'_i/{\rm d}x = \{A_x^{(i)}(\xi'_i,\eta'_i,x),\eta'_i\}$. It is straightforward to check that the generator of phase space dilations Eq. (26) is given by $\mathbf{A}_x(x) = \vec\xi{\cdot}\vec\eta/x = \vec\xi\,'(x)\cdot\vec\eta\,'(x)/x$. Then, Eq. (26) and Eq. (27) define a valid canonical transformation for all phase space variables. As before, we remark that the canonical $p$ is a collective variable that depends on both piston and gas DOFs.

The moving Hamiltonian is obtained by performing the substitution $\mathbf{q} = p - \mathbf{A}_x(x)$ in Eq. (22),

$$\mathbf{H}^{\rm M}(x,p) = \frac{(p-\mathbf{A}_x(x))^2}{2M} + \frac{1}{2}kx^2 + \mathbf{H}_{\rm int}(x) \qquad [28]$$

This Hamiltonian is precisely the classical version of Eq. (10). In the MBOA, we assume that the fast DOFs adiabatically follow equilibrium with respect to $\mathbf{H}^{\rm M}(x,p)$ at the instantaneous values of $(x,p)$.

For a piston initially at rest, taking the $N\to\infty$ limit while keeping the total mass of the gas $mN$ fixed, the emergent Hamiltonian reads,

$$\mathcal{H}^{\rm MBO}(x,p) = \frac{p^2}{2(M+\kappa)} + \frac{1}{2}kx^2 + E_0^{\rm fast}\left(\frac{x_0}{x}\right)^2, \qquad [29]$$

with $\kappa = mN/3$. The calculation is shown explicitly in Sec. 9 of S.I. Within the MBOA, the mass of the piston is thus renormalized as $M \to M+\kappa$, with each of the fast particles contributing a third of its own mass (see also Refs. (30, 41)). The mass renormalization modifies the frequency of oscillations of the piston about its equilibrium position, as shown in Fig. S7.

The MBOA predicts velocity and energy gradients for the fast particles in the gas. Consider displacing the piston away from its equilibrium position $x_*$. While the piston is in motion, the fast particles reach a state of moving equilibrium which features position-momentum correlations.

The lower-left panel of Fig. 5 shows a snapshot of the phase space distribution of fast positions $\xi_i$ and momenta $\eta_i$, for a numerical simulation carried out with $N = 4000$ and $mN/M = 1/5$. Each dot corresponds to one particle. The lower-right panel shows the MBOA prediction. Note that particles closer to the piston have, on average, a higher mean momentum and kinetic energy. The skew originates from the $p\vec{\xi}\cdot\vec{\eta}$ coupling in $\mathbf{H}^{\mathrm{M}}$, which favors larger/smaller values of the product $\vec{\xi}\cdot\vec{\eta}$ depending on the sign of $p$. In the lower inset, we sketch how the local kinetic energy of the gas varies in space. The particles nearest to the piston have a mean velocity equal to the average velocity of the piston in the MBO state $\langle\mathbf{v}\rangle = \langle\mathbf{q}\rangle/M = p/(M+\kappa)$ (see also the following discussion in Sec. D). Thus, the gradient is large when the speed of the piston is comparable to the rms speed of the gas particles.

As the piston oscillates, the energy gradient across the gas repeatedly vanishes and reforms in a near-reversible, long-lived state with minimal entropy production that is reminiscent of synchronized, low-dissipation states in periodically driven systems (47, 48). Interestingly, the synchronized state appears to violate thermodynamic (Boltzmann) entropy monotonic increase in the lab frame. The contradiction is removed if one instead analyzes the behavior of the thermodynamic entropy *in the moving frame*, which is nearly constant and slowly increasing in time, as shown in Fig. S7. As a final remark, in both the quantum and classical examples, the motion-induced interaction term is highly non-local leading to global synchronization between spins/particles in the MBO state, resembling the emergence of strong non-local effects in transport-induced non-equilibrium steady states (49).

**D. Momentum Fluctuations and Quantum Metric.** Eq. (7) and Eq. (27) encode an interesting relation between the fluctuations of the slow DOF and the Fubini-Study quantum metric tensor, which in turn is related to the quantum Fisher information. In particular, the fluctuations of the physical momentum $\hat{\mathbf{q}}$ in an MBO state are given by,

$$\sigma_{\mathbf{q}}^2(x,p) = \left\langle \psi_n^{\mathrm{MBO}} \right| \hat{\mathbf{A}}_x^2 \left| \psi_n^{\mathrm{MBO}} \right\rangle_c \equiv g_n^{\mathrm{M}}(x,p) \qquad [30]$$

where $g_n^{\mathrm{M}} \equiv \langle \hat{\mathbf{A}}_x^2 \rangle_c = \langle \hat{\mathbf{A}}_x^2 \rangle - \langle \hat{\mathbf{A}}_x \rangle^2$ is the Fubini-Study metric tensor. The superscript M highlights that the average is taken with respect to the MBO state, and hence $\sigma_{\mathbf{q}}$ implicitly depends on both $x$ and $p$. For classical systems, the analogous relation is found by removing hats and replacing quantum averages with phase space averages.

Let us understand the physical meaning of Eq. (30) in the context of the piston example. For a given trajectory with specific initial positions and momenta of the fast particles, the physical momentum $q$ of the piston evolves in time in a noisy fashion due to random collisions with the particles in the gas. In contrast, the canonical momentum $p = q + \vec{\xi}\cdot\vec{\eta}/x$ remains a smooth function of time. Indeed, it is easy to check that $p$ does not change during all particle-wall or particle-piston collisions. This is illustrated in Fig. 6(a), where the solid black line additionally shows the expectation value of $\mathbf{q}(x,p)$ in the instantaneous MBO equilibrium state $\langle\mathbf{q}\rangle = p/(1+\frac{\kappa}{M})$, and provides a good fit to the exact time-averaged $q(t)$. The two dotted black lines each show one standard deviation away from the mean $\mathbf{q}_\pm = \langle\mathbf{q}\rangle \pm \sigma_{\mathbf{q}}$. The fluctuations are directly related to the average particle energy and the mass ratio as $\sigma_{\mathbf{q}}^2 = \frac{E^{\mathrm{fast}}}{N}\frac{2\kappa}{1+\kappa/M}$. These microscopic fluctuations provide a measurable contribution to the total kinetic energy of the piston, which is not simply proportional to the square of the mean piston momentum,

$$K \equiv \frac{\langle\mathbf{q}^2\rangle}{2M} = \frac{\langle\mathbf{q}\rangle^2}{2M} + \frac{g_n^{\mathrm{M}}(x,p)}{2M}. \qquad [31]$$

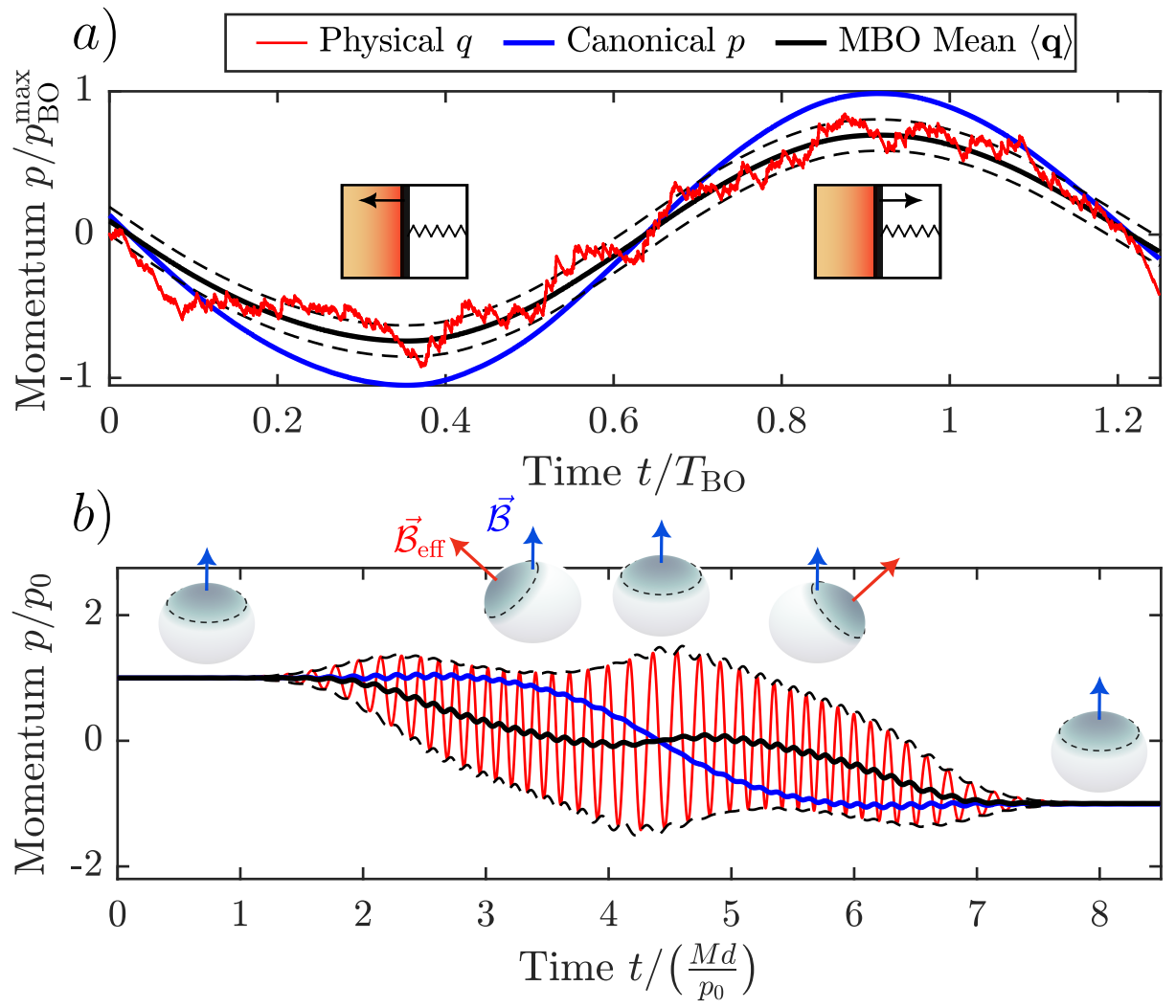

**Fig. 6.** Fluctuations of the physical momentum $q$ of the slow DOF and its relation to the Fubini-Study metric tensor. a) The physical momentum ($q$, red) and the canonical momentum ($p$, blue) during one oscillation of the piston about its equilibrium position. The time evolution of $q$ appears noisy due to the random collisions between the piston and the particles in the gas, while $p$ is a smooth function of time. The two dotted black lines show the MBO expectation values $\langle\mathbf{q}\rangle \pm \sigma_{\mathbf{q}}$, where $\sigma_{\mathbf{q}}^2 = g^{\mathrm{M}}(x,p)$ is the variance of $\mathbf{q}$ about its mean, and is given by the Fubini-Study metric tensor evaluated in the MBO equilibrium state. b) Reflection of a spin-1/2 particle from a region of rotating magnetic field. The curves correspond to a single semi-classical initial condition evolved within the TWA. As in the top panel, $q$ exhibits large fluctuations which manifest as quantum noise when averaging over different spin initial conditions. Numerical parameters are given in Secs. 6 and 9 of S.I.

In the BO limit, this correction is a classical analog of the diagonal Born-Huang (metric) correction. However, it is sub-extensive in $N$ and hence does not appear in Eq. (29).

These fluctuations are also manifest as quantum noise in the spin-1/2 model. Fig 6 shows a full Truncated Wigner simulation of an experiment where the particle reflects from a region of non-zero field rotation, as described in Sec. 2(A). The spin is initially polarized with the field, and its time evolution is sampled from four different initial conditions in a discrete phase-space (see Ref. (50) and Sec. 6 of S.I). Fig. 6 shows only a single trajectory corresponding to one of the initial conditions. As the spin enters a region of non-zero rotation $\theta'$, its physical momentum $q$ undergoes fast oscillations on the same timescale as the precession of the spin about the effective magnetic field. The canonical momentum $p = q - \hbar\theta'(x)S_x$, which is a collective variable of both spin and particle, instead varies comparatively smoothly in time. The MBO expectation value $\langle\mathbf{q}\rangle = p - \frac{\hbar\theta'}{2}\sin\phi$, with $\phi(x,p)$ the non-adiabatic tilt angle Eq. (17) provides a good fit to the time-averaged $q(t)$. Interestingly, the TWA is frame-dependent and it is important to work in the moving frame, where it is essentially exact.

## 3. Methods

In this section, we describe in detail the steps to implement the MBOA. A complete derivation of all results starting from the path integral is provided in Sec. 2 of S.I.

**Moving Frame Transformation and Dressed Operators.** The first step is to compute the general form of Weyl symbols in the moving frame. Let $\hat{O}$ be an arbitrary operator with Weyl symbol $\hat{\mathbf{O}}(x,p)$ and $\hat{U}_{\text{BO}}(\hat{x})$ the BO unitary with Weyl symbol $\hat{\mathbf{U}}_{\text{BO}}(x)$. The moving frame operator $\hat{O}'_{\text{M}}$ is found by conjugation under the BO unitary,

$$\hat{O}'_{\text{M}} \equiv \hat{U}^{\dagger}_{\text{BO}} \hat{O} \hat{U}_{\text{BO}}. \quad [32]$$

The Weyl symbol of Eq. (32) defines $\hat{\mathbf{O}}'_{\text{M}}$, the operator *dressed* by $\hat{U}_{\text{BO}}$. The dressed operator can be found through use of the Moyal product (28),

$$\hat{\mathbf{O}}'_{\text{M}}(x,p) = \hat{\mathbf{U}}^{\dagger}_{\text{BO}}(x) \star \hat{\mathbf{O}}(x,p) \star \hat{\mathbf{U}}_{\text{BO}}(x), \quad [33]$$

where the star symbol is defined as,

$$(f \star g)(x,p) = f(x,p) e^{\frac{i\hbar}{2}\Lambda} g(x,p), \quad [34]$$

with $\Lambda \equiv (\overleftarrow{\partial}_x \overrightarrow{\partial}_p - \overleftarrow{\partial}_p \overrightarrow{\partial}_x)$ the Poisson bracket operator.

Dressed operators can likewise be expressed in the original lab frame coordinates by performing the inverse unitary rotation in $\mathbb{H}^Q$,

$$\hat{\mathbf{O}}^{\text{M}}(x,p) = \hat{\mathbf{U}}_{\text{BO}}(x) \hat{\mathbf{O}}'_{\text{M}}(x,p) \hat{\mathbf{U}}^{\dagger}_{\text{BO}}(x), \quad [35]$$

Throughout the text, unprimed quantities refer to operators in the lab frame. For example, Eq. (10) is the Hamiltonian dressed by the BO unitary in the lab frame basis.

Let us emphasize that dressed operators, like any operators in quantum mechanics, can be represented in any frame without affecting the results for any physical observable. In this way, the pair $\hat{\mathbf{O}}^{\text{M}}$, and $\hat{\mathbf{O}}'_{\text{M}}$ represent the same operator written in different bases. For the spin example, the prime operators correspond to the choice of the Pauli matrices with $z$-axis along the instantaneous field, while for the non-prime lab-frame operators the coordinate system is fixed in space. However, the dressed $\hat{\mathbf{O}}^{\text{M}}$ is a different operator than the bare $\hat{\mathbf{O}}$, i.e. these operators are not related by some basis transformation. In constructing $\hat{\mathbf{O}}^{\text{M}}$, we have first conjugated the operator $\hat{O}$ by $\hat{U}_{\text{BO}}$, then taken its Wigner-Weyl transform, and afterwards performed the inverse unitary rotation in $\mathbb{H}^Q$ only. Because these two steps do not commute, the dressed operator $\hat{\mathbf{O}}^{\text{M}}$ will generically be different from $\hat{\mathbf{O}}$. The choice of the unitary $\hat{U}_{\text{BO}}$ thus sets how slow operators including $\hat{x}$ and $\hat{q}$ are dressed by the fast quantum DOFs.

For any operator that is polynomial in $\hat{q}$, the dressed $\hat{\mathbf{O}}^{\text{M}}$ can be straightforwardly found by expanding out the Moyal product Eq. (33), which truncates at finite order in $\hbar$. This yields the expressions for $\hat{\mathbf{q}}^{\text{M}}$, $\hat{\mathbf{x}}^{\text{M}}$, and $\hat{\mathbf{H}}^{\text{M}}$ given in Eq. (7) and Eq. (10). Concrete calculations and techniques for evaluating more complicated Weyl symbols, including Wigner functions, are given in Sec. 3 and Sec. 4 of S.I.

**Truncated Wigner Approximation.** To compute the time-evolution of an observable $\langle\hat{\Omega}(t)\rangle$ in the MBOA, one propagates the matrix elements of $\hat{\mathbf{\Omega}}^{\text{M}}$ according to classical trajectories. The eigenvalues $\mathcal{H}^{\text{MBO}}_n$ of the moving Hamiltonian $\hat{\mathbf{H}}^{\text{M}}$ provide classical Hamiltonians that generate the time-evolution of each trajectory. However, unlike the conventional TWA, one uses not only the eigenvalues, but also linear combinations of them, to describe coherence effects between different MBO states. The algorithm for computing $\hat{\mathbf{\Omega}}^{\text{M}}(t)$ is as follows:

Consider the matrix elements of the dressed observable $\hat{\mathbf{\Omega}}^{\text{M}}_{mn}(x,p)$ in the basis of eigenstates of $\hat{\mathbf{H}}^{\text{M}}(x,p)$. The matrix element $\hat{\mathbf{\Omega}}^{\text{M}}_{mn}$ evolves in time using the Hamiltonian $\mathcal{H}^{\text{MBO}}_{mn}$, which depends on both indices $m$ and $n$,

$$\mathcal{H}^{\text{MBO}}_{mn}(x,p) \equiv \frac{\mathcal{H}^{\text{MBO}}_m(x,p) + \mathcal{H}^{\text{MBO}}_n(x,p)}{2}. \quad [36]$$

Specifically, denote $x_{mn}(t)$ and $p_{mn}(t)$ as the solutions to the equations of motion satisfying $x_{mn}(0) = x_0$, $p_{mn}(0) = p_0$, and,

$$\dot{x}_{mn} = \frac{\partial \mathcal{H}^{\text{MBO}}_{mn}}{\partial p}, \qquad \dot{p}_{mn} = -\frac{\partial \mathcal{H}^{\text{MBO}}_{mn}}{\partial x}. \quad [37]$$

Then, the time-evolved Weyl symbol $\hat{\mathbf{\Omega}}^{\text{M}}(t)$ is given by,

$$\hat{\mathbf{\Omega}}^{\text{M}}_{mn}(t) = \mathrm{e}^{i\Phi_{mn}(t)} \hat{\mathbf{\Omega}}^{\text{M}}_{mn}(x_{mn}(t), p_{mn}(t)), \quad [38]$$

where we have defined the time-dependent phase,

$$\Phi_{mn}(t) = \frac{1}{\hbar}\int_0^t \mathrm{d}\tau \big[\mathcal{H}^{\text{MBO}}_m(x_{mn}(\tau), p_{mn}(\tau)) - \mathcal{H}^{\text{MBO}}_n(x_{mn}(\tau), p_{mn}(\tau))\big]. \quad [39]$$

The phase describes interference between different MBO eigenstates.

Finally, the expectation value of the observable $\hat{\Omega}$ is evaluated by averaging over a distribution of initial conditions,

$$\langle\hat{\Omega}(t)\rangle \approx \int \frac{\mathrm{d}x_0 \mathrm{d}p_0}{2\pi\hbar} \mathrm{Tr}\Big\{\hat{\mathbf{W}}^{\text{M}}(x_0,p_0)\hat{\mathbf{\Omega}}^{\text{M}}(t)\Big\}, \quad [40]$$

where $\hat{\mathbf{W}}^{\text{M}}$ is the dressed Weyl symbol of the density matrix. In order to implement the MBOA, one thus has to solve $D(D+1)/2$ classical equations of motion, where $D$ is the dimensionality of the fast Hilbert space $\mathbb{H}^Q$. However, often one is interested in dynamics starting from a single MBO state, for which the Wigner function expressed in the energy eigenbasis has only one non-zero matrix element. In that case, one needs only to propagate trajectories using a single eigenvalue $\mathcal{H}^{\text{MBO}}_n$ as discussed in Sec. 1. Indeed, note that for diagonal elements $m = n$ we have $\mathcal{H}^{\text{MBO}}_{nn} = \mathcal{H}^{\text{MBO}}_n$ and $\Phi_{nn} = 0$. However, the off-diagonal matrix elements are relevant when the initial density matrix contains coherent superpositions of the MBO states. They are necessary to describe phenomena that manifestly rely on the phase information between MBO surfaces like, e.g., the precession of the spin about the magnetic field (Sec. 6 of S.I.), or the coherent transient oscillations of $\langle\hat{q}(t)\rangle$ examined in Sec. 2.

## 4. Conclusion

We have developed a mixed quantum-classical framework, termed the Moving Born-Oppenheimer Approximation (MBOA), to describe both quantum and classical systems with slow and fast DOFs. The MBOA assumes fast DOFs equilibrate to a state that is non-perturbatively influenced by the motion of the slow DOFs, analogous to mechanical equilibrium in an accelerated or rotating frame. We derive our formalism using the path integral and an extension of the Truncated Wigner Approximation to matrix-valued Weyl symbols Eq. (40), with full details given in Sec. 2 of S.I.

The MBOA uncovers interesting dynamics beyond the reach of standard frameworks. As a specific example, we analyze the motion of a particle with spin degrees of freedom in a spatially varying magnetic field. The MBOA accurately describes non-trivial phase space trajectories, the generation of entangled and spin-squeezed states, and quantum interference effects manifest in coherent momentum oscillations and spin precession. We also apply the formalism to a purely classical problem of a heavy piston coupled to a gas of non-interacting fast particles. The MBOA correctly predicts momentum gradients in the gas and its dissipationless synchronization with the motion of the piston.

The MBOA should quantitatively capture coherent beyond-BOA dynamics and provide a means to generate quantum entanglement in a variety of settings. For example, it could describe wavepacket dynamics in crystals with modulated structures or slowly varying external potentials, non-perturbative effects in pump-probe experiments, phonon-mediated electron-electron interactions (as in the Cooper problem), entanglement generation in chemistry, and provide a framework for non-equilibrium quantum geometry.

Another avenue for future work is to systematically incorporate incoherent transitions between the dressed MBO states, which may be derivable from the path integral by going to higher orders in the Moyal expansion of the dressed Hamiltonian. These processes become important when MBO levels are dense, as the AGP acquires non-local terms that induce dissipation and drive the system toward equilibration. For example, the synchronized state of the piston and the gas appears to relax to local thermal equilibrium at long times. This is accompanied by an increase in the moving frame entropy of the fast particles, as seen in Fig. S7, and is beyond the MBOA.

*Note added:* After this work was completed, we became aware of a recent paper, Ref. (17), which develops a framework similar to the MBOA.

**ACKNOWLEDGMENTS.** We acknowledge fruitful discussions with D. Coker and P. Claeys in the early stages of this work, as well as valuable feedback from B. Levine and J. Richardson. A.P. and B.B. were supported by NSF Grant DMR-2103658 and the AFOSR Grant FA9550-21-1-0342. A.C. and B.B. acknowledge the hospitality of the Max Planck Institute for Complex Systems. B.B. thanks G. Delfino and N. Karve for useful discussions.

# Supporting Information for

## The Moving Born-Oppenheimer Approximation

**Bernardo Barrera, Daniel P. Arovas, Anushya Chandran and Anatoli Polkovnikov**

**Anatoli Polkovnikov.**
**E-mail: asp28@bu.edu**

### This PDF file includes:

Supporting text
Figs. S1 to S7
Table S1
SI References

## Supporting Information Text

## 1. Glossary of Notations and Definitions

This section introduces some convenient notation and definitions which are used throughout the Supplementary Information. The reader can skip directly to Sec. 2 for the derivation of the MBOA and refer back to Sec. 1, particularly Table S1, as a glossary of notation.

Let $\hat{U}$ be a unitary transformation parameterized as,

$$\hat{U}(\lambda) \equiv e^{-\frac{i}{\hbar}\lambda\hat{G}}, \tag{1}$$

where $\hat{G}$ is a Hermitian operator with Weyl symbol $\hat{\mathbf{G}}(x,p)$, termed the "generating function". For the purposes of this work, we restrict ourselves only to transformations where $\hat{G}$ is a well-defined operator with a well-defined $\hat{\mathbf{G}}$. The Weyl symbol of $\hat{U}(\lambda)$ is denoted as $\hat{\mathcal{U}}(x,p;\lambda) \equiv \hat{\mathcal{U}}(\lambda)$. Note the use of calligraphic (and not bold) font. The notation $\hat{\mathbf{U}}$ is instead reserved for a related quantity, $\hat{\mathbf{U}}(x,p;\lambda) \equiv \hat{\mathbf{U}}(\lambda) \equiv e^{-\frac{i}{\hbar}\lambda\hat{\mathbf{G}}(x,p)}$. In general, $\hat{\mathcal{U}}(\lambda) \neq \hat{\mathbf{U}}(\lambda)$, except when the generating function $\hat{\mathbf{G}}$ is only dependent on one of the phase space coordinates $x$ or $p$. This is an important distinction. While the former is unitary under the $\star$-product, $\hat{\mathcal{U}}^\dagger(\lambda) \star \hat{\mathcal{U}}(\lambda) = \hat{\mathbf{I}}$, the latter is unitary under the normal product, $\hat{\mathbf{U}}^\dagger(\lambda)\hat{\mathbf{U}}(\lambda) = \hat{\mathbf{I}}$. Omitting the $\lambda$ argument is taken to mean evaluation at $\lambda = 1$. That is, $\hat{U}(\lambda=1) \equiv \hat{U}$, $\hat{\mathcal{U}}(\lambda=1) \equiv \hat{\mathcal{U}}$, and $\hat{\mathbf{U}}(\lambda=1) \equiv \hat{\mathbf{U}}$.

Given some operator $\hat{\mathbf{Q}}$ in $\mathbb{H}^Q$, primes are used to denote its conjugation by $\hat{\mathbf{U}}(\lambda)$,

$$\hat{\mathbf{Q}}'(\lambda) \equiv \hat{\mathbf{U}}^\dagger(\lambda)\hat{\mathbf{Q}}\hat{\mathbf{U}}(\lambda). \tag{2}$$

As before, omitting the $\lambda$ argument means evaluating at $\lambda = 1$, such that $\hat{\mathbf{Q}}'(\lambda=1) \equiv \hat{\mathbf{Q}}'$.

Given some operator $\hat{O}$ in $\mathbb{H}^Q \otimes \mathbb{H}^C$, primes *and subscript* U are used to denote its conjugation by $\hat{U}(\lambda)$,

$$\hat{O}'_U(\lambda) \equiv \hat{U}^\dagger(\lambda)\hat{O}\hat{U}(\lambda). \tag{3}$$

The subscript emphasizes that, while $\hat{\mathbf{Q}}'$ and $\hat{\mathbf{Q}}$ are the same operators up to a basis choice, the Weyl symbol $\hat{\mathbf{O}}'_U$ is fundamentally a different operator than $\hat{\mathbf{O}}$. The latter are not related by a unitary transformation in $\mathbb{H}^Q$.

Taking the Wigner-Weyl transform of both sides of Eq. (3), we obtain two equivalent representations for the corresponding Weyl symbol,

$$\hat{\mathbf{O}}'_U(\lambda) = \hat{\mathcal{U}}^\dagger(\lambda) \star \hat{\mathbf{O}} \star \hat{\mathcal{U}}(\lambda) \equiv \hat{\mathbf{U}}^\dagger(\lambda)\hat{\mathbf{O}}_U(\lambda)\hat{\mathbf{U}}(\lambda). \tag{4}$$

The dressed operators $\hat{\mathbf{O}}_U$ are those which, by definition, are related to $\hat{\mathbf{O}}'_U$, i.e. Weyl symbols of $O'_U$, by conjugation under $\hat{\mathbf{U}}(\lambda)$. In general, the dressed $\hat{\mathbf{O}}_U$ is a different operator than the bare $\hat{\mathbf{O}}$. Using Eq. (4), $\hat{\mathbf{O}}_U$ can be found by first computing the Weyl symbol of the rotated operator $\hat{\mathbf{O}}'_U(\lambda)$ and then applying the inverse unitary rotation,

$$\hat{\mathbf{O}}_U(x,p;\lambda) = \hat{\mathbf{U}}(x,p;\lambda)\hat{\mathbf{O}}'_U(x,p;\lambda)\hat{\mathbf{U}}^\dagger(x,p;\lambda).$$

In the special case that the unitary corresponds to the BO unitary $\hat{U} = \hat{U}_{\mathrm{BO}}$, we equivalently use superscript M to denote dressing: $\hat{\mathbf{O}}_{U_{\mathrm{BO}}} \equiv \hat{\mathbf{O}}^{\mathrm{M}}$. In this way, the notation for dressed operators in the moving frame is consistent with the main text Eq. 7 and Eq. 10. The two notations are used interchangeably.

There are two unitary transformations of particular interest: the BO unitary $\hat{U}_{\mathrm{BO}}$ with generating function $\hat{\mathbf{G}}_{\mathrm{BO}}(x)$, and the MBO unitary $\hat{U}_{\mathrm{MBO}}$ with generating function $\hat{\mathbf{G}}_{\mathrm{MBO}}(x,p)$. The AGPs associated to the BO unitary are denoted by $\hat{\mathbf{A}}$, while those associated to the MBO one are denoted by $\hat{\mathbf{B}}$. That is,

$$\begin{aligned}
\hat{\mathbf{A}}_x(\lambda) &\equiv \big(i\hbar\partial_x\hat{\mathbf{U}}_{\mathrm{BO}}(\lambda)\big)\hat{\mathbf{U}}^\dagger_{\mathrm{BO}}(\lambda),\\
\hat{\mathbf{B}}'_x(\lambda) &\equiv \big(i\hbar\partial_x\hat{\mathbf{U}}_{\mathrm{MBO}}(\lambda)\big)\hat{\mathbf{U}}^\dagger_{\mathrm{MBO}}(\lambda),\\
\hat{\mathbf{B}}'_p(\lambda) &\equiv \big(i\hbar\partial_p\hat{\mathbf{U}}_{\mathrm{MBO}}(\lambda)\big)\hat{\mathbf{U}}^\dagger_{\mathrm{MBO}}(\lambda).
\end{aligned} \tag{5}$$

Omitting $\lambda$ means evaluating at 1: $\hat{\mathbf{A}}_x(x;\lambda=1) \equiv \hat{\mathbf{A}}_x(x)$ and $\hat{\mathbf{B}}'_{x,p}(x,p;\lambda=1) \equiv \hat{\mathbf{B}}'_{x,p}(x,p)$. The primes are used in the definition of $\hat{\mathbf{B}}'_{x,p}$ because the unitary $\hat{U}_{\mathrm{MBO}}$ is applied after $\hat{U}_{\mathrm{BO}}$ (see Eq. (49)). In this way, $\hat{\mathbf{B}}_{x,p} = \hat{\mathbf{U}}_{\mathrm{BO}}\hat{\mathbf{B}}'_{x,p}\hat{\mathbf{U}}^\dagger_{\mathrm{BO}}$.

Let us finally add a few words about the physical meaning of the operators $\hat{\mathbf{B}}'_x$ and $\hat{\mathbf{B}}'_p$. We set $\lambda = 1$ for the rest of this discussion. Because the unitary $\hat{\mathbf{U}}_{\mathrm{MBO}}$ diagonalizes the dressed Hamiltonian $\hat{\mathbf{H}}'_{U_{\mathrm{BO}}}$, then $\hat{\mathbf{B}}'_x$ serves as the adiabatic gauge potential with respect to $x$ and similarly $\hat{\mathbf{B}}'_p$ is the AGP with respect to $p$ *. As such, they are associated with the conserved operators $\hat{\mathbf{F}}'_x$ and $\hat{\mathbf{F}}'_p$ (2),

$$[\hat{\mathbf{F}}'_x, \hat{\mathbf{H}}'_{U_{\mathrm{BO}}}] = 0, \quad \hat{\mathbf{F}}'_x = \partial_x\hat{\mathbf{H}}'_{U_{\mathrm{BO}}} + \frac{i}{\hbar}[\hat{\mathbf{B}}'_x, \hat{\mathbf{H}}'_{U_{\mathrm{BO}}}] \tag{6}$$

$$[\hat{\mathbf{F}}'_p, \hat{\mathbf{H}}'_{U_{\mathrm{BO}}}] = 0, \quad \hat{\mathbf{F}}'_p = \partial_p\hat{\mathbf{H}}'_{U_{\mathrm{BO}}} + \frac{i}{\hbar}[\hat{\mathbf{B}}'_p, \hat{\mathbf{H}}'_{U_{\mathrm{BO}}}] \tag{7}$$

*Because $\hat{\mathbf{H}}_{U_{\mathrm{BO}}} = \hat{\mathbf{H}}^{\mathrm{M}}$ is the moving Hamiltonian, in the literature such transformations are called superadiabatic (1). Therefore $\hat{\mathbf{B}}_x$ and $\hat{\mathbf{B}}_p$ can be termed as superadiabatic gauge potentials.

The entries of these conserved operators are nothing but derivatives of the eigenvalues of $\hat{\mathbf{H}}_{U_{\rm BO}}$ with respect to $x$ and $p$ and hence have the meaning (up to a sign) of generalized forces. For the piston example, $-\langle 0|\hat{\mathbf{F}}_x|0\rangle$, where $|0\rangle$ is the ground state of $\hat{\mathbf{H}}_{U_{\rm BO}}$, is the ground state pressure in the moving frame. Using that,

$$\partial_x \hat{\mathbf{H}}'_{U_{\rm BO}} = \hat{\mathbf{U}}^\dagger_{\rm BO}\left(\partial_x \hat{\mathbf{H}}_{U_{\rm BO}} + \frac{i}{\hbar}[\hat{\mathbf{A}}_x, \hat{\mathbf{H}}_{U_{\rm BO}}]\right)\hat{\mathbf{U}}_{\rm BO} \tag{8}$$

we find,

$$\begin{aligned}\hat{\mathbf{F}}_x &= \partial_x \hat{\mathbf{H}}_{U_{\rm BO}} + \frac{i}{\hbar}[\hat{\mathbf{A}}_x + \hat{\mathbf{B}}_x, \hat{\mathbf{H}}_{U_{\rm BO}}],\\ \hat{\mathbf{F}}_p &= \partial_p \hat{\mathbf{H}}_{U_{\rm BO}} + \frac{i}{\hbar}[\hat{\mathbf{B}}_p, \hat{\mathbf{H}}_{U_{\rm BO}}].\end{aligned} \tag{9}$$

Thus, the quantity $\hat{\mathbf{A}}_x + \hat{\mathbf{B}}_x$ (and similarly for $p$) plays the role of the AGP for the moving Hamiltonian $\hat{\mathbf{H}}_{U_{\rm BO}}$ in the lab frame.

The table below summarizes important notations:

| | |
|---|---|
| $\hat{\Omega}$ | Operator in $\mathbb{H}^Q \otimes \mathbb{H}^C$. |
| $\mathbf{\Omega}$ | Weyl symbol in $\mathbb{H}^Q$ |
| $\hat{U}$ | Unitary transformation in $\mathbb{H}^Q \otimes \mathbb{H}^C$: $\hat{U}(\lambda) \equiv e^{-\frac{i}{\hbar}\lambda \hat{G}}$ |
| $\mathbf{G}$ | Generating function |
| $\hat{\mathcal{U}}$ | Weyl symbol of $\hat{U}(\lambda)$ |
| $\hat{\mathbf{U}}$ | Unitary transformation in $\mathbb{H}^Q$: $\hat{\mathbf{U}}(x,p;\lambda) \equiv e^{-\frac{i}{\hbar}\lambda \hat{\mathbf{G}}(x,p)}$ |
| $\hat{\mathbf{A}}_x$ | BO AGP $\hat{\mathbf{A}}_x(\lambda) \equiv \left(i\hbar\partial_x \hat{\mathbf{U}}_{\rm BO}(\lambda)\right)\hat{\mathbf{U}}^\dagger_{\rm BO}(\lambda)$ |
| $\hat{\mathbf{B}}'_{x,p}$ | MBO AGPs $\hat{\mathbf{B}}'_{x,p}(\lambda) \equiv \left(i\hbar\partial_{x,p} \hat{\mathbf{U}}_{\rm MBO}(\lambda)\right)\hat{\mathbf{U}}^\dagger_{\rm MBO}(\lambda)$ |
| $\hat{\mathbf{F}}_{x,p}$ | Generalized forces |

**Table S1. Glossary of notation and important definitions used throughout the Supplementary Information.**

## 2. Derivation of the MBOA

Here, we provide a formal derivation of all results presented in Sec. 3 of the main text. In this work, we restrict to examples where the slow coordinate $x$ is a one-dimensional, non-compact variable.

**Path Integral.** Consider the time evolution of a general observable $\hat{\Omega} = \Omega(\hat{x}, \hat{q})$ in the Heisenberg representation,

$$\langle \hat{\Omega}(t)\rangle = \mathrm{Tr}\left\{\hat{\rho}e^{\frac{i}{\hbar}t\hat{H}}\hat{\Omega}e^{-\frac{i}{\hbar}t\hat{H}}\right\}, \tag{10}$$

where $\hat{\rho}$ is the density matrix describing the initial state of a system of fast and slow DOFs.

We begin by deriving a path integral representation for Eq. (10). The time evolution operator is decomposed as an infinite product,

$$e^{-i\hat{H}t/\hbar} = \lim_{N\to\infty}\left(1 - \frac{i\Delta t}{\hbar}\hat{H}\right)^N, \tag{11}$$

with $\Delta t = t/N$, and in the usual manner, resolutions of the identity are inserted between time steps. The system is composed of a slow coordinate variable (C) coupled to a set of fast quantum variables (Q), for which the full Hilbert space is written as $\mathbb{H} = \mathbb{H}^C \otimes \mathbb{H}^Q$. The identity is resolved as,

$$\hat{I} = \sum_m \int \mathrm{d}x\, |x, m\rangle\,\langle x, m|\,, \tag{12}$$

where $|m\rangle$ labels an arbitrary fixed basis in $\mathbb{H}^Q$.

The Weyl symbol of the Hamiltonian $\hat{\mathbf{H}}$ appears naturally in the path integral through the use of the following identity,

$$\langle x, m|\,\hat{H}\,|y, n\rangle = \int_{-\infty}^{\infty}\frac{\mathrm{d}p}{2\pi\hbar}e^{\frac{i}{\hbar}p(x-y)}\hat{\mathbf{H}}_{mn}\left(\frac{x+y}{2}, p\right), \tag{13}$$

where $\hat{\mathbf{H}}_{mn}(x,p)$ is defined through the partial Wigner transform,

$$\hat{\mathbf{H}}_{mn}(x,p) = \int_{-\infty}^{\infty}\mathrm{d}\xi e^{\frac{i}{\hbar}p\xi}\left\langle x - \frac{\xi}{2}, m\right|\hat{H}\left|x + \frac{\xi}{2}, n\right\rangle. \tag{14}$$

In the continuum $N \to \infty$ limit, we obtain the path integral expression for the time-evolution operator,

$$\left\langle x^b_N, m_N\right| e^{-\frac{i}{\hbar}t\hat{H}}\left|x^b_0, m_0\right\rangle = \int D[x^b(\tau)]D[p^b(\tau)]e^{\frac{i}{\hbar}\int_0^t p^b(\tau)\dot{x}^b(\tau)\mathrm{d}\tau}\left[\mathcal{T}_t\exp\left\{-\frac{i}{\hbar}\int_0^t \mathrm{d}\tau \hat{\mathbf{H}}(x^b(\tau), p^b(\tau))\right\}\right]_{m_N m_0}$$

where $\mathcal{T}_t$ denotes the time-ordering operator, and the superscript $b$ stands for "backward" time evolution. Likewise, the path integral for $e^{i\hat{H}t/\hbar}$ sums over "forward" paths,

$$\left\langle x_0^f, m_0\right| e^{\frac{i}{\hbar}t\hat{H}} \left|x_N^f, m_N\right\rangle = \int D[x^f(\tau)]D[p^f(\tau)]e^{-\frac{i}{\hbar}\int_0^t p^f(\tau)\dot{x}^f(\tau)\mathrm{d}\tau}\left[\mathcal{T}_t \exp\left\{-\frac{i}{\hbar}\int_0^t \mathrm{d}\tau \hat{\mathbf{H}}(x^f(\tau), p^f(\tau))\right\}\right]^\dagger_{m_0 m_N}$$

Next, it is convenient to switch to mean and difference coordinates between the forward and backward paths,

$$x = \frac{x^f + x^b}{2} \quad p = \frac{p^f + p^b}{2} \quad \delta x = x^f - x^b \quad \delta p = p^f - p^b. \qquad [15]$$

The mean coordinates play the role of the classical path, while the differences encode the quantum fluctuations (3). Combining the forward and backward expressions in Eq. (10) yields,

$$\langle\hat{\Omega}(t)\rangle = \int D[x]D[p]D[\delta x]D[\delta p]e^{\frac{i}{\hbar}\int_0^t \mathrm{d}\tau(\delta x(\tau)\dot{p}(\tau) - \delta p(\tau)\dot{x}(\tau))}\mathrm{Tr}\Bigg\{\hat{\mathbf{W}}(x(0), p(0))$$
$$\left[\mathcal{T}_t e^{-\frac{i}{\hbar}\int_0^t \mathrm{d}\tau\hat{\mathbf{H}}\left(x(\tau)+\frac{\delta x(\tau)}{2}, p(\tau)+\frac{\delta p(\tau)}{2}\right)}\right]^\dagger \hat{\mathbf{\Omega}}(x(t), p(t))\left[\mathcal{T}_t e^{-\frac{i}{\hbar}\int_0^t \mathrm{d}\tau\hat{\mathbf{H}}\left(x(\tau)-\frac{\delta x(\tau)}{2}, p(\tau)-\frac{\delta p(\tau)}{2}\right)}\right]\Bigg\}, \qquad [16]$$

where $\hat{\mathbf{W}}(x,p)$ and $\hat{\mathbf{\Omega}}(x,p)$ are the Weyl symbols associated to the density matrix and observable, respectively. The finite $N$ representation of Eq. (16), is schematically illustrated in Fig. S1.

In standard TWA, the classical $\hbar \to 0$ limit is obtained from Eq. (16) by taking a saddle point approximation (3). However, this technique cannot be immediately applied, since the Hamiltonian Weyl symbol $\hat{\mathbf{H}}$ is not a scalar, but an operator in $\mathbb{H}^Q$. We now introduce the MBOA as a technique to map Eq. (16) back into a standard scalar path integral.

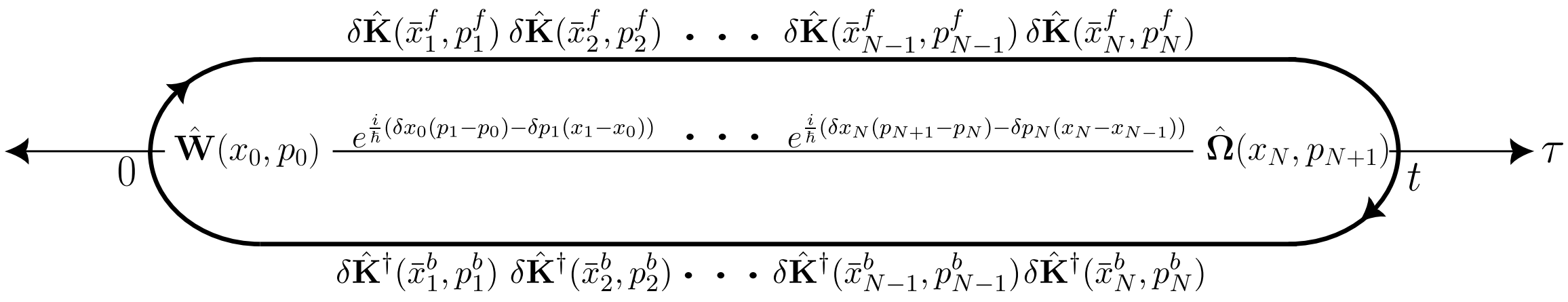

**Fig. S1.** Finite $N$ representation of Eq. (16). We have abbreviated $\delta\hat{\mathbf{K}}(x,p) = \left(1 + \frac{i\Delta t}{\hbar}\hat{\mathbf{H}}(x,p)\right)$ and $\bar{x}_i = (x_i + x_{i-1})/2$. Arrows indicate the direction of time, with the top branch corresponding to forward and the bottom to backward time evolution. The quantity shown in the center is the total phase accumulated over the contour.

**Unitary Transformation and Dressed Operators.** Eq. (10) is invariant under arbitrary basis transformation of $\hat{H}$, $\hat{\Omega}$, and $\hat{\rho}$. In particular, let $\hat{U}$ be an arbitrary unitary operator. Then, Eq. (10) is invariant under transformations where each operator appearing under the trace is mapped to,

$$\hat{O}'_U \equiv \hat{U}^\dagger\hat{O}\hat{U}. \qquad [17]$$

The unitary is parameterized as $\hat{U} = e^{-\frac{i}{\hbar}\hat{G}}$, where $\hat{G}$ is a Hermitian operator with Weyl symbol $\hat{\mathbf{G}}(x,p)$, and the factor $1/\hbar$ is introduced such that $\hat{\mathbf{G}}$ has a well defined classical limit.

The Weyl symbol of Eq. (17) defines $\hat{\mathbf{O}}'_U$, the operator *dressed* by $\hat{U}$. The dressed operator can be found through use of the Moyal product (3),

$$\hat{\mathbf{O}}'_U(x,p) = \hat{\mathcal{U}}^\dagger(x,p) \star \hat{\mathbf{O}}(x,p) \star \hat{\mathcal{U}}(x,p), \qquad [18]$$

where $\hat{\mathcal{U}}(x,p)$ is the Weyl symbol of $\hat{U}$ † and the star symbol is defined as,

$$(f \star g)(x,p) \equiv f(x,p)e^{\frac{i\hbar}{2}\Lambda}g(x,p), \qquad [19]$$

with $\Lambda$ the Poisson bracket operator,

$$\Lambda = (\overleftarrow{\partial}_x\overrightarrow{\partial}_p - \overleftarrow{\partial}_p\overrightarrow{\partial}_x). \qquad [20]$$

Dressed operators can likewise be expressed in the original lab frame coordinates by performing the inverse unitary rotation in $\mathbb{H}^Q$,

$$\hat{\mathbf{O}}_U(x,p) = \hat{\mathbf{U}}(x,p)\hat{\mathbf{O}}'_U(x,p)\hat{\mathbf{U}}^\dagger(x,p), \qquad [21]$$

† The calligraphic font is added to reserve the notation $\hat{\mathbf{U}}(x,p)$ for another operator introduced in Eq. (21)

where $\hat{\mathbf{U}}(x,p) \equiv e^{-\frac{i}{\hbar}\hat{\mathbf{G}}(x,p)}$. Throughout the text, unprimed quantities refer to operators in the lab frame. For example, Eq. 10 of the main text is the Hamiltonian dressed by the BO unitary in the lab frame basis.

We emphasize that $\hat{\mathbf{O}}_U$ and $\hat{\mathbf{O}}'_U$ represent the same operator written in different bases. For the spin example, the prime operators correspond to the choice of the Pauli matrices with $z$-axis along the instantaneous field, while for the non-prime lab-frame operators the coordinate system is fixed in space. At the same time, we stress that the dressed $\hat{\mathbf{O}}_U$ is a fundamentally different operator than the bare $\hat{\mathbf{O}}$, i.e. these operators are not related by some basis transformation. In constructing $\hat{\mathbf{O}}_U$, we have first conjugated the operator $\hat{O}$ by $\hat{U}$, then taken its Wigner-Weyl transform, and afterwards performed the inverse unitary rotation in $\mathbb{H}^Q$ only. Because these two steps do not commute, the dressed operator $\hat{\mathbf{O}}_U$ depends on the choice of $\hat{U}$.

The path integral in Eq. (16) is invariant under the substitution: $\hat{\mathbf{H}} \to \hat{\mathbf{H}}'_U$, $\hat{\mathbf{\Omega}} \to \hat{\mathbf{\Omega}}'_U$, and $\hat{\rho} \to \hat{\rho}'_U$. In this way, the gauge freedom of basis choice in Eq. (10) is mapped to the gauge freedom of non-linear transformations of Weyl symbols Eq. (18). Indeed, as we show in Sec. 3, in the classical $\hbar \to 0$ limit for the fast DOF, the basis transformation reduces to a canonical transformation.

The key idea is to pick the unitary $\hat{U}$ such that the dressed Hamiltonian $\hat{\mathbf{H}}'_U(x,p)$ is diagonal for all $(x,p)$. Note that this problem is non-linear and can generally only be solved approximately by some perturbative expansion or by iterations (see Sec. 3). A common approach in the literature is to employ so-called Moyal perturbation theory (4), where the Moyal product is truncated to some order in an adiabatic parameter (commonly a dimensionless parameter involving the ratio of the electronic and nuclear masses) or expansion in $\hbar$ (5–10). Once this unitary is found, either exactly or approximately, the path integral in Eq. (16) can be evaluated using standard saddle-point methods because the time-ordered exponentials reduce to diagonal time-dependent phase factors evaluated along the forward and backward phase space trajectories.

A few comments are in order. Let us point out an important subtlety highlighting the difference between the unitary $\hat{U}$ and its Weyl symbol $\hat{\mathcal{U}}(x,p)$. The former is *time-independent*, since we consider Hamiltonians that are constant in time. The latter implicitly depends on time because in the path integral $x$ and $p$ become time-dependent variables. This is a key difference with the work done in Ref. (11), who likewise employ a path integral approach, but work with *time-dependent* unitary transformations. Fixing $x(t)$ to be a smooth, time-dependent variable misses the metric term associated with fluctuations of the momentum of the slow DOF (see discussion after Eq. 3.13 of Ref. (11) and Fig. 6 of our main text). This term is responsible for capturing the emergent interactions between fast DOFs in the moving frame leading to spin entanglement and squeezing.

We also remark on the distinction between $\hat{\mathcal{U}}$ and $\hat{\mathbf{U}}$. While the former is a unitary operator under the star product $\hat{\mathcal{U}}^\dagger \star \hat{\mathcal{U}} = \hat{\mathbf{I}}$, the latter is unitary under the normal product $\hat{\mathbf{U}}^\dagger\hat{\mathbf{U}} = \hat{\mathbf{I}}$. In the special case when the unitary transformation is exclusively dependent on either $x$ or $p$, such as those considered up to now, we have $\hat{\mathcal{U}} = \hat{\mathbf{U}}$.

We now discuss two approximations for $\hat{U}$: BOA and MBOA.

**Born-Oppenheimer Approximation.** The BOA uses the unitary $\hat{U} \equiv U_{\mathrm{BO}}(\hat{x})$ which diagonalizes the interaction part of the Hamiltonian, and neglects the difference between $\hat{\mathbf{H}}(x,p)$ and $\hat{\mathbf{H}}_{U_{\mathrm{BO}}}(x,p) \equiv \hat{\mathbf{H}}^{\mathrm{M}}(x,p)$. That is, the Moyal product in Eq. (18) is replaced by the ordinary product,

$$\hat{\mathbf{H}}'_U = \hat{\mathcal{U}}^\dagger_{\mathrm{BO}} \star \hat{\mathbf{H}} \star \hat{\mathcal{U}}_{\mathrm{BO}} \approx \hat{\mathbf{U}}^\dagger_{\mathrm{BO}} \hat{\mathbf{H}} \hat{\mathbf{U}}_{\mathrm{BO}}. \tag{22}$$

The Hamiltonian $\hat{\mathbf{H}}'_U(x,p)$ is then by construction diagonal for any $x$ and $p$ and the BOA follows from a standard saddle-point analysis of the path integral in Eq. (16).

We remark that the BOA neglects the dressing of, not only the Hamiltonian, but all other operators: $\hat{\mathbf{O}}_{U_{\mathrm{BO}}}(x,p) \approx \hat{\mathbf{O}}(x,p)$. In particular, both the coordinate $\hat{\mathbf{x}}$ and the momentum $\hat{\mathbf{q}}$ remain diagonal operators in the fast space, i.e. they can be treated as scalars. This fails to capture centrifugal and inertial forces in the moving frame and violates canonical commutation relations between $\hat{\mathbf{x}}$ and $\hat{\mathbf{q}}$. For classical systems, it analogously violates canonical Poisson bracket relations between fast and slow variables, as discussed for the piston example.

**Moving Born-Oppenheimer Approximation.** To derive the MBOA, we first dress operators with the unitary $U_{\mathrm{BO}}(\hat{x})$. Because $U_{\mathrm{BO}}(\hat{x})$ only depends on the coordinate, one can evaluate all Moyal products exactly and recover the dressed operators in Eqs. 7 and 10 of the main text. To be consistent with the notation used in the main text, from now on we add the superscript M to denote operators dressed specifically by the BO unitary. That is, we define $\hat{\mathbf{O}}^{\mathrm{M}} \equiv \hat{\mathbf{O}}_{U_{\mathrm{BO}}}$.

The BO unitary diagonalizes the interacting part of the Hamiltonian, but the dressing introduces new off-diagonal terms in $\hat{\mathbf{H}}^{\mathrm{M}}$. In order to proceed, we introduce a second unitary transformation $\hat{U}_{\mathrm{MBO}}$ diagonalizing, not the bare Hamiltonian $\hat{\mathbf{H}}$, but the dressed Hamiltonian $(\hat{\mathbf{H}}^{\mathrm{M}})'$. The full unitary is written as $\hat{U} = \hat{U}_{\mathrm{BO}}\hat{U}_{\mathrm{MBO}}$. The MBOA neglects further dressing associated to this second unitary transformation, i.e. it applies the BOA but to the dressed Hamiltonian,

$$\hat{\mathbf{H}}'_U = \hat{\mathcal{U}}^\dagger_{\mathrm{MBO}} \star \hat{\mathbf{H}}'_{U_{\mathrm{BO}}} \star \hat{\mathcal{U}}_{\mathrm{MBO}} \approx \hat{\mathbf{U}}^\dagger_{\mathrm{MBO}} \hat{\mathbf{H}}'_{U_{\mathrm{BO}}} \hat{\mathbf{U}}_{\mathrm{MBO}}. \tag{23}$$

The formal corrections that control this approximation are discussed in Sec. 5.

Practically, the MBOA allows one to write the path integral in the eigenbasis of $\hat{\mathbf{H}}^{\mathrm{M}}$ and neglect any off-diagonal terms in the time-ordered exponentials, such that Eq. (16) becomes,

$$\langle\hat{\Omega}(t)\rangle = \sum_{m,n}\int D[x(\tau)]D[p(\tau)]D[\delta x(\tau)]D[\delta p(\tau)]$$
$$e^{\frac{i}{\hbar}\int \mathrm{d}\tau(\dot{p}\delta x - \dot{x}\delta p + \mathcal{H}^{\mathrm{MBO}}_m(x+\frac{\delta x}{2},p+\frac{\delta p}{2}) - \mathcal{H}^{\mathrm{MBO}}_n(x-\frac{\delta x}{2},p-\frac{\delta p}{2}))}\hat{\mathbf{W}}^{\mathrm{M}}_{nm}(x(0),p(0))\hat{\mathbf{\Omega}}^{\mathrm{M}}_{mn}(x(t),p(t)). \tag{24}$$

The indices $m$ and $n$ denote matrix elements in the basis of eigenstates of the dressed $\hat{\mathbf{H}}^{\mathrm{M}}$, with $\mathcal{H}_m^{\mathrm{MBO}}$ the corresponding eigenvalues. Note that they explicitly depend on $x$ and $p$.

Following the standard TWA procedure (3), we expand the action in the exponential to linear order in the quantum fluctuations $\delta x$ and $\delta p$. Integrating them out yields a delta function enforcing the classical equation of motion for $x(t)$ and $p(t)$ with an additional subtlety: these equations depend on both indices $n$ and $m$,

$$\dot{x} = \frac{\partial \mathcal{H}_{mn}^{\mathrm{MBO}}}{\partial p}, \qquad \dot{p} = -\frac{\partial \mathcal{H}_{mn}^{\mathrm{MBO}}}{\partial x}, \tag{25}$$

where we have defined,

$$\mathcal{H}_{mn}^{\mathrm{MBO}}(x,p) \equiv \frac{\mathcal{H}_m^{\mathrm{MBO}}(x,p) + \mathcal{H}_n^{\mathrm{MBO}}(x,p)}{2}. \tag{26}$$

Solving these equations for each pair of indices $m$ and $n$ and different initial conditions $x(0)$ and $p(0)$, which are sampled from the matrix elements of the Wigner function $\hat{\mathbf{W}}_{nm}^{\mathrm{M}}(x,p)$, yields unique classical trajectories $x_{mn}(t)$ and $p_{mn}(t)$. The sub-indices $m$ and $n$ denote the particular Hamiltonian Eq. (26) used for time evolution.

Note that the classical trajectories are generated not only by the eigenvalues $\mathcal{H}_n^{\mathrm{MBO}}$, but also by linear combinations of them. This happens when the slow DOF evolves in the forward path with a different eigenvalue $\mathcal{H}_n^{\mathrm{MBO}}$ than in the backward path. Hence, the $\mathcal{H}_{mn}^{\mathrm{MBO}}$ with $m \neq n$ are responsible for the dynamics of the off-diagonal elements of $\hat{\mathbf{\Omega}}^{\mathrm{M}}$. They are important when describing superpositions of the MBO states for which the dressed Wigner function $\hat{\mathbf{W}}^{\mathrm{M}}$ has non-zero off-diagonal elements. We are also left with a non-vanishing phase factor,

$$\Phi_{mn}(t) \equiv \frac{1}{\hbar}\int_0^t \mathrm{d}\tau \left[\mathcal{H}_m^{\mathrm{MBO}}(x_{mn}(\tau), p_{mn}(\tau)) - \mathcal{H}_n^{\mathrm{MBO}}(x_{mn}(\tau), p_{mn}(\tau))\right], \tag{27}$$

that can be attached to the time-evolution of the observable,

$$\hat{\mathbf{\Omega}}_{mn}^{\mathrm{M}}(t) \equiv \mathrm{e}^{i\Phi_{mn}(t)}\hat{\mathbf{\Omega}}_{mn}^{\mathrm{M}}(x_{mn}(t), p_{mn}(t)). \tag{28}$$

Finally, we obtain,

$$\langle \hat{\Omega}(t) \rangle \approx \int \frac{\mathrm{d}x_0 \mathrm{d}p_0}{2\pi\hbar} \mathrm{Tr}\left\{\hat{\mathbf{W}}^{\mathrm{M}}(x_0,p_0)\hat{\mathbf{\Omega}}^{\mathrm{M}}(t)\right\}. \tag{29}$$

## 3. General Expressions for Weyl Symbols

This section develops a formal procedure for computing Weyl symbols of operators $\hat{\mathbf{O}}'_U$ of the form given by Eq. (4). In the special case when $\hat{\mathcal{U}}$ is only $x$- ($p$-)dependent, the expansion of the $\star$-product in powers of $\Lambda$ terminates at finite order for any observable $\hat{\mathbf{O}}$ that is polynomial in $p$ ($x$). However, such an expansion does not truncate for a general unitary $\hat{\mathcal{U}}(x,p)$. Instead, we discuss an expression for $\hat{\mathbf{O}}'_U$ which admits an iterative gradient expansion in terms of the AGPs of $\hat{\mathbf{U}}$. These results are used in Sections 5 and 6 of the supplementary information to derive the next order corrections to the MBOA.

Let us establish a procedure for computing the leading classical term in the Weyl symbol of operator $\hat{O}'_U(\lambda)$,

$$\hat{O}'_U(\lambda) = \hat{U}^\dagger(\lambda)\hat{O}\hat{U}(\lambda) = e^{\frac{i}{\hbar}\lambda\hat{G}}\hat{O}e^{-\frac{i}{\hbar}\lambda\hat{G}}. \tag{30}$$

We are interested in $\hat{\mathbf{O}}'_U(\lambda = 1) \equiv \hat{\mathbf{O}}'_U$, but given that $\hat{\mathbf{O}}'_U(\lambda = 0) = \hat{\mathbf{O}}$ is known, we set up a differential equation describing the flow of $\hat{\mathbf{O}}'_U(\lambda)$ as a function of $\lambda$. This allows one to compute the desired $\hat{\mathbf{O}}'_U$ iteratively. Differentiating Eq. (30) and taking the Wigner-Weyl transform yields [‡],

$$\frac{\mathrm{d}\hat{\mathbf{O}}'_U(\lambda)}{\mathrm{d}\lambda} = \frac{i}{\hbar}\left(\hat{\mathbf{G}} \star \hat{\mathbf{O}}'_U(\lambda) - \hat{\mathbf{O}}'_U(\lambda) \star \hat{\mathbf{G}}\right). \tag{31}$$

Expanding the Moyal product to second order,

$$\frac{\mathrm{d}\hat{\mathbf{O}}'_U(\lambda)}{\mathrm{d}\lambda} = \frac{1}{i\hbar}\left[\hat{\mathbf{O}}'_U(\lambda), \hat{\mathbf{G}}\right] + \{\hat{\mathbf{O}}'_U(\lambda), \hat{\mathbf{G}}\}_{\mathrm{SPB}} + \mathcal{O}(\hbar), \tag{32}$$

where $[.]$ is the standard commutator, and $\{.\}_{\mathrm{SPB}}$ denotes the symmetrized Poisson bracket operator,

$$\{\hat{\mathbf{A}}, \hat{\mathbf{B}}\}_{\mathrm{SPB}} \equiv \frac{1}{2}\left(\frac{\partial \hat{\mathbf{A}}}{\partial x}\frac{\partial \hat{\mathbf{B}}}{\partial p} + \frac{\partial \hat{\mathbf{B}}}{\partial p}\frac{\partial \hat{\mathbf{A}}}{\partial x} - \frac{\partial \hat{\mathbf{A}}}{\partial p}\frac{\partial \hat{\mathbf{B}}}{\partial x} - \frac{\partial \hat{\mathbf{B}}}{\partial x}\frac{\partial \hat{\mathbf{A}}}{\partial p}\right). \tag{33}$$

We neglect $\mathcal{O}(\hbar)$ and higher terms, which vanish in the classical limit for the slow coordinate. Note the resemblance of Eq. (32) to a standard canonical transformation. If the classical limit is also taken for the fast quantum variable, the right hand side reduces to the full Poisson bracket $\{.\}_{\mathrm{fast}} + \{.\}_{\mathrm{slow}}$. Thus, one can see that $\hat{\mathbf{G}}$ reduces to the generator of a family of continuous canonical transformations parameterized by $\lambda$ (2). Observe the coexistence of the Poisson bracket for the slow coordinate subspace, and the commutator for the fast quantum subspace, highlighting the mixed quantum-classical nature of our approximation.

[‡] For brevity, we suppress the $x$ and $p$-dependence of Weyl symbols.

The solution to Eq. (32) can be written self-consistently as,

$$\hat{\mathbf{O}}'_U(\lambda) = e^{\frac{i}{\hbar}\lambda\hat{\mathbf{G}}}\left(\hat{\mathbf{O}}\int_0^\lambda \mathrm{d}\lambda' e^{-\frac{i}{\hbar}\lambda'\hat{\mathbf{G}}}\left(\{\hat{\mathbf{O}}'_U(\lambda'),\hat{\mathbf{G}}\}_{\mathrm{SPB}}\right)e^{\frac{i}{\hbar}\lambda'\hat{\mathbf{G}}}\right)e^{-\frac{i}{\hbar}\lambda\hat{\mathbf{G}}}, \tag{34}$$

and can be solved iteratively. Starting from the ansatz $(\hat{\mathbf{O}}'_U)^{(0)}(\lambda) = e^{\frac{i}{\hbar}\lambda\hat{\mathbf{G}}}\hat{\mathbf{O}}e^{-\frac{i}{\hbar}\lambda\hat{\mathbf{G}}}$, one plugs it into the right-hand-side of Eq. (34) to obtain a next iteration $(\hat{\mathbf{O}}'_U)^{(1)}(\lambda)$. Repeating this process yields a recursive series of Weyl symbols $(\hat{\mathbf{O}}'_U)^{(n)}(\lambda)$,

$$(\hat{\mathbf{O}}'_U)^{(n)}(\lambda) = e^{\frac{i}{\hbar}\lambda\hat{\mathbf{G}}}\left(\hat{\mathbf{O}}\int_0^\lambda \mathrm{d}\lambda' e^{-\frac{i}{\hbar}\lambda'\hat{\mathbf{G}}}\left(\{(\hat{\mathbf{O}}'_U)^{(n-1)}(\lambda'),\hat{\mathbf{G}}\}_{\mathrm{SPB}}\right)e^{\frac{i}{\hbar}\lambda'\hat{\mathbf{G}}}\right)e^{-\frac{i}{\hbar}\lambda\hat{\mathbf{G}}}, \tag{35}$$

which, if convergent, reduce to the exact solution in the limit $n\to\infty$. Formally, this procedure is equivalent to a Dyson series expansion of $\hat{\mathbf{O}}'_U(\lambda)$,

$$\hat{\mathbf{O}}'_U(\lambda) = e^{\frac{i}{\hbar}\lambda\hat{\mathbf{G}}}\left(\mathcal{T}_\lambda \exp\left\{\int_0^\lambda \mathrm{d}\lambda'\mathcal{L}(\lambda',\hat{\mathbf{G}})\right\}\hat{\mathbf{O}}\right)e^{-\frac{i}{\hbar}\lambda\hat{\mathbf{G}}}, \tag{36}$$

where the operator $\mathcal{L}(\lambda,\hat{\mathbf{G}})$ is defined as,

$$\mathcal{L}(\lambda,\hat{\mathbf{G}})\hat{\mathbf{O}} = e^{-\frac{i}{\hbar}\lambda\hat{\mathbf{G}}}\left(\{e^{\frac{i}{\hbar}\lambda\hat{\mathbf{G}}}\hat{\mathbf{O}}e^{-\frac{i}{\hbar}\lambda\hat{\mathbf{G}}},\hat{\mathbf{G}}\}_{\mathrm{SPB}}\right)e^{\frac{i}{\hbar}\lambda\hat{\mathbf{G}}} = e^{-\frac{i}{\hbar}\lambda\hat{\mathbf{G}}}\left(\{\hat{\mathbf{O}}'(\lambda),\hat{\mathbf{G}}\}_{\mathrm{SPB}}\right)e^{\frac{i}{\hbar}\lambda\hat{\mathbf{G}}}, \tag{37}$$

and $\mathcal{T}_\lambda$ is the $\lambda$-ordering operator. One can verify Eq. (34) and Eq. (36) by noting that they satisfy the correct boundary condition $\hat{\mathbf{O}}'_U(\lambda=0)=\hat{\mathbf{O}}$, and that differentiating them with respect to $\lambda$ yields Eq. (32).

Finally, evaluating at the desired $\lambda=1$ and defining $\hat{\mathbf{U}}(x,p;\lambda=1)\equiv\hat{\mathbf{U}}(x,p)$, we recover,

$$\hat{\mathbf{O}}'_U(x,p) = \hat{\mathbf{U}}^\dagger(x,p)\hat{\mathbf{O}}_U(x,p)\hat{\mathbf{U}}(x,p), \tag{38}$$

where the dressed operator $\hat{\mathbf{O}}_U(x,p)$ is obtained as,

$$\hat{\mathbf{O}}_U \equiv \mathcal{T}_\lambda \exp\left\{\int_0^1 \mathrm{d}\lambda\mathcal{L}(\lambda,\hat{\mathbf{G}})\right\}\hat{\mathbf{O}}. \tag{39}$$

## 4. Moving Frame Transformation and Dressed Operators

In this section, we derive the dressing of operators due to the transformation to the BO moving frame. We work in the basis given by $\hat{U} = U_{\mathrm{BO}}(\hat{x}) \equiv e^{-\frac{i}{\hbar}\hat{G}_{\mathrm{BO}}}$, such that the transformed Weyl symbols are,

$$\hat{\mathbf{O}}'_{U_{\mathrm{BO}}} = \hat{\mathcal{U}}^\dagger_{\mathrm{BO}}\star\hat{\mathbf{O}}\star\hat{\mathcal{U}}_{\mathrm{BO}} = \hat{\mathbf{U}}^\dagger_{\mathrm{BO}}\hat{\mathbf{O}}_{U_{\mathrm{BO}}}\hat{\mathbf{U}}_{\mathrm{BO}}. \tag{40}$$

Recall that the superscript M is used to denote operators dressed specifically by the BO unitary. That is, the notation $\hat{\mathbf{O}}_{U_{\mathrm{BO}}}\equiv\hat{\mathbf{O}}^{\mathrm{M}}$ is used interchangeably.

Eq. (40) can be evaluated by (i) expanding the $\star$-product, (ii) iterating Eq. (35), or (iii) directly evaluating Eq. (39). For illustration, all three approaches are shown to compute the dressing of the operators $\hat{\mathbf{x}}$, $\hat{\mathbf{q}}$, and $\hat{\mathbf{q}}^2$. We remark that the $n$-th order Taylor expansion of Eq. (39) is equivalent to the $n$-th iteration of Eq. (35).

**Moyal Product.**

- Computation of $\hat{\mathbf{x}}^{\mathrm{M}}$: Since both the BO unitary and the Weyl symbol $\hat{\mathbf{x}} = x\hat{\mathbf{I}}$ are $p$-independent, one can replace all $\star$-products with normal multiplication, $\hat{\mathcal{U}}^\dagger_{\mathrm{BO}}\star\hat{\mathbf{x}}\star\hat{\mathcal{U}}_{\mathrm{BO}} = \hat{\mathbf{U}}^\dagger_{\mathrm{BO}}\hat{\mathbf{x}}\hat{\mathbf{U}}_{\mathrm{BO}} = \hat{\mathbf{x}}$, resulting in $\hat{\mathbf{x}}^{\mathrm{M}} = \hat{\mathbf{x}}$.
- Computation of $\hat{\mathbf{q}}^{\mathrm{M}}$: Since the Weyl symbol $\hat{\mathbf{q}} = p\hat{\mathbf{I}}$ is linear in momentum, one can expand the star product up to linear order in $\Lambda$,

$$(\hat{\mathbf{q}}^{\mathrm{M}})' = \hat{\mathbf{U}}^\dagger_{\mathrm{BO}}(1+\frac{i\hbar}{2}\Lambda)\hat{\mathbf{q}}(1+\frac{i\hbar}{2}\Lambda)\hat{\mathbf{U}}_{\mathrm{BO}} = \hat{\mathbf{q}} + (i\hbar\partial_x\hat{\mathbf{U}}^\dagger_{\mathrm{BO}})\hat{\mathbf{U}}_{\mathrm{BO}}, \tag{41}$$

  where we have used $(\partial_x\hat{\mathbf{U}}^\dagger_{\mathrm{BO}})\hat{\mathbf{U}}_{\mathrm{BO}} = -\hat{\mathbf{U}}^\dagger_{\mathrm{BO}}(\partial_x\hat{\mathbf{U}}_{\mathrm{BO}})$ and $\hat{\mathbf{q}}' = \hat{\mathbf{q}}$. From this, one finds $\hat{\mathbf{q}}^{\mathrm{M}} = \hat{\mathbf{U}}_{\mathrm{BO}}(\hat{\mathbf{q}}^{\mathrm{M}})'\hat{\mathbf{U}}^\dagger_{\mathrm{BO}} = \hat{\mathbf{q}} - (i\hbar\partial_x\hat{\mathbf{U}}_{\mathrm{BO}})\hat{\mathbf{U}}^\dagger_{\mathrm{BO}} = \hat{\mathbf{q}} - \hat{\mathbf{A}}_x$.
- Computation of $(\hat{\mathbf{q}}^2)^{\mathrm{M}}$: One can use the previous result to obtain $(\hat{\mathbf{q}}^2)'_{U_{\mathrm{BO}}} = \hat{\mathbf{q}}'_{U_{\mathrm{BO}}}\star\hat{\mathbf{q}}'_{U_{\mathrm{BO}}} = (\hat{\mathbf{q}}'_{U_{\mathrm{BO}}})^2$, where we have used $\hat{\mathbf{q}}'_{U_{\mathrm{BO}}}\Lambda\hat{\mathbf{q}}'_{U_{\mathrm{BO}}} = 0$. Rotating back to the lab frame and using superscript M notation, $(\hat{\mathbf{q}}^2)^{\mathrm{M}} = (\hat{\mathbf{q}}^{\mathrm{M}})^2 = (\hat{\mathbf{q}} - \hat{\mathbf{A}}_x)^2$.

As shown above, direct calculation of the dressed operators is straightforward for the BO unitary. However, for a general unitary which depends on both $\hat{x}$ and $\hat{q}$, we have to resort to the Dyson series to obtain dressed operators order by order. This will be relevant for finding corrections to the MBOA. We therefore now illustrate how one can recover the results above using the Dyson series and iterative method.

**Dyson Series.**

- Computation of $\hat{\mathbf{x}}^{\mathrm{M}}$: Observe that $\mathcal{L}(\lambda, \hat{\mathbf{G}}_{\mathrm{BO}})\hat{\mathbf{x}} = 0$ due to the fact that the BO generating function is only $x$-dependent. Hence, only the zero-th order term survives in the Taylor expansion of Eq. (39), resulting in $\hat{\mathbf{x}}^{\mathrm{M}} = \hat{\mathbf{x}}$.

- Computation of $\hat{\mathbf{q}}^{\mathrm{M}}$: Note that,

$$\mathcal{L}(\lambda_1, \hat{\mathbf{G}}_{\mathrm{BO}})\hat{\mathbf{q}} = e^{-\frac{i}{\hbar}\lambda_1 \hat{\mathbf{G}}_{\mathrm{BO}}} \left(\{p\,\hat{\mathbf{I}}, \hat{\mathbf{G}}_{\mathrm{BO}}\}_{\mathrm{SPB}}\right) e^{\frac{i}{\hbar}\lambda_1 \hat{\mathbf{G}}_{\mathrm{BO}}} = -e^{-\frac{i}{\hbar}\lambda_1 \hat{\mathbf{G}}_{\mathrm{BO}}} \partial_x \hat{\mathbf{G}}_{\mathrm{BO}} e^{\frac{i}{\hbar}\lambda_1 \hat{\mathbf{G}}_{\mathrm{BO}}}. \quad [42]$$

Since both $\hat{\mathbf{G}}_{\mathrm{BO}}$ and the RHS of Eq. (42) are only $x$-dependent, any higher order term vanishes, $\mathcal{L}(\lambda_2, \hat{\mathbf{G}}_{\mathrm{BO}})\mathcal{L}(\lambda_1, \hat{\mathbf{G}}_{\mathrm{BO}})\hat{\mathbf{q}} = 0$. Hence, one only needs Taylor expand Eq. (39) to first order, obtaining,

$$\hat{\mathbf{q}}^{\mathrm{M}} = \hat{\mathbf{q}} - \int_0^1 \mathrm{d}\lambda_1 e^{-\frac{i}{\hbar}\lambda_1 \hat{\mathbf{G}}_{\mathrm{BO}}} \partial_x \hat{\mathbf{G}}_{\mathrm{BO}} e^{\frac{i}{\hbar}\lambda_1 \hat{\mathbf{G}}_{\mathrm{BO}}} = \hat{\mathbf{q}} - \hat{\mathbf{A}}_x, \quad [43]$$

through application of the identity (12),

$$\partial_\lambda \hat{\mathbf{A}}_x(\lambda) = e^{-\frac{i}{\hbar}\lambda \hat{\mathbf{G}}_{\mathrm{BO}}} \partial_x \hat{\mathbf{G}}_{\mathrm{BO}} e^{\frac{i}{\hbar}\lambda \hat{\mathbf{G}}_{\mathrm{BO}}}. \quad [44]$$

$\hat{\mathbf{A}}_x(\lambda)$ is defined as in Eq. (5).

- Computation of $(\hat{\mathbf{q}}^2)^{\mathrm{M}}$: Similarly for $\hat{\mathbf{q}}^2$, Eq. (39) is Taylor expanded up to second order, noticing that all higher order terms vanish. We obtain,

$$(\hat{\mathbf{q}}^2)^{\mathrm{M}} = \hat{\mathbf{q}}^2 - 2p\int_0^1 \mathrm{d}\lambda_1 \frac{\partial}{\partial \lambda_1}\left(\hat{\mathbf{A}}_x(\lambda_1)\right) + \frac{1}{2}\int_0^1\int_0^1 \mathrm{d}\lambda_2 \mathrm{d}\lambda_1 \left[\frac{\partial}{\partial \lambda_1}\left(\hat{\mathbf{A}}_x(\lambda_1)\right), \frac{\partial}{\partial \lambda_2}\left(\hat{\mathbf{A}}_x(\lambda_2)\right)\right]_+, \quad [45]$$

where $[\cdot,\cdot]_+$ denotes the anti-commutator. Finally, $(\hat{\mathbf{q}}^2)^{\mathrm{M}} = \hat{\mathbf{q}}^2 - 2p\hat{\mathbf{A}}_x + \hat{\mathbf{A}}_x^2 = (\hat{\mathbf{q}} - \hat{\mathbf{A}}_x)^2$.

**Iterations.**

- Computation of $\hat{\mathbf{x}}^{\mathrm{M}}$: Starting from the ansatz $\hat{\mathbf{x}}^{(0)}(\lambda) = x\,\hat{\mathbf{I}}$ [§], applying Eq. (35) yields $\hat{\mathbf{x}}^{(1)} = \hat{\mathbf{x}}^{(0)}$. This follows from the vanishing Poisson bracket $\{\hat{\mathbf{x}}^{(0)}, \hat{\mathbf{G}}_{\mathrm{BO}}\}_{\mathrm{SPB}} = 0$, since the BO generating function is $p$-independent. Hence, $\hat{\mathbf{x}}^{(0)}$ is a fixed point of the iterative scheme and $\hat{\mathbf{x}}^{\mathrm{M}} = \hat{\mathbf{U}}_{\mathrm{BO}}\hat{\mathbf{x}}^{(0)}(\lambda = 1)\hat{\mathbf{U}}_{\mathrm{BO}}^\dagger = x\,\hat{\mathbf{I}}$.

- Computation of $\hat{\mathbf{q}}^{\mathrm{M}}$: Starting from the ansatz $\hat{\mathbf{q}}^{(0)}(\lambda) = p\,\hat{\mathbf{I}}$, the first iteration gives,

$$\hat{\mathbf{q}}^{(1)}(\lambda) = e^{\frac{i}{\hbar}\lambda \hat{\mathbf{G}}_{\mathrm{BO}}}(p\,\hat{\mathbf{I}} - \hat{\mathbf{A}}_x(\lambda))e^{-\frac{i}{\hbar}\lambda \hat{\mathbf{G}}_{\mathrm{BO}}}. \quad [46]$$

Here, we have used the fact that $\{\hat{\mathbf{q}}^{(0)}, \hat{\mathbf{G}}_{\mathrm{BO}}\}_{\mathrm{SPB}} = -\partial_x \hat{\mathbf{G}}_{\mathrm{BO}}$ and applied Eq. (44). Then, noting that $\{\hat{\mathbf{q}}^{(1)}(\lambda), \hat{\mathbf{G}}_{\mathrm{BO}}\}_{\mathrm{SPB}} = -\partial_x \hat{\mathbf{G}}_{\mathrm{BO}}$, the second iteration produces no corrections, $\hat{\mathbf{q}}^{(2)}(\lambda) = \hat{\mathbf{q}}^{(1)}(\lambda)$. Thus, $\hat{\mathbf{q}}^{(1)}(\lambda)$ is the fixed point of iterations. From this, we obtain $\hat{\mathbf{q}}^{\mathrm{M}} = \hat{\mathbf{U}}_{\mathrm{BO}}\hat{\mathbf{q}}^{(1)}(\lambda = 1)\hat{\mathbf{U}}_{\mathrm{BO}}^\dagger = p\,\hat{\mathbf{I}} - \hat{\mathbf{A}}_x(\lambda = 1)$.

- Computation of $(\hat{\mathbf{q}}^2)^{\mathrm{M}}$: Begin with the ansatz $(\hat{\mathbf{q}}^2)^{(0)}(\lambda) = p^2\,\hat{\mathbf{I}}$. By noting that $\{(\hat{\mathbf{q}}^2)^{(0)}, \hat{\mathbf{G}}_{\mathrm{BO}}\}_{\mathrm{SPB}} = -2p\partial_x \hat{\mathbf{G}}_{\mathrm{BO}}$ and using Eq. (44),

$$(\hat{\mathbf{q}}^2)^{(1)}(\lambda) = e^{\frac{i}{\hbar}\lambda \hat{\mathbf{G}}_{\mathrm{BO}}}(p^2\,\hat{\mathbf{I}} - 2p\hat{\mathbf{A}}_x(\lambda))e^{-\frac{i}{\hbar}\lambda \hat{\mathbf{G}}_{\mathrm{BO}}}. \quad [47]$$

Iterating again and using $\{(\hat{\mathbf{q}}^2)^{(1)}, \hat{\mathbf{G}}_{\mathrm{BO}}\}_{\mathrm{SPB}} = -2p\partial_x \hat{\mathbf{G}}_{\mathrm{BO}} + [e^{\frac{i}{\hbar}\lambda \hat{\mathbf{G}}_{\mathrm{BO}}}\hat{\mathbf{A}}_x(\lambda)e^{-\frac{i}{\hbar}\lambda \hat{\mathbf{G}}_{\mathrm{BO}}}, \partial_x \hat{\mathbf{G}}_{\mathrm{BO}}]_+$, where $[\cdot,\cdot]_+$ denotes the anti-commutator,

$$\begin{aligned}(\hat{\mathbf{q}}^2)^{(2)}(\lambda) &= e^{\frac{i}{\hbar}\lambda \hat{\mathbf{G}}_{\mathrm{BO}}}\left(p^2\,\hat{\mathbf{I}} - 2p\hat{\mathbf{A}}_x(\lambda) + \int_0^\lambda \mathrm{d}\lambda' \left[\hat{\mathbf{A}}_x(\lambda'), \frac{\partial}{\partial \lambda'}\left(\hat{\mathbf{A}}_x(\lambda')\right)\right]_+\right)e^{-\frac{i}{\hbar}\lambda \hat{\mathbf{G}}_{\mathrm{BO}}}, \\ &= e^{\frac{i}{\hbar}\lambda \hat{\mathbf{G}}_{\mathrm{BO}}}\left(p^2\,\hat{\mathbf{I}} - 2p\hat{\mathbf{A}}_x(\lambda) + \hat{\mathbf{A}}_x^2(\lambda)\right)e^{-\frac{i}{\hbar}\lambda \hat{\mathbf{G}}_{\mathrm{BO}}}.\end{aligned} \quad [48]$$

In the last line, we have identified the total derivative $\partial_\lambda \hat{\mathbf{A}}_x^2(\lambda) = [\hat{\mathbf{A}}_x(\lambda), \partial_\lambda \hat{\mathbf{A}}_x(\lambda)]_+$.

Finally, a third iteration gives $\hat{\mathbf{q}}^{(3)} = \hat{\mathbf{q}}^{(2)}$. Thus, $(\hat{\mathbf{q}}^2)^{\mathrm{M}} = \hat{\mathbf{U}}_{\mathrm{BO}}((\hat{\mathbf{q}}^2)^{(2)}(\lambda = 1))\hat{\mathbf{U}}_{\mathrm{BO}}^\dagger = (p\,\hat{\mathbf{I}} - \hat{\mathbf{A}}_x)^2$.

[§] Throughout this subsection, we abbreviate $\hat{\mathbf{O}}^{(n)}(\lambda) \equiv (\hat{\mathbf{O}}'_{U_{\mathrm{BO}}})^{(n)}(\lambda)$.

## 5. Moving Born-Oppenheimer Approximation and Leading Corrections

The MBOA consists of a specific ansatz for $\hat{U}$. It is composed of two rotations,

$$\hat{U} = \hat{U}_{\text{BO}}\hat{U}_{\text{MBO}} = e^{-\frac{i}{\hbar}\hat{G}_{\text{BO}}}e^{-\frac{i}{\hbar}\hat{G}_{\text{MBO}}}, \tag{49}$$

where $\hat{U}_{\text{MBO}}$ is constructed to diagonalize the moving Hamiltonian. Its generating function $\hat{\mathbf{G}}_{\text{MBO}}(x,p)$ is chosen such that $\hat{\mathbf{U}}^{\dagger}_{\text{MBO}}(\hat{\mathbf{U}}^{\dagger}_{\text{BO}}\hat{\mathbf{H}}_{U_{\text{BO}}}\hat{\mathbf{U}}_{\text{BO}})\hat{\mathbf{U}}_{\text{MBO}} \equiv \hat{\mathbf{U}}^{\dagger}_{\text{MBO}}\hat{\mathbf{H}}'_{U_{\text{BO}}}\hat{\mathbf{U}}_{\text{MBO}}$ is diagonal ¶.

The MBOA neglects further dressing of all operators due to $\hat{U}_{\text{MBO}}$. In particular, we make the approximation,

$$\hat{\mathbf{O}}'_{U} = \hat{\mathcal{U}}^{\dagger}_{\text{MBO}} \star \hat{\mathbf{O}}'_{U_{\text{BO}}} \star \hat{\mathcal{U}}_{\text{MBO}} \approx \hat{\mathbf{U}}^{\dagger}_{\text{MBO}}\hat{\mathbf{O}}'_{U_{\text{BO}}}\hat{\mathbf{U}}_{\text{MBO}}. \tag{50}$$

Note the similarity to Eq. (22). Indeed, the MBOA can be understood as the Born-Oppenheimer approximation in a moving frame.

In this section, we calculate the leading correction to Eq. (50) and use it to give a criterion for the validity of MBOA and the regime where it breaks down. We evaluate Eq. (50) using the first order term in the Taylor expansion of Eq. (39),

$$\hat{\mathbf{O}}'_{U} = \hat{\mathcal{U}}^{\dagger}_{\text{MBO}} \star \hat{\mathbf{O}}'_{U_{\text{BO}}} \star \hat{\mathcal{U}}_{\text{MBO}} \approx \hat{\mathbf{U}}^{\dagger}_{\text{MBO}}\hat{\mathbf{O}}'_{U_{\text{BO}}}\hat{\mathbf{U}}_{\text{MBO}} + \hat{\mathbf{U}}^{\dagger}_{\text{MBO}}\left(\int_0^1 \mathrm{d}\lambda \mathcal{L}(\lambda, \hat{\mathbf{G}}_{\text{MBO}})\hat{\mathbf{O}}'_{U_{\text{BO}}}\right)\hat{\mathbf{U}}_{\text{MBO}} + \mathcal{O}(\mathcal{L}^2). \tag{51}$$

Note that this expression is formally the first term in a gradient expansion of $\hat{\mathbf{O}}'_U$, and is controlled by the proximity of $\hat{\mathbf{G}}_{\text{MBO}}$ to an operator which is close to a constant in the phase space of the slow DOF, such that its $x$ and $p$ derivatives are small. The subscript MBO is dropped from the generating function for the remainder of the derivation, such that $\hat{\mathbf{G}} \equiv \hat{\mathbf{G}}_{\text{MBO}}$. Expanding the Poisson bracket gives,

$$\int_0^1 \mathrm{d}\lambda \mathcal{L}(\lambda, \hat{\mathbf{G}})\hat{\mathbf{O}}'_{U_{\text{BO}}} = \frac{1}{2}\int_0^1 \mathrm{d}\lambda\, e^{-\frac{i}{\hbar}\lambda\hat{\mathbf{G}}}\left[\partial_x\left(e^{\frac{i}{\hbar}\lambda\hat{\mathbf{G}}}\hat{\mathbf{O}}'_{U_{\text{BO}}}e^{-\frac{i}{\hbar}\lambda\hat{\mathbf{G}}}\right), \partial_p\hat{\mathbf{G}}\right]_+ e^{\frac{i}{\hbar}\lambda\hat{\mathbf{G}}} - \big(x \leftrightarrow p\big), \tag{52}$$

where $[\cdot,\cdot]_+$ denotes the anti-commutator, and the shorthand $(x \leftrightarrow p)$ denotes the same term, but where $x$ and $p$ have been interchanged. One can in turn rewrite this expression in terms of the AGPs of $\hat{\mathbf{U}}_{\text{MBO}}$,

$$\int_0^1 \mathrm{d}\lambda \mathcal{L}(\lambda, \hat{\mathbf{G}})\hat{\mathbf{O}}'_{U_{\text{BO}}} = \frac{1}{2}\int_0^1 \mathrm{d}\lambda\big[\partial_x\hat{\mathbf{O}}'_{U_{\text{BO}}} + \frac{i}{\hbar}[\hat{\mathbf{B}}'_x(\lambda), \hat{\mathbf{O}}'_{U_{\text{BO}}}], \partial_\lambda\hat{\mathbf{B}}'_p(\lambda)\big]_+ - \big(x \leftrightarrow p\big),$$

where $\hat{\mathbf{B}}'_x(\lambda)$ and $\hat{\mathbf{B}}'_p(\lambda)$ are defined as in Eq. (5) and we have made use of the identity in Eq. (44). Keeping only terms linear in $\hat{\mathbf{B}}'_x(\lambda)$ and $\hat{\mathbf{B}}'_p(\lambda)$ gives,

$$\int_0^1 \mathrm{d}\lambda \mathcal{L}(\lambda, \hat{\mathbf{G}})\hat{\mathbf{O}}'_{U_{\text{BO}}} \approx \frac{1}{2}\big[\partial_x\hat{\mathbf{O}}'_{U_{\text{BO}}}, \hat{\mathbf{B}}'_p\big]_+ - (x \leftrightarrow p), \tag{53}$$

with $\hat{\mathbf{B}}'_x \equiv \hat{\mathbf{B}}'_x(\lambda = 1)$ and $\hat{\mathbf{B}}'_p \equiv \hat{\mathbf{B}}'_p(\lambda = 1)$.

Let us now discuss the leading order correction for a few key operators. The correction to the dressing of the Hamiltonian is especially important, since it determines the regime of validity of the MBOA. Switching back to the lab frame and the superscript M notation, we obtain,

$$\hat{\mathbf{H}}^{\text{SA}} \approx \hat{\mathbf{H}}^{\text{M}} + \frac{1}{2}\left[\partial_x\hat{\mathbf{H}}^{\text{M}} + \frac{i}{\hbar}[\hat{\mathbf{A}}_x, \hat{\mathbf{H}}^{\text{M}}], \hat{\mathbf{B}}_p\right]_+ - \frac{1}{2}\left[\partial_p\hat{\mathbf{H}}^{\text{M}}, \hat{\mathbf{B}}_x\right]_+, \tag{54}$$

through application of Eq. (8). Here, the notation $\hat{\mathbf{O}}^{\text{SA}} \equiv \hat{\mathbf{O}}_U$ is introduced to highlight operators that have been dressed by $\hat{\mathbf{B}}_x$ and $\hat{\mathbf{B}}_p$, the generators of so-called super-adiabatic transformations (1).

The MBOA is valid when the extra dressing in $\hat{\mathbf{H}}^{\text{SA}}$ is negligible. This is justified when $\hat{\mathbf{B}}_x$ and $\hat{\mathbf{B}}_p$ are small, i.e. when the unitary $\hat{\mathbf{U}}_{\text{MBO}}(x,p)$ is nearly independent of $x$ and $p$. This implies that the eigenstates of the moving Hamiltonian $\hat{\mathbf{H}}^{\text{M}}(x,p)$ vary slowly with $x$ and $p$. Such validity conditions are of course expected, as the MBOA is nothing but the BOA in the moving frame.

Finally, other operators can also acquire super-adiabatic dressing. In particular, the leading order correction to the position and momentum operators is,

$$\begin{aligned}\hat{\mathbf{q}}^{\text{SA}} &\approx p\,\hat{\mathbf{I}} - \hat{\mathbf{A}}_x - \hat{\mathbf{B}}_x - \frac{1}{2}\big[\partial_x\hat{\mathbf{A}}_x, \hat{\mathbf{B}}_p\big]_+,\\ \hat{\mathbf{x}}^{\text{SA}} &\approx x\,\hat{\mathbf{I}} + \hat{\mathbf{B}}_p.\end{aligned} \tag{55}$$

The dressing of these operators can be non-negligible even when the dressing of the Hamiltonian is negligible. In particular, Sec. 6 illustrates how the corrections in Eq. (55) are important to obtain agreement with exact dynamics in the spin-1/2 toy model.

¶ In this section, we opt for the notation $\hat{\mathbf{H}}_{U_{\text{BO}}}$ instead of $\hat{\mathbf{H}}^{\text{M}}$

## 6. Explicit Calculations for Spin Toy Model

In this section, we describe explicit calculations for the spin-1/2 toy model. We consider a Hamiltonian of the form in Eq. 3 of the main text, with $\hat{V} = 0$, $\hat{H}_{\rm int}$ given by Eq. 4 of the main text, and $\vec{\mathcal{B}}(x) = \mathcal{B}(x)(\cos\theta(x)\hat{z} + \sin\theta(x)\hat{y})$. The two generating functions described in Sec 5 are given by,

$$\hat{\mathbf{G}}_{\rm BO}(x) = -\frac{\hbar\theta(x)}{2}\hat{\sigma}_x, \quad \hat{\mathbf{G}}_{\rm MBO}(x,p) = -\frac{\hbar\phi(x,p)}{2}\hat{\sigma}_y, \tag{56}$$

with,

$$\phi(x,p) = \tan^{-1}\left(\frac{\hbar\theta'(x)p}{2M\mu\mathcal{B}(x)}\right). \tag{57}$$

The moving Hamiltonian is given by Eq. 15 in the main text, which we repeat here for completeness:

$$\hat{\mathbf{H}}^{\rm M}(x,p) = \frac{p^2}{2M} + \frac{\hbar^2(\theta'(x))^2}{8M} - \mu\vec{\mathcal{B}}(x)\cdot\hat{\vec{\sigma}} + \frac{\hbar}{2M}p\,\theta'(x)\hat{\sigma}_\perp, \tag{58}$$

with eigenvalues,

$$\mathcal{H}_\pm^{\rm MBO} = \frac{p^2}{2M} + \frac{(\hbar\theta'(x))^2}{8M} \pm \sqrt{(\mu\mathcal{B}(x))^2 + \left(\frac{\hbar\theta'(x)p}{2M}\right)^2}. \tag{59}$$

**Reflection and Dynamical Trapping.** We show agreement of exact and mixed quantum-classical dynamics using the MBOA for the setup described in Sec. 2 of the main text. The magnetic field profile is given by,

$$\theta(x) = \theta_0(1 + \mathrm{erf}(x/d)), \tag{60}$$

where $\mathrm{erf}(x)$ is the error function, such that the field rotates by a total angle of $2\theta_0$, and $d$ determines the width of the rotation region. Then, $\theta'(x) = 2\theta_0/(\sqrt{\pi}d)\exp\left[-x^2/d^2\right]$. The field strength $\mathcal{B}(x) = \mathcal{B}$ is assumed to be constant.

The phase space portraits of the Hamiltonians $\mathcal{H}_\pm^{\rm MBO}$ are qualitatively different as a function of the dimensionless parameter,

$$\zeta = \frac{\hbar^2\theta_0^2}{\pi M\mu\mathcal{B}d^2}. \tag{61}$$

Consider the MBO ground state $\mathcal{H}_-^{\rm MBO}$. When $\zeta < 1$, i.e. when the field rotates slowly, the energy minimum of $\mathcal{H}_-^{\rm MBO}(x,p)$ is at $p = 0$ for any value of $x$. This corresponds to the perturbative regime, where one can think of the particle as acquiring a space-dependent mass renormalization coming from the Taylor expansion of the square root in Eq. (59). Conversely, for $\zeta > 1$ there is a bifurcation and $\mathcal{H}_-^{\rm MBO}(x,p)$ is minimal at nonzero momentum $p$ when $x$ is sufficiently close to the origin, i.e. in the region of the fastest rotation of the magnetic field (see Fig. S2). At the point $\zeta = 1$, the mass renormalization at the origin $x = 0$ becomes exactly identical to the bare mass. Interestingly, the dispersion relation for $\zeta > 1$ is identical to the one in systems with Spin-Orbit coupling (see e.g. Fig. 2 in Ref. (13)). One can view this toy model as a setup where the spin-orbit coupling is turned on and then turned off as the particle travels through space.

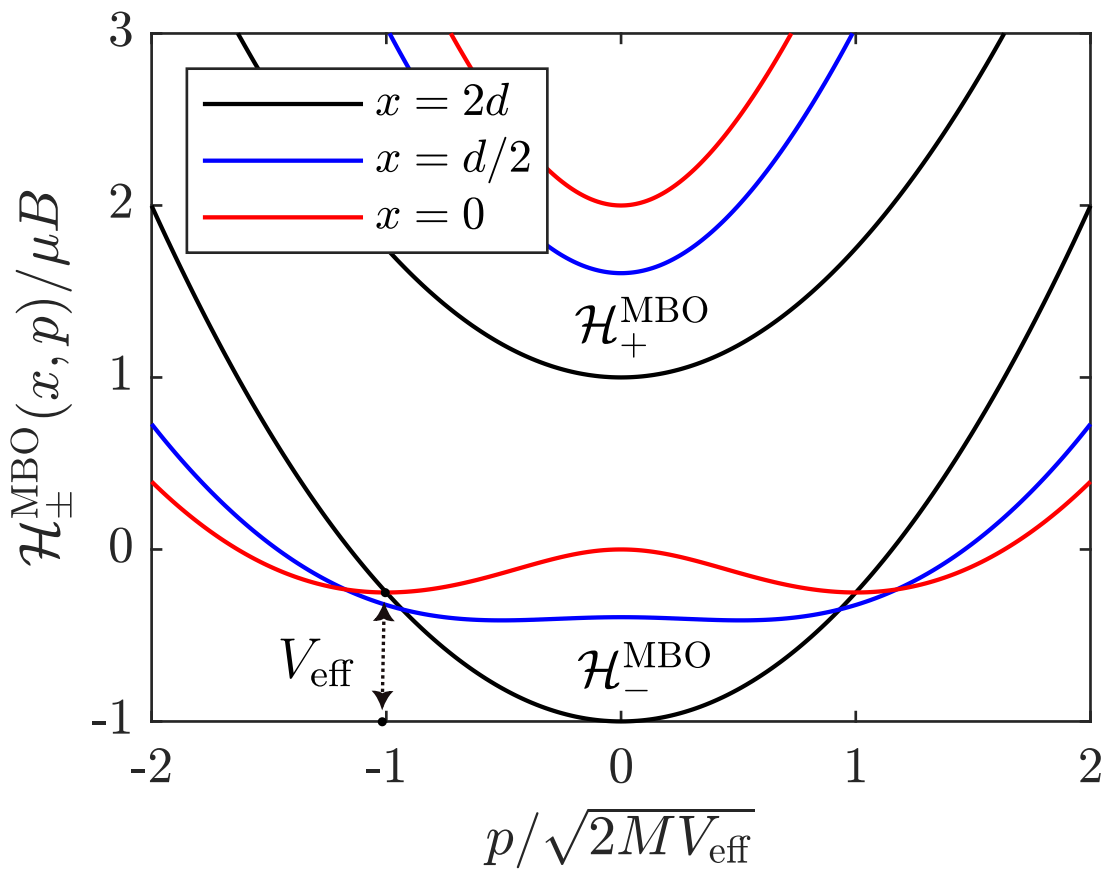

**Fig. S2.** Momentum dispersions of $\mathcal{H}_\pm^{\rm MBO}(x,p)$ at different slices of $x$ for $\zeta = 2$. For small $|x|/d$, the bottom branch $\mathcal{H}_-^{\rm MBO}$ is minimal at non-zero $p$, resulting in the double-well dispersion shown in red. The difference between the energy minima at $|x|/d \gg 1$ and $|x|/d = 0$ functions as an effective energetic barrier of height $V_{\rm eff}$ at the origin.

Now, consider a setup where the particle is prepared in the BO ground state at $x \to -\infty$, where $\theta(x)$ is constant, and moves towards the origin with a mean momentum $p_0$. The energy minimum of $\mathcal{H}_-^{\mathrm{MBO}}(x,p)$ at $x=0$ is denoted as $\mathcal{E}_{\min}$, and given by,

$$\mathcal{E}_{\min} = \begin{cases} \mu\mathcal{B}\big(\frac{\zeta}{2}-1\big), & \zeta < 1, \\ -\frac{\mu\mathcal{B}}{2\zeta}, & \zeta > 1. \end{cases} \qquad [62]$$

The field rotation creates an effective energetic barrier of height $V_{\mathrm{eff}} \equiv \mathcal{E}_{\min} + \mu\mathcal{B}$ at the origin. If the initial kinetic energy of the particle is greater than this energetic barrier $p_0^2/(2M) > V_{\mathrm{eff}}$, it is able to pass through the region of field rotation. Conversely, if $p_0^2/(2M) < V_{\mathrm{eff}}$, then it is reflected back.

Note that the effective barrier for the case $\zeta < 1$ can be understood from the metric correction to the BO surface. However, this is not the case for the non-perturbative regime $\zeta > 1$. As we increase the rate of rotation more and more at fixed value of $\mathcal{B}$ such that $\zeta \gg 1$, the effective barrier stops increasing with the rotation rate, as suggested by the diagonal Born-Huang (metric) correction, and instead plateaus to a constant value of $V_{\mathrm{eff}} = \mu\mathcal{B}$ corresponding to $\mathcal{E}_{\min} = 0$. This can be understood intuitively by noticing that this barrier corresponds to the gain in potential energy that a spin initially aligned with the magnetic field experiences as it tilts behind the field at the maximum angle of $\phi = \pi/2$. To obtain the correct value for the effective barrier, it is thus key that the average $\langle \hat{\mathbf{H}}^{\mathrm{M}} \rangle$ is computed with respect to the MBO, and not BO, ground state. In this section we work in the latter regime with $\zeta = 2$, corresponding to the phase space energy contours shown in Fig. 3(a).

To compare MBOA with exact simulations, we study the time evolution of a wavepacket where the spin is initially aligned with the local magnetic field,

$$|\Psi(t=0)\rangle \propto \int \mathrm{d}x e^{-(x-x_0)^2/4\sigma_x^2} e^{\frac{i}{\hbar}p_0 x} |x\rangle \otimes \left|\psi_-^{\mathrm{BO}}(x)\right\rangle . \qquad [63]$$

Here, $\left|\psi_-^{\mathrm{BO}}(x)\right\rangle$ denotes a spin aligned with $\vec{\mathcal{B}}(x)$, i.e. a spin in the local BO ground state. We choose $x_0$ to be far away from any field rotation - concretely at $x_0 = -3.5d$. In Fig. S3(a), we show the exact time evolution of $|\Psi(t)\rangle$ and compare it to the MBOA, which correctly predicts the reflection from the origin. In the two lower panels, Figs. S3(c) and S3(d), we compare, for the same setup, the velocity and force computed from the equations of motion, $\dot{x} = \langle \partial_p \hat{\mathbf{H}}^{\mathrm{M}} \rangle$ and $\dot{p} = -\langle \partial_x \hat{\mathbf{H}}^{\mathrm{M}} \rangle$, where the average is taken with respect to the BO (solid) or MBO (dashed) ground state. The red curve shows exact numerical values. For the velocity, averaging with respect to the BO state in our chosen gauge gives $\langle \partial_p \hat{\mathbf{H}}^{\mathrm{M}} \rangle_{\mathrm{BO}} = \frac{p}{M}$, while with respect to the MBO state yields $\langle \partial_p \hat{\mathbf{H}}^{\mathrm{M}} \rangle_{\mathrm{MBO}} = \frac{p}{M} - \frac{\hbar\theta'(x)}{2M}\sin\phi(x,p)$. This distinction is important to recover the correct velocity and force on the particle.

Finally let us investigate the dynamics of the trapped orbits shown in red in Fig. 3 of the main text. We consider an initial state of the form of Eq. (63), but with $x_0 = p_0 = 0$. The width of the wavepacket is chosen as $\sigma_x = d/10$, such that the packet is well within the separatrix of $\mathcal{H}_-^{\mathrm{MBO}}$. Within the BOA, we expect the packet to spread on a timescale set by $\tau \sim M\sigma_x^2/\hbar$. However, in the MBOA, only trajectories in $\mathcal{H}_+^{\mathrm{MBO}}$ spread; orbits in $\mathcal{H}_-^{\mathrm{MBO}}$ remain bounded at the origin. We therefore expect a fraction of the probability density $|\Psi(x)|^2$ to remain localized near the origin even after long times. Semiclassically, the trapped probability can be estimated by calculating the projection of the (dressed) Wigner function,

$$\hat{\mathbf{W}}^{\mathrm{M}}(x,p) = W(x,p) \begin{pmatrix} \cos^2\left(\frac{\phi(x,p)}{2}\right) & \frac{\sin\phi(x,p)}{2} \\ \frac{\sin\phi(x,p)}{2} & \sin^2\left(\frac{\phi(x,p)}{2}\right) \end{pmatrix}, \qquad [64]$$

onto the MBO ground state manifold. Here, we have defined $W(x,p) = 2\exp\left(-\frac{(x-x_0)^2}{2\sigma_x^2} - \frac{2\sigma_x^2(p-p_0)^2}{\hbar^2}\right)$, and explicitly shown the matrix form of $\hat{\mathbf{W}}^{\mathrm{M}}(x,p)$ in the basis of MBO states. Note that for non-zero $\phi$, the non-vanishing off-diagonal elements show that an initial BO state is a superposition of the dressed states.

The projection of the Wigner function onto the MBO ground state manifold is then found as,

$$P_{\mathrm{trapped}} \approx \int \frac{\mathrm{d}x\mathrm{d}p}{2\pi\hbar} W(x,p)\cos^2\left(\frac{\phi(x,p)}{2}\right). \qquad [65]$$

In the panel S3(b), we plot the computed total probability of a particle to be contained within the region $x \in (-d/2, d/2)$, with $P_{\mathrm{trapped}} = \int_{-d/2}^{d/2} \mathrm{d}x |\Psi(x)|^2$. The red curve shows exact dynamics, which confirms that $P_{\mathrm{trapped}}$ quickly decays as the wavefunction in $\mathcal{H}_+^{\mathrm{MBO}}$ disperses, stabilizing close to the value predicted by Eq. (65) at long times.

**Interference.** We now study time evolution for the second setup discussed in Sec. 2 of the main text. The field profile has a constant rate of rotation $\theta'(x) = \theta'$ and magnitude $\mathcal{B}(x) = \mathcal{B}$. The initial state is given by Eq. (63). For non-zero $\theta'$ and $p$, the BO ground state is a superposition of the two MBO states at each $(x,p)$.

According to Eq. 38 of the main text, we evaluate the classical trajectories $(x_{mn}(t), p_{mn}(t))$ as the solution to Hamilton's equations of motion starting at initial condition $(x,p)$ and evolved with the Hamiltonian,

$$\mathcal{H}_{mn}^{\mathrm{MBO}} = \frac{\mathcal{H}_m^{\mathrm{MBO}} + \mathcal{H}_n^{\mathrm{MBO}}}{2}, \qquad [66]$$

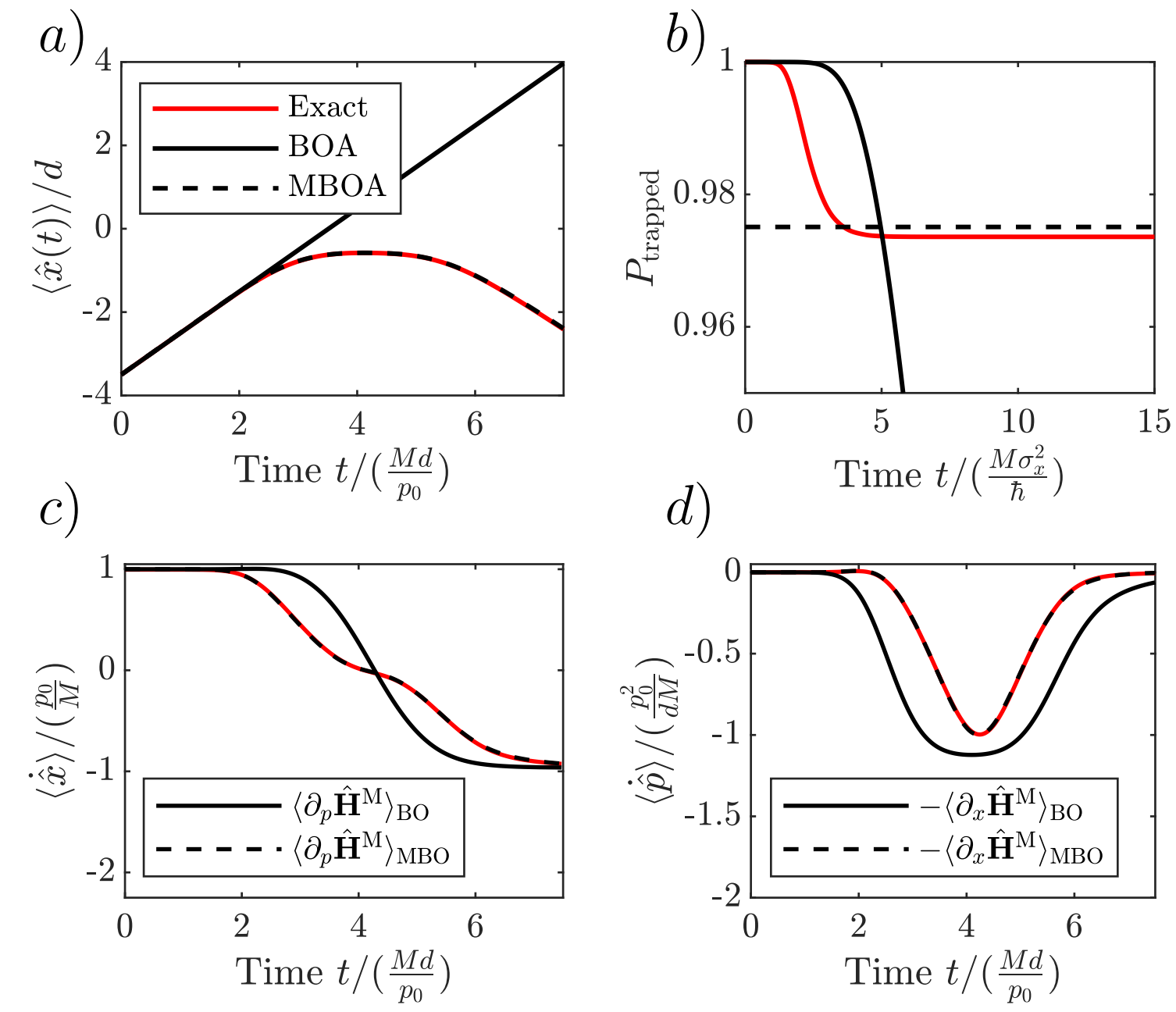

**Fig. S3.** Numerical comparison between exact and MBO dynamics for the setup shown in Sec. 2 of the main text. a) Reflection of a wavepacket from the origin. b) Dynamical trapping of a wavepacket within the region of bounded orbits in $\mathcal{H}_-^{\rm MBO}$. The quantity $P_{\rm trapped}$ is defined as the total probability contained in the interval $x \in (-d/2, d/2)$. c) For the setup shown in panel (a), comparison of exact mean velocity $\mathrm{d}\langle\hat{x}\rangle/\mathrm{d}t$ (red line) with that obtained by averaging $\partial_p \hat{\mathbf{H}}^{\rm M}$ with respect to the MBO (dotted black line) or BO (solid black line) ground state. d) Same as panel c) but for $\mathrm{d}\langle\hat{p}\rangle/\mathrm{d}t$. The reflection experiment was carried out with a magnetic field profile with $\theta_0 = 20\sqrt{2\pi}$ and $\zeta = 2$. The initial wavepacket has position $x_0 = -3.5d$, momentum $p_0 = (1/\sqrt{2\pi})\hbar\theta_0/d$, and width $\sigma_x = d/4$. The dynamical trapping experiment was carried out with the same magnetic field profile and initial wavepacket parameters $x_0 = p_0 = 0$ and $\sigma_x = d/10$.

with $\mathcal{H}_m^{\rm MBO}$ given by Eq. (59). For constant field magnitude $\mathcal{B}$ and rate of rotation $\theta'$, the diagonal $m = n$ solutions are given by,

$$\dot{x}_{--}(t) \equiv \dot{x}_-(t) = \frac{p_-(t)}{M} - \frac{\hbar\theta'}{2M}\sin\phi(p_-(t)), \qquad \dot{p}_{--}(t) \equiv \dot{p}_-(t) = 0,$$
$$\dot{x}_{++}(t) \equiv \dot{x}_+(t) = \frac{p_+(t)}{M} + \frac{\hbar\theta'}{2M}\sin\phi(p_+(t)), \qquad \dot{p}_{++}(t) \equiv \dot{p}_+(t) = 0, \qquad [67]$$

with $\phi(p)$ given by Eq. (57). The off-diagonal $m \neq n$ solutions satisfy,

$$\dot{x}_{-+}(t) = \frac{p_{-+}(t)}{M}, \quad \dot{x}_{+-}(t) = \frac{p_{+-}(t)}{M}, \quad \dot{p}_{-+}(t) = \dot{p}_{+-}(t) = 0. \qquad [68]$$

Since the momenta are conserved for all time, $p_{mn}(t) = p$, the integral for the phase of the Weyl symbol is trivial,

$$\Phi_{+-}(t) = \frac{1}{\hbar}\int_0^t \mathrm{d}\tau \left[\mathcal{H}_+^{\rm MBO}(x_{+-}(\tau), p_{+-}(\tau)) - \mathcal{H}_-^{\rm MBO}(x_{+-}(\tau), p_{+-}(\tau))\right], \qquad [69]$$
$$= \frac{2t}{\hbar}\sqrt{(\mu\mathcal{B})^2 + \left(\frac{\hbar\theta' p}{2M}\right)^2} \equiv \frac{t\Delta\mathcal{H}(p)}{\hbar},$$

and $\Phi_{-+}(t) = -\Phi_{+-}(t)$. We have abbreviated $\Delta\mathcal{H}(p) = \mathcal{H}_+^{\rm MBO}(p) - \mathcal{H}_-^{\rm MBO}(p)$.

Finally, we obtain the time evolution of the dressed momentum Weyl symbol $\hat{\mathbf{q}}^{\rm M} = p\hat{\mathbf{I}} - \hat{\mathbf{A}}_x$. We explicitly show it in matrix form, where the matrix elements are shown in the eigenbasis of $\hat{\mathbf{H}}^{\rm M}$,

$$\hat{\mathbf{q}}^{\rm M}(t) = \begin{pmatrix} p - \frac{\hbar\theta'}{2}\sin\phi(p) & e^{-\frac{it\Delta\mathcal{H}(p)}{\hbar}}\frac{\hbar\theta'}{2}\cos\phi(p) \\ e^{\frac{it\Delta\mathcal{H}(p)}{\hbar}}\frac{\hbar\theta'}{2}\cos\phi(p) & p + \frac{\hbar\theta'}{2}\sin\phi(p) \end{pmatrix}.$$

Plugging back into Eq. 40 of the main text and using Eq. (64), we obtain,

$$\langle\hat{q}(t)\rangle \approx p_0 + \frac{\hbar\theta'}{4}\int \frac{\mathrm{d}x\mathrm{d}p}{2\pi\hbar} W(x,p)\sin(2\phi(p))\left(\cos\left(\frac{t}{\hbar}\Delta\mathcal{H}(p)\right) - 1\right), \qquad [70]$$

which results in Fig. 3(b). The solution at long times is obtained approximately as $\lim_{t\to\infty}\langle\hat{q}(t)\rangle \approx p_0 - \frac{\hbar\theta'}{4}\sin(2\phi(p_0))$, when the oscillating phases $\sim \cos(t\Delta\mathcal{H}/\hbar)$ coming from different initial conditions $(x,p)$ dephase, and where we have replaced $\langle\sin(2\phi(p))\rangle \approx \sin(2\phi(p_0))$.

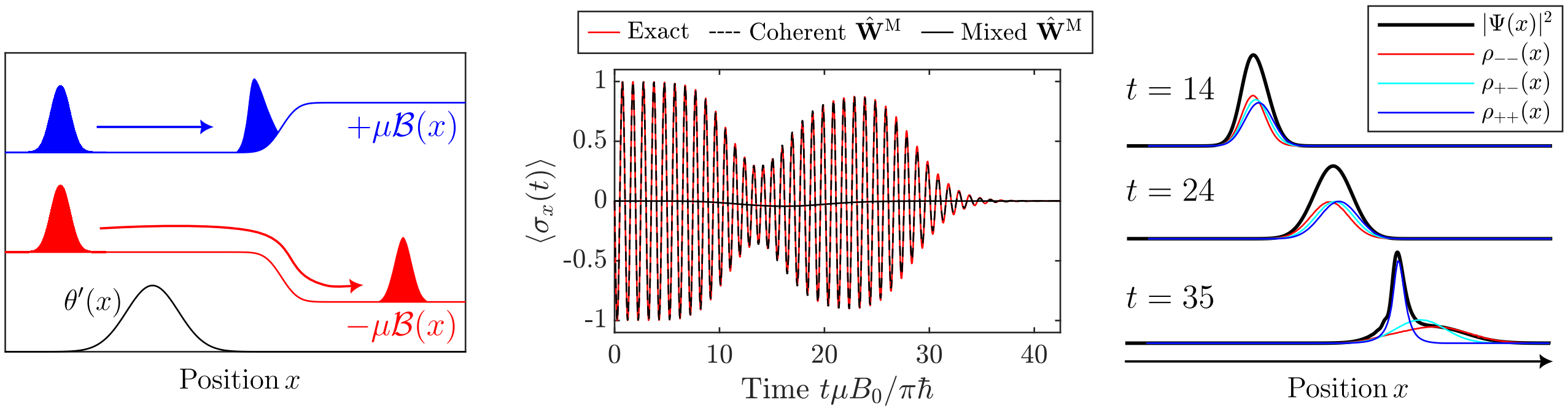

**Fig. S4.** Spin precession and decoherence. Left panel: schematic description of the experiment. The spin is initially polarized in a direction perpendicular to the magnetic field, such that the particle is in a superposition of both adiabatic surfaces. The particle first crosses a region with non-zero field rotation $\theta'(x)$, then enters a zone where $\mathcal{B}(x)$ increases smoothly. The upper-surface component is reflected while the lower component is transmitted. Middle panel: corresponding dynamics of the spin component along the direction of motion $\langle\hat{\sigma}_x(t)\rangle$ for the setup on the left. While the wavepackets on each adiabatic surface have spatial overlap, the spin exhibits coherent Larmor precession. The amplitude of the oscillations is reduced as the spin tilts away from magnetic field plane while crossing the region of non-zero rotation. At the reflection point, the Larmor oscillations decohere as the two wavepackets separate in space. Right panel: snapshots of the spatial probability density of the particle at selected times. The black curve shows the exact probability density $|\Psi(x)|^2$, while the colored curves show semiclassical probability distributions obtained from the $x_{mn}(t)$ trajectories sampled from the Wigner function.

**Spin Precession.** We now demonstrate that the off–diagonal $(m \neq n)$ contributions in Eq. (28) are essential for capturing coherent phenomena such as spin precession about a magnetic field. Consider a setup inspired by Fig. 2 of Ref. (14) and Tully's third problem (15). The particle is initially prepared in a wavepacket far from any field rotation, where the magnetic field points along the $z$–axis. The spin is polarized along $-y$, so the initial state is a coherent superposition of both BO surfaces (see the left panel of Fig. S4). The wavepacket first traverses a region with non-zero field rotation $\theta'(x)$, which couples the BO surfaces. It then enters a region where the field magnitude increases slowly in space, so that the packet on the ground MBO surface is transmitted while the packet on the excited one is reflected. The specific magnetic–field profile and parameters are,

$$\mathcal{B}(x) = \mathcal{B}_0 + \frac{\mathcal{B}_0}{2}\left(1 + \operatorname{erf}\left(\frac{2x}{d}\right)\right), \qquad \theta'(x) = \frac{2\theta_0}{\sqrt{\pi}d}\exp\left\{-\frac{(x - x^0_{\theta'})^2}{d^2}\right\}, \tag{71}$$

with $x^0_{\theta'} = -3.5d$ and $\theta_0 = 5\pi$. The wavepacket is prepared in a Gaussian with,

$$|\Psi(t=0)\rangle \propto \int \mathrm{d}x e^{-(x-x_0)^2/4\sigma_x^2} e^{\frac{i}{\hbar}p_0 x}\,|x\rangle \otimes \left(\frac{1}{\sqrt{2}}\left(\left|\psi^{\mathrm{BO}}_-(x)\right\rangle - i\left|\psi^{\mathrm{BO}}_+(x)\right\rangle\right)\right), \tag{72}$$

and $\sigma_x = d/4$, $x_0 = -6d$, $p_0 \approx 1.59\hbar\theta_0/d$, and $p_0^2/2M \approx 0.7812\mu\mathcal{B}_0$.

The middle panel of Fig. S4 shows the time evolution of the spin component along the direction of motion, $\langle\hat{\sigma}_x(t)\rangle$. At short times we observe sharp oscillations with period $T_S = \pi\hbar/(\mu\mathcal{B}_0)$, reflecting spin precession about $\vec{\mathcal{B}}$. As the particle enters the region with $\theta'(x) \neq 0$, the amplitude of the $x$-component of the Larmor oscillation is reduced. The spin lags behind the real magnetic field $\vec{\mathcal{B}}$ and instead precesses about an effective momentum-dependent magnetic field $\vec{\mathcal{B}}_{\mathrm{eff}}$ that is tilted in the $x$-direction (see discussion around Eq. 15 of the main text). At the reflection point, the oscillations of $\langle\hat{\sigma}_x(t)\rangle$ decay as the ground and excited wavepackets separate in space. At long times, the two wavepackets are fully separated, the reduced spin density matrix is maximally mixed, and phase coherence is lost.

We now show how the spin dynamics is captured within the MBOA. Recall that the MBO unitary is given by $\hat{\mathbf{U}}_{\mathrm{MBO}}(x,p) = \exp\left\{i\frac{\phi(x,p)}{2}\hat{\sigma}_y\right\}$, with $\phi$ defined in Eq. (57), so that $\hat{\mathbf{U}}^\dagger_{\mathrm{MBO}}\,\hat{\mathbf{H}}^{\mathrm{M}}\,\hat{\mathbf{U}}_{\mathrm{MBO}}$ is diagonal. Because $\hat{\sigma}_x$ is a $p$–independent operator, it does not acquire dressing upon transforming to the moving frame; that is, $\hat{\sigma}_x^{\mathrm{M}} = \hat{\sigma}_x$. Using that $\hat{\mathbf{U}}^\dagger_{\mathrm{MBO}}\hat{\sigma}_x\hat{\mathbf{U}}_{\mathrm{MBO}} = \hat{\sigma}_x\cos\phi(x,p) - \hat{\sigma}_z\sin\phi(x,p)$ and applying Eq. (28), the time-evolved matrix elements of $\hat{\sigma}_x(t)$ in the basis of MBO states read,

$$\hat{\sigma}_x(t) = \begin{pmatrix} -\sin\phi(x_{--}(t), p_{--}(t)) & e^{i\Phi_{-+}(t)}\cos\phi(x_{-+}(t), p_{-+}(t)) \\ e^{i\Phi_{+-}(t)}\cos\phi(x_{+-}(t), p_{+-}(t)) & \sin\phi(x_{++}(t), p_{++}(t)) \end{pmatrix}, \tag{73}$$

with the equations of motion for $(x_{mn}(t), p_{mn}(t))$ and $\Phi_{mn}(t)$ given by Eqs. (18, 36, 37, 39) of the main text. To obtain the full expectation value $\langle\hat{\sigma}_x(t)\rangle$, we average over initial conditions sampled from the (dressed) Wigner function $\hat{\mathbf{W}}^{\mathrm{M}}(x,p)$ as in Eq. (29). In the MBO basis,

$$\hat{\mathbf{W}}^{\mathrm{M}}(x,p) = \frac{W(x,p)}{2}\begin{pmatrix} 1 & i \\ -i & 1 \end{pmatrix}, \tag{74}$$

where $W(x,p) = 2\exp\left(-\frac{(x-x_0)^2}{2\sigma_x^2} - \frac{2\sigma_x^2(p-p_0)^2}{\hbar^2}\right)$. Finally,

$$\langle\hat{\sigma}_x(t)\rangle \approx \int\frac{\mathrm{d}x\mathrm{d}p}{2\pi\hbar}\mathrm{Tr}\left\{\hat{\mathbf{W}}^{\mathrm{M}}(x,p)\hat{\sigma}_x(t)\right\} = \int\frac{\mathrm{d}x\mathrm{d}p}{2\pi\hbar}W(x,p)\left(\frac{1}{2}\left(\sin[\phi_{++}(t)] - \sin[\phi_{--}(t)]\right) - \sin[\Phi_{+-}(t)]\cos[\phi_{+-}(t)]\right), \tag{75}$$

where we have abbreviated $\phi_{mn}(t) \equiv \phi(x_{mn}(t), p_{mn}(t))$. The first two terms are simply the average of the trajectories moving independently on each MBO surface. The second term contains the oscillating phase factor $\Phi_{+-}(t) = -\Phi_{-+}(t)$ that is responsible for capturing the interference between adiabatic surfaces leading to spin precession.

To visualize the different trajectories $x_{mn}(t)$, the right panel of Fig. S4 shows the spatial densities at select instants of time obtained from both the exact wavefunction $|\Psi(x,t)|^2$ and the distribution of semiclassical trajectories $\rho_{mn}(t)$ sampled from the initial Wigner function and propagated with Hamiltonian $\mathcal{H}^{\mathrm{MBO}}_{mn}$. The ensemble of $x_{--}(t)$ trajectories, corresponding to evolution on the ground MBO surface, are transmitted through the origin, while the $x_{++}(t)$ trajectories are reflected. The trajectories $x_{+-}(t) = x_{-+}(t)$ are non-trivial solutions which evolve according to the Hamiltonian $\mathcal{H}^{\mathrm{MBO}}_{+-} = p^2/2M + (\hbar\theta'(x))^2/8M$. Interestingly, they are not observable through the particle position $\langle \hat{x}(t)\rangle$. Indeed, since the position operator remains undressed upon transforming to the moving frame $\hat{\mathbf{x}}^{\mathrm{M}} = x\,\hat{\mathbf{I}}$, its time evolution within the MBOA is simply given by $\langle \hat{x}(t)\rangle = \langle x_{--}(t) + x_{++}(t)\rangle/2$. This can be seen clearly from the snapshot at time $t = 35$. However, the $x_{+-}(t)$ trajectories do play a role when computing time evolution of *spin-dependent* observables, including the dressed momentum of the particle as in Fig. 3 of the main text. They evolve somewhere in between the ground and excited surface trajectories, so that the phase $\Phi_{+-}(t)$ is evaluated roughly along the mean position of the total wavepacket, as shown in the right panel of Fig. S4.

The resulting $\langle \hat{\sigma}_x(t)\rangle$ computed from the MBOA is shown as the dotted black curve in the middle panel of Fig. S4. The decrease in the amplitude of the $x$-component of the Larmor precession at around $t \approx 14$ appears in the MBOA calculation as a decrease in the value of $\cos\phi_{+-}(t)$ as the $x_{+-}(t)$ trajectories enter the region of non-zero $\theta'$. At around $t \approx 35$, the excited state is reflected, causing spatial separation of the wavepackets and a decoherence in the Larmor oscillations. In the MBOA calculation, this appears as a dephasing of the $\Phi_{+-}(t)$ associated to different initial conditions, which average out to zero at $t \gtrsim 35$.

For comparison, if we set the off–diagonal elements of $\hat{\mathbf{W}}^{\mathrm{M}}$ to zero, i.e. $\hat{\mathbf{W}}^{\mathrm{M}}(x,p) = (W(x,p)/2)\,\hat{\mathbf{I}}$, the initial state is an incoherent mixture and the evolution is simply the weighted sum of the dynamics on the two adiabatic surfaces, without the last term containing $\Phi_{+-}(t)$ in Eq. (75). This "mixed" $\hat{\mathbf{W}}^{\mathrm{M}}$ fails to reproduce the oscillations of the spin as it precesses about the magnetic field. In contrast, the "coherent" $\hat{\mathbf{W}}^{\mathrm{M}}$ (keeping the off–diagonal terms) captures both the initial oscillations and their eventual decay.

**Corrections Beyond the MBOA.** In this section, we discuss the implications of further super-adiabatic corrections to the dressing of operators Eq. (51) in the spin-1/2 toy model. Using the explicit expression for $\hat{\mathbf{G}}_{\mathrm{MBO}}$ in Eq. (56), the AGPs associated with $\hat{\mathbf{U}}_{\mathrm{MBO}}$ are,

$$\hat{\mathbf{B}}'_x = (i\hbar\partial_x \hat{\mathbf{U}}_{\mathrm{MBO}})\hat{\mathbf{U}}^\dagger_{\mathrm{MBO}} = -\frac{\hbar\partial_x\phi(x,p)}{2}\hat{\sigma}_y, \qquad [76]$$

$$\hat{\mathbf{B}}'_p = (i\hbar\partial_p \hat{\mathbf{U}}_{\mathrm{MBO}})\hat{\mathbf{U}}^\dagger_{\mathrm{MBO}} = -\frac{\hbar\partial_p\phi(x,p)}{2}\hat{\sigma}_y. \qquad [77]$$

Assuming constant field magnitude $\mathcal{B}(x) = \mathcal{B}$ and rate of rotation $\theta'(x) = \theta'$, we compute the leading order correction to the dressing of $\hat{\mathbf{H}}^{\mathrm{SA}}$, $\hat{\mathbf{x}}^{\mathrm{SA}}$, and $\hat{\mathbf{q}}^{\mathrm{SA}}$. Using $\partial_x\hat{\mathbf{A}}_x = \hat{\mathbf{B}}_x = 0$, Eq. (54) and Eq. (55) reduce to,

$$\hat{\mathbf{H}}^{\mathrm{SA}} = \hat{\mathbf{H}}^{\mathrm{M}}, \quad \hat{\mathbf{q}}^{\mathrm{SA}} = \hat{\mathbf{q}}^{\mathrm{M}}, \quad \hat{\mathbf{x}}^{\mathrm{SA}} = \hat{\mathbf{x}}^{\mathrm{M}} + \hat{\mathbf{B}}_p = x\,\hat{\mathbf{I}} + \hat{\mathbf{B}}_p. \qquad [78]$$

Note that the Hamiltonian and momentum do not acquire additional super-adiabatic dressing, but the coordinate does. In particular, $\hat{\mathbf{x}}^{\mathrm{SA}}$ becomes an operator in $\mathbb{H}^Q$. Interestingly, all higher order corrections vanish, and Eq. (78) is exact. Since $\hat{\mathbf{H}}^{\mathrm{SA}} = \hat{\mathbf{H}}^{\mathrm{M}}$, we conclude that the MBOA is exact for constant field magnitude and rate of rotation. This is not surprising, since the system has discrete translational symmetry, and the Hamiltonian can be exactly block-diagonalized through application of Bloch's theorem.

To illustrate the implications of the super-adiabatic dressing on dynamics, we compute the time evolution of $\hat{\mathbf{x}}^{\mathrm{SA}}$ and $\hat{\mathbf{q}}^{\mathrm{SA}}$. Since $\hat{\mathbf{H}}^{\mathrm{SA}} = \hat{\mathbf{H}}^{\mathrm{M}}$, the solutions to the equations of motion remain the same and are given by Eq. (67). Applying Eq. 38 of the main text starting at initial condition $(x,p)$, we obtain,

$$\hat{\mathbf{x}}^{\mathrm{SA}}(t) = \begin{pmatrix} x + \left(\frac{p}{M} - \frac{\hbar\theta'}{2M}\sin\phi\right)t & ie^{-\frac{it\Delta\mathcal{H}}{\hbar}}\frac{\hbar\phi'}{2} \\ -ie^{\frac{it\Delta\mathcal{H}}{\hbar}}\frac{\hbar\phi'}{2} & x + \left(\frac{p}{M} + \frac{\hbar\theta'}{2M}\sin\phi\right)t \end{pmatrix}, \qquad [79]$$

and,

$$\hat{\mathbf{q}}^{\mathrm{SA}}(t) = \begin{pmatrix} p - \frac{\hbar\theta'}{2}\sin\phi & e^{-\frac{it\Delta\mathcal{H}}{\hbar}}\frac{\hbar\theta'}{2}\cos\phi \\ e^{\frac{it\Delta\mathcal{H}}{\hbar}}\frac{\hbar\theta'}{2}\cos\phi & p + \frac{\hbar\theta'}{2}\sin\phi \end{pmatrix}. \qquad [80]$$

Here, all matrix elements are shown in the basis of MBO states. Because the position operator also acquires off-diagonal matrix elements, the transient oscillations of the quantity $\langle \hat{q}(t)\rangle$ described in Fig. 3(b) are also present in the time-evolution of $\langle \hat{x}(t)\rangle$. This is to be expected, since it is required from the Heisenberg equation,

$$\frac{\mathrm{d}\hat{x}(t)}{\mathrm{d}t} = \frac{\hat{q}(t)}{M}.$$

One can now verify that, by including super-adiabatic dressing, the solutions Eq. (79) and Eq. (80) correspondingly satisfy $\mathrm{d}\hat{\mathbf{x}}^{\mathrm{SA}}/\mathrm{d}t = \hat{\mathbf{q}}^{\mathrm{SA}}/M$. This equality is violated without the contribution from the super-adiabatic dressing of the coordinate. In the regime of applicability of the MBOA, the coordinate oscillations are, however, relatively weaker than the momentum oscillations, as they are suppressed by an additional factor of $M$.

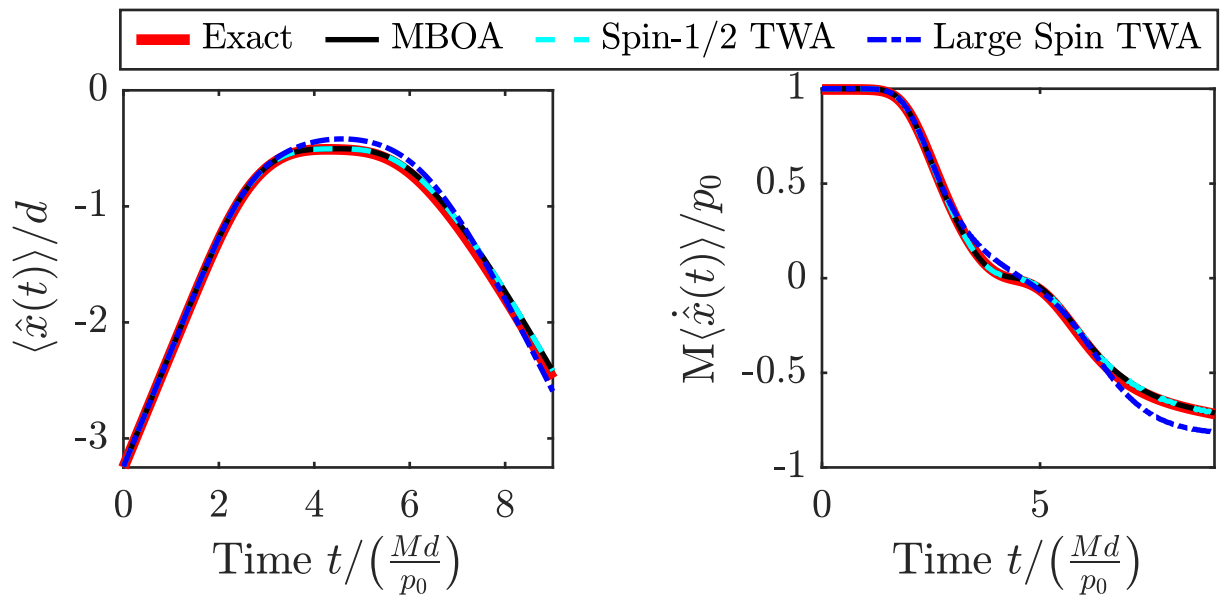

**Fig. S5.** Reflection of a spin-1/2 particle from a region of magnetic field rotation. The curves show dynamics obtained from different approximation schemes. In particular, while the spin-1/2 TWA generated by Eq. (84) agrees with the MBOA and exact dynamics, the large spin TWA given by Eq. (83) is slightly off.

**Fluctuations and Quantum Metric.** This section details the calculations for Fig. 6(b). The experiment corresponds to the same setup as in Fig. S3, with slightly different parameters. The magnetic field has a constant magnitude $\mathcal{B}(x) = \mathcal{B}$ and a rotation profile $\theta(x) = \theta_0(1 + \mathrm{erf}(x/d))$, with $\zeta = 2.5$ as defined in Eq. (61) and $\theta_0 = 50.5$. The spin is prepared initially polarized in the field direction and in a wavepacket with mean position $x_0 = -3.25d$, momentum $p_0 = \hbar\theta_0/(\sqrt{2\pi}d)$, and width $\sigma_x = d/4$.

To visualize the quantum fluctuations in the time evolution of the physical momentum $\hat{q}$, Fig. S5 shows dynamics carried out using the full Truncated Wigner Approximation. Generally the TWA is justified in the large spin $S$ limit, therefore it is natural to extend the Hamiltonian to an arbitrary spin $S$. Within the TWA the the full quantum dynamics under the Hamiltonian $\hat{H} = \hat{q}^2/2M - \mu\,\vec{\mathcal{B}}(\hat{x}) \cdot \hat{\vec{S}}/S$ can be approximated by the dynamics of a classical dimensionless spin $\vec{S}$ and coordinate $(x, q)$ with equations of motion,

$$\dot{x} = \frac{\partial H}{\partial q}, \quad \dot{q} = -\frac{\partial H}{\partial x}, \quad \hbar\dot{\vec{S}} = -\vec{S} \times \frac{\partial H}{\partial \vec{S}}, \tag{81}$$

where $H(x, q, \vec{S}) = q^2/2M - \mu\vec{\mathcal{B}}(x) \cdot (\vec{S}/S)$ [‖]. Note that within the TWA the magnitude of the classical spin vector is $|\vec{S}| = \sqrt{S(S+1)}$, where an extra $S$ comes from the transversal spin quantum fluctuations. Let us point out that if we treat $\hat{q}$ and $\hat{x}$ classically, like it is done in the BOA, the TWA approximation is exact for any spin $S$ because the corresponding Hamiltonian is linear in spin operators (3). However, for quantum $\hat{q}$ and $\hat{x}$ this is not the case, and the TWA is generally exact only in the large $S$ limit.

One can also write the classical equations of motion given by Eq. (81) in the moving frame. To do so, we define a transformation of the lab frame spin and momentum into the new canonical variables $p$ and $\vec{S}'(x)$ such that,

$$p = q - \hbar S_x\theta'(x), \quad \vec{S}'(x) = R(\theta(x))\vec{S}. \tag{82}$$

Here, $R(\theta(x))$ denotes a matrix that rotates the spin by an angle $\theta(x)$ in the $y$-$z$ plane, in such a way that the $z$ component of the primed spin coincides with the direction of the local magnetic field $\vec{\mathcal{B}}(x)$. Eq. (82) preserves both coordinate and spin Poisson brackets. The Hamiltonian written in the moving coordinates reads,

$$H^{\mathrm{M}}_{S\to\infty}(x, p, \vec{S}') = \frac{(p + \hbar S'_x\theta'(x))^2}{2M} - \mu\mathcal{B}(x)\frac{S'_z}{S}. \tag{83}$$

Notice that this Hamiltonian is no longer linear in the spin and hence, unlike in the BOA, the TWA is not expected to be exact for small $S$. The resulting dynamics generated by Eq. (83) is labelled as Large Spin TWA in Fig. S5.

Interestingly, one can observe that if we first perform the unitary transformation to the moving frame while keeping all operators quantum mechanical, and only after replace spin operators with classical spin vectors, we obtain a different classical Hamiltonian. Indeed, from Eq. 15 of the main text it follows that,

$$H^{\mathrm{M}}_{S=1/2}(x, p, \vec{S}') = \frac{p^2}{2M} + \frac{(\hbar\theta'(x))^2}{8M} - \mu\mathcal{B}(x)\frac{S'_z}{S} + \frac{\hbar S'_x\theta'(x)p}{M}. \tag{84}$$

Because for spin-1/2 the moving Hamiltonian is again linear in spin, the TWA in the moving frame is exact as long as we can treat $p$ and $x$ classically. Notice that in contrast to Eq. (83), there is no $(S'_x)^2$ term in the moving TWA Hamiltonian Eq. (84). Therefore, this Hamiltonian in general produces different dynamics, which are shown under the label of spin-1/2 TWA in Fig. S5 and provide a better fit to the exact quantum result. This simple example thus illustrates a general result that the accuracy of the semiclassical approximation generally depends on the choice of reference frame.

To carry out the TWA, both the slow DOF and spin start from some initial condition $(x(0), p(0))$ and $\vec{S}(0)$, and evolve according to their coupled equation of motion generated by either Eq. (83) or Eq. (84). Fig 6(b) shows a single trajectory corresponding to the initial condition $(x(0), p(0)) = (x_0, p_0)$ and $\vec{S}(0) = (\frac{1}{2}, \frac{1}{2}, \frac{1}{2})$, evolved with Eq. (84). To reproduce the full quantum mechanical result, we must however average over a distribution of initial conditions capturing quantum

[‖] Note that the factor of $\hbar$ appearing in the equations of motion comes from the fact that the classical angular momentum is $\vec{L} = \hbar\vec{S}$ and equations for $\vec{L}$ do not contain $\hbar$. In all numerical examples we set $\hbar = 1$.

noise. For a given $(x(0), p(0))$, the spin is sampled from 2 different initial conditions, $\vec{S}_{\pm}(0) = (\pm\frac{1}{2}, \pm\frac{1}{2}, \frac{1}{2})$, which are averaged with equal weight. The $\{(x(0), p(0))\}$ are then separately averaged over a distribution given by the Wigner function of the initial state. For a Gaussian wavepacket with width $\sigma_x$, the initial conditions are sampled from the distribution $W(x,p) = 2\exp\left(-\frac{(x-x_0)^2}{2\sigma_x^2} - \frac{2\sigma_x^2(p-p_0)^2}{\hbar^2}\right)$, producing Fig. S5. The TWA carried out using $H^{\mathrm{M}}_{S=1/2}$ both agrees with exact quantum dynamics and the MBOA result, while that obtained with $H^{\mathrm{M}}_{S\to\infty}$ is slightly off.

**Higher Dimensions and Berry Curvature.** We briefly comment on how the MBOA formalism can be extended to higher dimensions and general magnetic field profiles. In the example considered so far, the particle was limited to 1-d motion along the $x$-axis, and the magnetic field was confined to the $y$-$z$ plane. The motion of a spin-1/2 particle in a general spatially varying magnetic field has been studied before, and it is well-known that the leading corrections beyond the BOA produce a Lorentz-type force arising from the Berry curvature of the spin (8, 16–18). As we show below, the Lorentz-type force is recovered within the MBOA from the leading order in $p$ correction to the adiabatic state using the Kato AGP.

For a spin-$\frac{1}{2}$ in a general magnetic field profile $\vec{\mathcal{B}}(\vec{x})$, the components of the Kato AGP can be written compactly as,

$$\hat{\mathbf{A}}_{x_i} = \frac{\hbar}{2}[\vec{n}(\vec{x}) \times \partial_{x_i}\vec{n}(\vec{x})] \cdot \hat{\vec{\sigma}},$$

where $\vec{n}(\vec{x})$ is the unit vector of the magnetic field at position $\vec{x}$. The AGP is clearly in the Kato form as it contains only the spin components orthogonal to $\vec{n}$, which have zero expectation values for either of the two BO eigenstates. Thus, within the MBOA, the spin follows an effective momentum-dependent magnetic field given by,

$$\vec{\mathcal{B}}_{\mathrm{eff}}(\vec{x},\vec{p}) = \vec{\mathcal{B}}(\vec{x}) - \frac{\hbar}{2M\mu}\Big[\vec{n}(\vec{x}) \times (\vec{p}\cdot\nabla)\vec{n}(\vec{x})\Big] \equiv \vec{\mathcal{B}}(\vec{x}) - \delta\vec{\mathcal{B}}(\vec{x},\vec{p}) \qquad [85]$$

Note that $\delta\vec{\mathcal{B}}\cdot\vec{\mathcal{B}} = 0$. Treating $\delta\vec{\mathcal{B}}$ to first order in perturbation theory, we obtain the leading-order correction to the ground state expectation value,

$$\langle\hat{\vec{\sigma}}\rangle(\vec{x},\vec{p}) = \vec{n}(\vec{x}) - \left(\delta\vec{\mathcal{B}}/\mathcal{B}\right)(\vec{x},\vec{p}). \qquad [86]$$

The second term in Eq. (86) describes the $\vec{p}$-dependent tilt away from the instantaneous direction of the magnetic field $\vec{n}(\vec{x})$. There is thus a $\vec{p}$-dependent correction to the BO force,

$$\begin{aligned} F_i(\vec{x},\vec{p}) = \mu\langle\partial_{x_i}\vec{\mathcal{B}}(\vec{x})\cdot\hat{\vec{\sigma}}\rangle = \mu\partial_{x_i}\vec{\mathcal{B}}(\vec{x})\cdot\langle\hat{\vec{\sigma}}\rangle = \mu\partial_{x_i}\mathcal{B}(\vec{x}) - \frac{\hbar p_j}{2M}\partial_{x_i}\vec{n}\cdot(\vec{n}\times\partial_{x_j}\vec{n}) = \mu\partial_{x_i}\mathcal{B}(\vec{x}) - \frac{\hbar p_j}{2M}\vec{n}\cdot(\partial_{x_j}\vec{n}\times\partial_{x_i}\vec{n}), \\ \equiv \mu\partial_{x_i}\mathcal{B}(\vec{x}) + \Omega_{ij}\frac{p_j}{M}, \end{aligned} \qquad [87]$$

where we have identified $\Omega_{ij} \equiv \hbar\vec{n}\cdot(\partial_{x_i}\vec{n}\times\partial_{x_j}\vec{n})/2$ as the Berry curvature tensor with $\Omega_{ij} = -\Omega_{ji}$. Finally, rewriting in terms of the Berry curvature vector $\Omega_{ij} = \varepsilon_{ijk}\mathcal{F}_k$, which plays the role of a pseudo-magnetic field, we obtain the form,

$$\vec{F}(\vec{x},\vec{p}) = \mu\langle\nabla(\vec{\mathcal{B}}(\vec{x})\cdot\hat{\vec{\sigma}})\rangle = \mu\nabla\mathcal{B}(\vec{x}) + \frac{\vec{p}}{M}\times\vec{\mathcal{F}}(\vec{x}) \qquad [88]$$

The first term is the normal BO force, while the second is the Lorentz-like force containing the Berry curvature. We thus show how we recover this well-known result as the leading order correction in momentum of the MBOA. Within MBOA one can also go beyond the linearized regime and study Berry curvature effects in the non-perturbative regime. However, such an analysis is beyond the scope of this work.

## 7. Bell Pair Protocol

The aim of this section is to provide a proof of concept for motion-generated Bell pairs. In particular, we describe a specific protocol that implements the adiabatic path shown in Fig. 4(a). As discussed in main text Sec. 2, it is composed of 3 parts. The spins are initially polarized in the $|\uparrow\uparrow\rangle$ ground state in a region of non-rotating field with magnitude $\mathcal{B}_0$, with initial center of mass momentum $p_0$. They cross through a region of space (near $x = 0$) where the rate of field rotation $\theta'(x)$ is adiabatically ramped up to some $\theta'_0$ while keeping the field magnitude large, $\zeta = \frac{(\hbar\theta'_0)^2}{4M\mu\mathcal{B}_0} \ll 1$. They finally enter a region where the field strength is slowly reduced to zero as $x \to \infty$. For concreteness, we parameterize the the magnetic field as,

$$\mathcal{B}(x) = \frac{\mathcal{B}_0}{2}\left(1 + \mathrm{erf}\left(\frac{x^0_{\mathcal{B}} - x}{d_{\mathcal{B}}}\right)\right), \qquad [89]$$

$$\theta'(x) = \frac{\theta'_0}{2}\left(1 + \mathrm{erf}\left(\frac{x - x^0_{\theta'}}{d_{\theta'}}\right)\right). \qquad [90]$$

Here, the length scales $d_{\mathcal{B}}$ and $d_{\theta'}$ should be sufficiently large to ensure adiabaticity of the MBO states, as controlled by the leading corrections to the MBOA discussed in Sec 5. The spins are initially prepared in a region arbitrarily far away from any field rotation, $x \to -\infty$. The Bell pair is recovered asymptotically as $x \to \infty$.

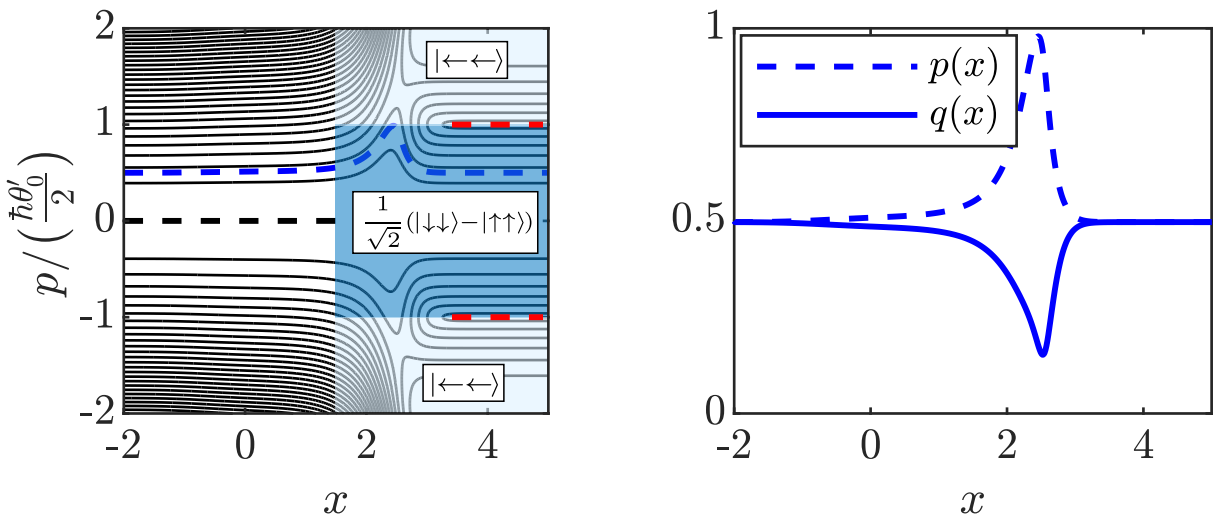

**Fig. S6.** Proof of concept for motion-generated Bell states. Left: Phase space contours of the MBO ground state $\mathcal{H}_0^{\rm MBO}(x,p)$. Trajectories with small initial momentum are able to realize the desired path to the outgoing entangled region, implementing the adiabatic protocol shown in red in Fig. 4(a). Right: Physical $q = m\dot{x}$ and canonical $p$ momentum as a function of position for the trajectory shown in blue on the left. The choice of parameters for this plot is $\mathcal{B}_0 = 10$, $\theta_0' = 1$, $d_{\mathcal{B}} = 1$, $d_{\theta'} = 1$, $x_B^0 = 1$, $x_{\theta'}^0 = -1$, with $m = \hbar = \mu = 1$.

The subtlety in constructing such a protocol is that the phase space path $(x(t), p(t))$ is not independently controlled, but is set by the MBOA equations of motion Eq. 13 and Eq. 14 of the main text. In particular, we must ensure that, i) there exists a valid path in phase space connecting $x \to -\infty$ and $x \to \infty$, and ii), that the outgoing momentum satisfies the condition $|\hbar\theta_0'/p| > 2$, which implies that the quadratic $\hat{\mathbf{S}}_x^2$ term dominates over the linear one in Eq. 20 of the main text. The latter ensures that the outgoing state lies in the entangled regime of Fig. 4(a).

One possible way to satisfy these constraints is to introduce an external scalar potential $V(x)$. The presence of $V(x)$ does not alter the eigenstates of $\hat{\mathbf{H}}^{\rm M}(x,p)$, but allows more freedom in controlling the dynamics of the particle. The choice we use here is given by,

$$V(x) = \sqrt{(2\mu\mathcal{B}(x))^2 + \left(\frac{(\hbar\theta'(x))^2}{4M}\right)^2} - \frac{(\hbar\theta'(x))^2}{4M}, \quad [91]$$

which ensures that a spin with vanishingly small initial momentum $p \to 0$ is able to realize a path from $x \to -\infty$ to $x \to \infty$, shown as a dashed black line. The resulting MBO ground state phase space contours $\mathcal{H}_0^{\rm MBO}(x,p)$ are shown in the left panel of Fig. S6.

The two dotted red lines on the right separate the two different regimes i) $|\hbar\theta_0'/p| > 2$ (between the lines) and ii) $|\hbar\theta_0'/p| < 2$ (outside the lines) of the ground state phase diagram. Outgoing states in the first regime will remain entangled as $x \to \infty$, while outgoing states in the second regime will not. We observe that spins with initial momenta close to $p_0 = 0$ are able to realize the desired protocol. In particular, the blue line shows a contour for which the initial momentum is set to be $p_0 = \hbar\theta_0'/4$. In Fig. S6(b), we plot both the physical $q(x) = m\dot{x}$ and moving momentum $p(x)$ for this particular contour. We remark that this choice of potential is not a unique way of generating entanglement, but merely illustrative.

## 8. Spin Squeezing

In this section, we describe in detail the calculations for spin squeezing in the main text Sec. 2. We begin by noting that, for the multi-spin model, the AGP is simply the sum of the single-spin AGPs, $\hat{\mathbf{A}}_x(x) = -\hbar\theta'(x)\hat{\mathbf{S}}_x$, with $\hat{\mathbf{S}}_x = \sum_{i=1}^N \hat{\sigma}_x^i/2$. Then, the moving Hamiltonian reads,

$$(\hat{\mathbf{H}}^{\rm M})' = \frac{p^2}{2M} - 2\mu\mathcal{B}\hat{\mathbf{S}}_z + \frac{p\hbar\theta'}{M}\hat{\mathbf{S}}_x + \frac{(\hbar\theta')^2}{2M}\hat{\mathbf{S}}_x^2. \quad [92]$$

Recall that the primes signify that Eq. (92) is written in the BO basis, where the magnetic field is aligned along the $z$-direction. Here, $\hat{\mathbf{S}}_z$ is the total spin operator along $z$, $\hat{\mathbf{S}}_z = \sum_{i=1}^N \hat{\sigma}_z^i/2$.

We now make a series of approximations to extract properties of the ground state of Eq. (92). The ground state lies in the collective spin $S = N/2$ subspace, and the analysis can be restricted to this spin sector. Let us additionally consider the regime $|p| \ll N|\hbar\theta'|$, where the linear in momentum term can be dropped out such that Eq. (92) reduces to,

$$(\hat{\mathbf{H}}^{\rm M})' \cong \frac{p^2}{2M} - 2\mu\mathcal{B}\left(\hat{\mathbf{S}}_z - \frac{\chi}{N}\hat{\mathbf{S}}_x^2\right), \quad [93]$$

and $\chi = N(\hbar\theta')^2/4M\mu\mathcal{B}$.

We consider the large $N$ limit of the Holstein-Primakoff transformation (19), which defines the mapping from spin to boson operators,

$$\hat{\mathbf{S}}_+ \cong \sqrt{N}a, \quad \hat{\mathbf{S}}_- \cong \sqrt{N}a^\dagger, \quad \hat{\mathbf{S}}_z = \frac{N}{2} - a^\dagger a. \quad [94]$$

Eq. (93) then reads,

$$(\hat{\mathbf{H}}^{\rm M})' \cong \frac{p^2}{2M} - \mu\mathcal{B}N + 2\mu\mathcal{B}\left(a^\dagger a + \frac{\chi}{4}(a+a^\dagger)^2\right). \quad [95]$$

Note that Eq. (95) is quadratic in creation and annihilation operators, and is diagonalized by the squeezing transformation,

$$\hat{\mathbf{U}}_{\mathrm{MBO}} = \exp\left\{\frac{r}{2}(a^2 - a^{\dagger 2})\right\}, \quad r = \frac{1}{4}\ln\left(1+\chi\right), \tag{96}$$

such that,

$$\hat{\mathbf{U}}^{\dagger}_{\mathrm{MBO}}(\hat{\mathbf{H}}^{\mathrm{M}})'\hat{\mathbf{U}}_{\mathrm{MBO}} = \frac{p^2}{2M} - \mu\mathcal{B}N + 2\mu\mathcal{B}\sqrt{1+\chi}a^{\dagger}a + \mu\mathcal{B}\left(\sqrt{1+\chi}-1\right). \tag{97}$$

The MBO ground state can thus be written as $\left|\psi_0^{\mathrm{MBO}}\right\rangle = \hat{\mathbf{U}}_{\mathrm{MBO}}\left|0\right\rangle$, where $\left|0\right\rangle$ is the vacuum state. To see that $\left|\psi_0^{\mathrm{MBO}}\right\rangle$ is indeed spin-squeezed along the $x$ direction, it is enough to compute the variances,

$$\begin{aligned} \Delta\hat{\mathbf{S}}_x^2 &= \frac{N}{4}\left\langle\psi_0^{\mathrm{MBO}}\right|(a+a^{\dagger})^2\left|\psi_0^{\mathrm{MBO}}\right\rangle = \frac{N}{4}\frac{1}{\sqrt{1+\chi}}, \\ \Delta\hat{\mathbf{S}}_y^2 &= -\frac{N}{4}\left\langle\psi_0^{\mathrm{MBO}}\right|(a-a^{\dagger})^2\left|\psi_0^{\mathrm{MBO}}\right\rangle = \frac{N}{4}\sqrt{1+\chi}, \end{aligned} \tag{98}$$

yielding the uncertainty ratio shown in Eq. 21 of the main text.

The truncation of the Holstein-Primakoff transformation Eq. (94) breaks down when $\chi$ becomes of order $N^2$. One way to argue this is to note that the ground state spin fluctuations along the $y$ direction, $\Delta\hat{\mathbf{S}}_y = \frac{\sqrt{N}}{2}(1+\chi)^{1/4}$, become of order $\Delta\hat{\mathbf{S}}_y \sim N/2 = S$ for $\chi \sim N^2$. Since the noise along $y$ becomes of the same order as the total spin, one can no longer argue that the spin Wigner function is closely localized about the $z$-axis pole (see Fig. 4(b)). After this point, for $\chi > N^2$ the Wigner function stops resembling a squeezed coherent state. Indeed, in the $\chi \to \infty$ limit, the ground state reduces to the $S_x = 0$ eigenstate. Therefore, the maximum quadrature ratio achievable, while still remaining in the squeezed regime, is of order $\Delta\hat{\mathbf{S}}_x/\Delta\hat{\mathbf{S}}_y \sim 1/N$.

## 9. Many Particle Piston

In this section, we provide explicit calculations for the piston example. Consider the system defined by the Hamiltonian in the main text Eq. 22 and Eq. 23, with $V(x) = kx^2/2$. We first discuss the solution within the BOA and afterwards within the MBOA.

**BOA Microcanonical Ensemble.** The BOA assumes that the fast DOFs adiabatically follow a microcanonical distribution $\rho_E(x,p) = \delta(E(x,p) - \mathbf{H}(x,p))$, with the energy $E(x,p)$ ** fixed such that the microcanonical entropy ††,

$$\begin{aligned} S(x,p,E) &= \ln\Omega(x,p,E), \\ \Omega(x,p,E) &= \int \mathrm{d}\vec{\xi}\mathrm{d}\vec{\eta}\,\delta(E - \mathbf{H}(x,p)), \end{aligned} \tag{99}$$

is $x$-independent. Recall that the bold font signifies a function of the fast phase space variables $\vec{\xi}$ and $\vec{\eta}$, with the slow ones treated as external parameters. We drop $\{\vec{\xi},\vec{\eta}\}$ from the arguments of bold functions like in the quantum case. Additionally, we reserve the symbol $p$ for the canonical and $q$ for the physical momentum. In the lab frame the two are equal, such that $\mathbf{q} = p$, in analogy to the Wigner-Weyl mapping $\hat{\mathbf{q}} = p\,\hat{\mathbf{I}}$.

The microcanonical entropy Eq. (99) (assuming that $N$ is even) can be computed explicitly for $N \gg 1$ as,

$$S(x,p,E) = \frac{N}{2}\ln[2\pi m E^{\mathrm{fast}}x^2] - \ln[(N/2)!] = N\ln\left[x\sqrt{\frac{4\pi m E^{\mathrm{fast}}}{N}}\right] + \frac{N}{2} + \mathcal{O}(\ln N), \tag{100}$$

where we have defined $E^{\mathrm{fast}}$ as the total kinetic energy of the fast particles,

$$E^{\mathrm{fast}} \equiv E - \frac{p^2}{2M} - \frac{1}{2}kx^2. \tag{101}$$

From Eq. (100), the condition for constant $S$ is given by,

$$x^2E^{\mathrm{fast}}(x) = \mathrm{const.} \tag{102}$$

That is, the position of the piston fixes the energy of the fast particles. The gas heats up as the piston contracts and cools down as it expands.

The effective BO Hamiltonian describing the motion of the piston is obtained by averaging the full Hamiltonian of the system over the fast variables along the constant entropy contour,

$$\mathcal{H}^{\mathrm{BO}}(x,p) = \frac{1}{\Omega(x,p,E)}\int \mathrm{d}\vec{\xi}\mathrm{d}\vec{\eta}\,\delta(E(x,p) - \mathbf{H}(x,p))\mathbf{H}(x,p) = \frac{p^2}{2m} + \frac{1}{2}kx^2 + E_0^{\mathrm{fast}}\left(\frac{x_0^2}{x^2}\right), \tag{103}$$

where $E_0^{\mathrm{fast}}$ denotes the initial kinetic energy of the fast particles and $x_0$ the initial position of the piston.

** The constraint that $E(x,p)$ is conserved in time is not imposed, but follows from the emergent equations of motion for the slow variables.

†† For the remainder of the section we work in units where $k_B = 1$.

**BOA Canonical Ensemble.** As the canonical and microcanonical ensembles are equivalent in the many-particle limit, it is instructive to see how one can recover Eq. (103) from the computationally simpler canonical framework. We assume that the fast DOFs follow a Gibbs distribution,

$$\rho_\beta(x,p) = \frac{1}{Z(x,p,\beta)} e^{-\beta(x,p)\mathbf{H}(x,p)},$$

with inverse temperature $\beta(x,p)$ fixed such that the entropy,

$$S(x,p,\beta) = \beta E(x,p,\beta) + \ln Z(x,p,\beta), \qquad [104]$$

is conserved. Here $E(x,p,\beta)$ is the mean energy of the system in the Gibbs ensemble. The entropy can be exactly evaluated as,

$$S(x,p,\beta) = N \ln\left(x\sqrt{\frac{2\pi m}{\beta}}\right) + \frac{N}{2}. \qquad [105]$$

The constant entropy condition then reads,

$$\frac{x^2}{\beta(x,p)} = \text{const}, \qquad [106]$$

which is the analogous statement for Eq. (102) in the canonical ensemble. The expectation value of the Hamiltonian over the fast variables along the constant entropy contour then yields,

$$\mathcal{H}^{\rm BO}(x,p) = \frac{1}{Z(x,p,\beta)} \int \mathrm{d}\vec{\xi}\mathrm{d}\vec{\eta}\, e^{-\beta(x,p)\mathbf{H}(x,p)}\mathbf{H}(x,p) = \frac{p^2}{2m} + \frac{1}{2}kx^2 + \frac{N}{2\beta_0}\left(\frac{x_0^2}{x^2}\right). \qquad [107]$$

By identifying $E_0^{\rm fast} = N/(2\beta_0)$, Eq. (103) and Eq. (107) agree.

**MBOA Slow Hamiltonian.** The MBOA assumes that the fast DOFs follow instantaneous equilibrium with respect to the moving Hamiltonian $\mathbf{H}^{\rm M}(x,p)$. As in the quantum case, one now has to distinguish the dressed physical momentum $\mathbf{q}$, which also depends on fast coordinates $\vec{\xi}$, from the canonical momentum $p$, which is the proper canonical variable. In particular, $\mathbf{q} = p - \mathbf{A}_x(x)$.

As in the BOA, the microcanonical and canonical descriptions are equivalent at large $N$. We work with the latter because it is computationally simpler. The fast DOFs are assumed to follow a Gibbs distribution,

$$\rho_\beta^{\rm M}(x,p) = \frac{1}{Z^{\rm M}(x,p,\beta^{\rm M})} e^{-\beta^{\rm M}(x,p)\,\mathbf{H}^{\rm M}(x,p)},$$

where we introduced notation $\beta^{\rm M}$ for the temperature of the moving equilibrium to distinguish it from the BO temperature. The moving Hamiltonian is given by,

$$\mathbf{H}^{\rm M}(x,p) = \frac{(p-\mathbf{A}_x(x))^2}{2M} + \frac{1}{2}kx^2 + \frac{|\vec{\eta}|^2}{2m} + U_0\sum_{i=1}^{N}(\Theta(-\xi_i) + \Theta(\xi_i - x)), \quad \mathbf{A}_x(x) = \frac{\vec{\xi}\cdot\vec{\eta}}{x}, \qquad [108]$$

In passing, note that writing Eq. (108) instead using dilated coordinates $\vec{\xi}' = \vec{\xi}/x$ and $\vec{\eta}' = \vec{\eta}x$ is equivalent to working in the basis of instantaneous BO states in the quantum language.

It is convenient to express the Hamiltonian as a quadratic form in the fast momenta,

$$\mathbf{H}^{\rm M}(x,p) = \frac{p^2}{2M} + \frac{1}{2}kx^2 + \frac{\vec{\eta}^T\cdot\hat{\mu}^{-1}(\vec{\xi})\cdot\vec{\eta}}{2} + \vec{b}(\vec{\xi})^T\cdot\vec{\eta}, \qquad [109]$$

where,

$$\hat{\mu}^{-1}(\vec{\xi}) = \frac{1}{m}\hat{1} + \frac{\vec{\xi}\cdot\vec{\xi}^T}{Mx^2}, \qquad [110]$$

is a position-dependent inverse mass matrix, and,

$$\vec{b}(\vec{\xi}) = -\frac{p}{M}\frac{\vec{\xi}}{x}, \qquad [111]$$

is a position-dependent momentum bias. Here the superscript T denotes the vector transpose and hats denote matrix operators. Note that since the vector outer product $\vec{\xi}\cdot\vec{\xi}^T$ is of rank one, the matrix $\hat{\mu}^{-1}(\vec{\xi})$ has $N-1$ trivial eigenvalues equal to $1/m$, and only a single non-trivial one equal to $1/m + |\vec{\xi}|^2/(x^2M)$. This simplifies the evaluation of the Gaussian integral over the momentum, yielding the expression for the partition function,

$$Z^{\rm M}(x,p,\beta^{\rm M}) = \int \mathrm{d}\vec{\xi}\mathrm{d}\vec{\eta}\, e^{-\beta^{\rm M}\mathbf{H}^{\rm M}(x,p)} = x^N\left(\frac{2\pi m}{\beta^{\rm M}}\right)^{N/2} e^{-\frac{\beta^{\rm M}kx^2}{2}} \int_0^1 \mathrm{d}\vec{\xi}\, \frac{e^{-\frac{\beta^{\rm M}p^2}{2M}\frac{1}{1+\frac{m}{M}|\vec{\xi}|^2}}}{\sqrt{1+\frac{m}{M}|\vec{\xi}|^2}}. \qquad [112]$$

Before proceeding, we briefly comment on the evaluation of this type of integral, which is recurrent in this calculation. It is convenient to rewrite the integral into the form,

$$\int_0^1 \mathrm{d}\vec{\xi}\, f(|\vec{\xi}|^2) = \int_{-\infty}^{\infty} \mathrm{d}u f(uN)P(u), \tag{113}$$

with,

$$P(u) = \int_0^1 \mathrm{d}\vec{\xi}\, \delta\big(u - |\vec{\xi}|^2/N\big), \quad f(uN) = \frac{e^{-\frac{\beta^{\mathrm{M}} p^2}{2M}\frac{1}{1+\frac{mN}{M}u}}}{\sqrt{1+\frac{mN}{M}u}}. \tag{114}$$

We now take the $N \to \infty$ limit, while keeping the total mass of the gas $mN$ fixed. Using the central limit theorem, we find,

$$\lim_{N\to\infty} P(u) = \delta(u - 1/3), \tag{115}$$

and,

$$\lim_{N\to\infty} \int_0^1 \mathrm{d}\vec{\xi}\, f(|\vec{\xi}|^2) = \int_{-\infty}^{\infty} \mathrm{d}u f(uN)\delta(u - 1/3) = f(N/3). \tag{116}$$

Corrections to this approximation come as $\mathcal{O}(1/N)$ and are thus sub-extensive. This result gives the partition function in the large $N$ limit,

$$Z^{\mathrm{M}}(x,p,\beta^{\mathrm{M}}) \overset{N\to\infty}{=} x^N \left(\frac{2\pi m}{\beta^{\mathrm{M}}}\right)^{N/2} \frac{e^{-\beta^{\mathrm{M}}\left(\frac{p^2}{2(M+\kappa)}+\frac{kx^2}{2}\right)}}{\sqrt{1+\frac{\kappa}{M}}}, \tag{117}$$

where,

$$\kappa = \frac{mN}{3},$$

is the correction to the mass of the piston due to its dressing by fast particles (18). Similarly, the mean energy is found as,

$$\begin{aligned} E^{\mathrm{M}}(x,p,\beta^{\mathrm{M}}) &= \int \mathrm{d}\vec{\xi}\mathrm{d}\vec{\eta}\, \frac{e^{-\beta^{\mathrm{M}}\mathbf{H}^{\mathrm{M}}(x,p)}}{Z^{\mathrm{M}}(x,p,\beta)} \mathbf{H}^{\mathrm{M}}(x,p) = \frac{p^2}{2M} \frac{\int \mathrm{d}\vec{\xi}\, e^{-\frac{\beta^{\mathrm{M}}p^2}{2M}\frac{1}{1+\frac{m}{M}|\vec{\xi}|^2}} (1+\frac{m}{M}|\vec{\xi}|^2)^{-3/2}}{\int \mathrm{d}\vec{\xi}\, e^{-\frac{\beta^{\mathrm{M}}p^2}{2M}\frac{1}{1+\frac{m}{M}|\vec{\xi}|^2}} (1+\frac{m}{M}|\vec{\xi}|^2)^{-1/2}} + \frac{1}{2}kx^2 + \frac{N}{2\beta^{\mathrm{M}}}, \\ &\overset{N\to\infty}{=} \frac{p^2}{2(M+\kappa)} + \frac{1}{2}kx^2 + \frac{N}{2\beta^{\mathrm{M}}}, \end{aligned} \tag{118}$$

where in the last line, the same $N \to \infty$ is taken for both integrals in the numerator and denominator while holding the product $mN$ fixed.

Eq. (104) provides the expression for the entropy of the moving equilibrium state,

$$S^{\mathrm{M}}(x,p,\beta^{\mathrm{M}}) = N \ln\left(x\sqrt{\frac{2\pi m}{\beta^{\mathrm{M}}(x,p)}}\right) + \frac{N}{2} - \frac{1}{2}\ln\Big(1+\frac{\kappa}{M}\Big). \tag{119}$$

Apart from the last constant and sub-extensive term, the expression for the Gibbs entropy of the MBO Gibbs ensemble is identical to Eq. (105) for the BO Gibbs ensemble.

We now need to connect the moving entropy, which is conserved within the MBOA, with initial conditions. There are two natural possibilities coming from different experimental realizations. Suppose the piston is initially at some fixed position and zero (for concreteness) momentum enforced by e.g. an external potential which does not allow the piston to move, effectively making its mass infinite. This large potential is an analogue of a large magnetic field $\mathcal{B}$ in the spin example. It makes the BO approximation essentially exact. The first option is that this potential is released on a time scale which is short compared to the period of oscillations of the piston, but long enough for fast particles to re-equilibrate in the moving frame. In this case, we can find the initial temperature $\beta_0^{\mathrm{M}}$ from equating the entropies given by Eq. (105) and Eq. (119),

$$\beta_0^{\mathrm{M}} = \frac{\beta_0}{(1+\frac{\kappa}{M})^{\frac{1}{N}}} \tag{120}$$

In the limit $\kappa \ll M$ this expression gives $\beta_0^{\mathrm{M}} \approx \beta_0(1 - m/(3M))$ and in the opposite limit $\kappa \gg M$ we have $\beta_0^{\mathrm{M}} \approx \beta_0(1 - \ln(\kappa/M)/N)$.

Another possibility, which is realized in our numerical experiment, is to suddenly remove the additional trapping potential. Then, in order to find the moving temperature, one should compute the energy of fast particles in the moving frame and equate it to the initial energy,

$$E_0^{\mathrm{fast}} = \int \mathrm{d}\vec{\xi}\mathrm{d}\vec{\eta}\, \frac{e^{-\beta_0^{\mathrm{M}}\mathbf{H}^{\mathrm{M}}(x,0)}}{Z(x,0,\beta_0^{\mathrm{M}})} \frac{|\vec{\eta}|^2}{2m} \overset{N\to\infty}{=} \frac{N}{2\beta_0^{\mathrm{M}}}\left(1 - \frac{1}{N}\frac{\kappa}{M+\kappa}\right), \tag{121}$$

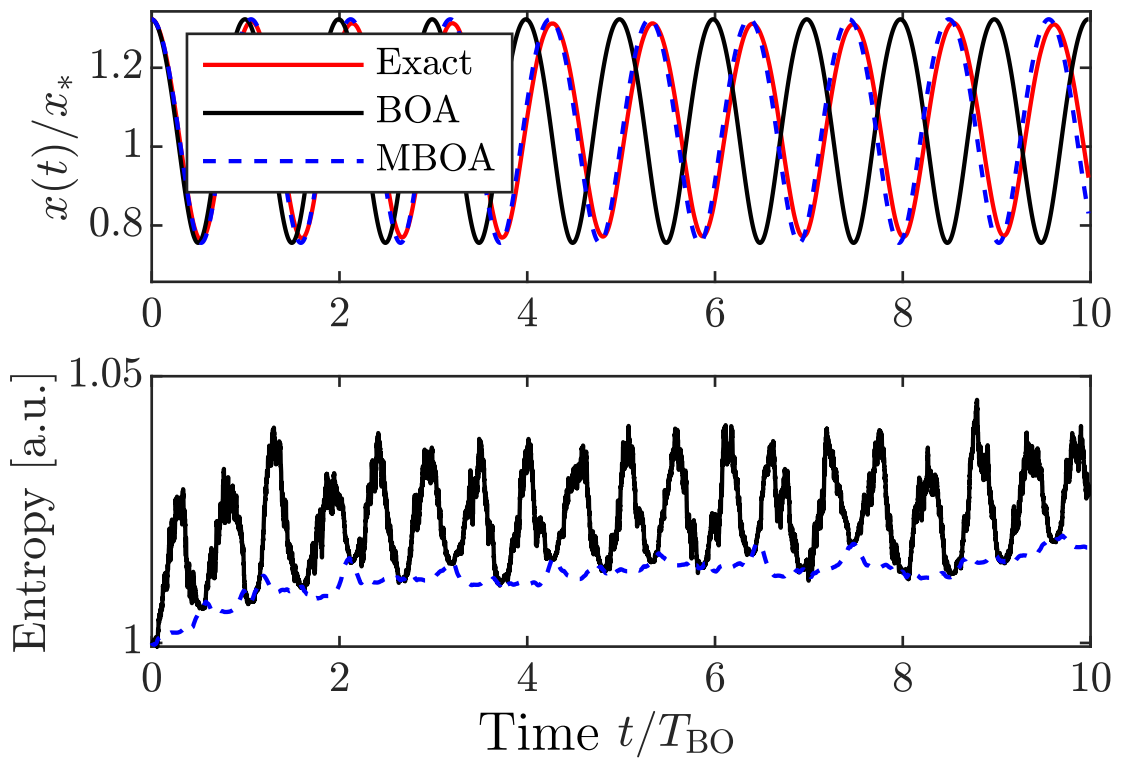

**Fig. S7.** Oscillations of the piston about a displacement from its equilibrium position $x_*$. Upper panel: Due to the mass renormalization $M \to M+\kappa$, the observed period of oscillations is longer than that predicted by the BOA, which is denoted as $T_{\rm BO}$. Lower panel: Time evolution of the BO and MBO entropies given by Eq. (129). While the BO entropy exhibits reversible oscillations, the MBO entropy is monotonically increasing. The synchronized long-lived state between the fast particles and the piston can therefore be understood from entropy conservation in a moving frame. The dynamics was carried out with $N = 2000$ fast particles and a mass renormalization of $\kappa/M \equiv Nm/(3M) = 2/15$. The piston is initially diabatically displaced from the associated BO equilibrium position at the initial temperature of the gas: $x(t=0) = 1.75x_*(\beta_0) = 1.75\sqrt{N/(k\beta_0)}$. Note that $x_*(\beta_0)$ is different from the new equilibrium position $x_*$ of the piston after the quench: $x_* = \sqrt{1.75}x_*(\beta_0)$.

where $x$ is the initial position of the piston. Equating this expression to the one in the initial BO ensemble we find that,

$$\beta_0^{\rm M} = \beta_0 \left(1 - \frac{1}{N}\frac{\kappa}{M+\kappa}\right). \qquad [122]$$

While this expression is not identical to Eq. (120), it has very similar small and large $\kappa$ behavior. Depending on the initialization of the piston, one should use either Eq. (120) or Eq. (122) to relate initial moving temperature of the gas to the temperature of the gas when the piston is in static equilibrium. The correction is, however, sub-extensive.

The MBO slow Hamiltonian can be obtained by averaging $\mathbf{H}^{\rm M}(x,p)$ over fast variables at a constant entropy $S^{\rm M}$,

$$\begin{aligned}\mathcal{H}^{\rm MBO}(x,p) &= \frac{p^2}{2(M+\kappa)} + \frac{1}{2}kx^2 + \frac{N}{2\beta_0^{\rm M}}\left(\frac{x_0^2}{x^2}\right), \\ &\approx \frac{p^2}{2(M+\kappa)} + \frac{1}{2}kx^2 + \frac{N}{2\beta_0}\left(\frac{x_0^2}{x^2}\right)\left(1+\frac{1}{N}\frac{\kappa}{\kappa+M}\right),\end{aligned} \qquad [123]$$

where in the second line we have Taylor expanded $1/\beta_0^{\rm M}$ using Eq. (122) to linear order in $1/N$. Note that the fast particles renormalize the mass of the piston, with each contributing a third of their own mass. The correction to the mass can be detected by studying the oscillations of the piston about the equilibrium position. Figure S7 shows a comparison with exact dynamics. Due to the mass renormalization, the piston oscillates at a slower frequency than that predicted by the BOA. In addition, there is a subextensive correction to the BO potential which can be understood as coming from microscopic fluctuations of the piston momentum. This is directly related to the metric tensor and is discussed in more detail in the following section and in Fig. 6 of the main text. In the limit of large particle number $N$, this correction can be neglected.

**Canonical vs. Physical Momentum.** The canonical momentum $p$ is simply related to the MBO average of the physical momentum $\langle \mathbf{q}\rangle$ by a multiplicative constant,

$$\langle \mathbf{q}\rangle = \left\langle p - \frac{\vec{\xi}\cdot\vec{\eta}}{x}\right\rangle = p - \frac{1}{Z^{\rm M}}\int d\vec{\xi}d\vec{\eta}\,\frac{\vec{\xi}\cdot\vec{\eta}}{x}e^{-\beta^{\rm M}\mathbf{H}^{\rm M}} = p\left(1 - \int d\vec{\xi}\frac{\frac{m}{M}|\vec{\xi}|^2}{1+\frac{m}{M}|\vec{\xi}|^2}\right) \overset{N\to\infty}{=} \frac{p}{1+\frac{\kappa}{M}}. \qquad [124]$$

Interestingly, while the instantaneous physical momentum of the piston $q$ is a noisy quantity due to microscopic noise from particle-piston collisions, the canonical momentum $p$ evolves smoothly in time. In particular, note that the quantity $p = q + \vec{\xi}\cdot\vec{\eta}/x$ is exactly conserved in all particle-wall and particle-piston collisions. This is illustrated in Fig. 6 of the main text, which shows a comparison between the two from exact numerics for a simulation carried out with $N = 500$ particles, $\kappa/M = mN/3M = 5/12$, and with the piston initially displaced from the BO equilibrium position associated to the initial temperature of the gas: $x(t=0) = 1.25x_*(\beta_0) = 1.25\sqrt{N/(k\beta_0)}$. The expression derived in Eq. (124) provides a good fit to the time averaged $q$. The $y$-axis is shown in units of $p_{\rm BO}^{\rm max}$ - the BO prediction for the maximum absolute value of $p$ during the period of oscillation.

The dotted black lines show the expected fluctuations of $\mathbf{q}$ within the MBO Gibbs ensemble. In particular, each line is one standard deviation $\sigma_{\mathbf{q}}$ away from the mean value, where,

$$\sigma_{\mathbf{q}}^2 = \langle \mathbf{q}^2\rangle - \langle \mathbf{q}\rangle^2 = \langle \mathbf{A}_x^2\rangle - \langle \mathbf{A}_x\rangle^2 \overset{N\to\infty}{=} \frac{\kappa M}{\kappa+M}\frac{1}{\beta^{\rm M}}. \qquad [125]$$

Note that these fluctuations are directly related to the variance of the AGP, which is the classical analogue of the quantum metric tensor. However, the noise is purely classical, arising from fluctuations within the MBO state of the fast DOFs. The extra contribution to the piston kinetic energy arising from these fluctuations provides precisely the subextensive correction to the BO potential shown in Eq. (123).

**MBOA Fast Equilibrium State.** Let us now develop some intuition for the moving equilibrium state of the gas particles. Although the gas particles become effectively interacting in the moving frame due to the all-to-all coupling from the $\mathbf{A}_x^2$ term, one can obtain an effective single-particle Hamiltonian by treating the $\mathbf{A}_x^2$ term within a mean-field approximation (see Sec. 11). To do so, we first compute the expectation value of the AGP in the MBO state,

$$\langle \mathbf{A}_x \rangle = \frac{1}{Z^{\mathrm{M}}} \int d\vec{\xi} d\vec{\eta}\, \frac{\vec{\xi}\cdot\vec{\eta}}{x} e^{-\beta^{\mathrm{M}} \mathbf{H}^{\mathrm{M}}} = \int d\vec{\xi}\, \frac{p \frac{m}{M} |\vec{\xi}|^2}{1 + \frac{m}{M}|\vec{\xi}|^2} \overset{N\to\infty}{=} \frac{p\kappa}{\kappa + M}. \qquad [126]$$

Plugging this result into Eq. (146), we obtain the effective single-particle Hamiltonian,

$$\mathbf{H}^{\mathrm{M}}_{\mathrm{mf}} = \frac{\eta^2}{2m} - \frac{p}{M+\kappa}\frac{\xi\eta}{x} + U_0(\Theta(-\xi) + \Theta(\xi - x)), \qquad [127]$$

where we have omitted a constant, i.e. $\vec{\xi}$ and $\vec{\eta}$-independent, energy shift. The single-particle phase space distribution $\rho^{\mathrm{M}}_{\mathrm{mf}} \propto e^{-\beta^{\mathrm{M}} \mathbf{H}^{\mathrm{M}}_{\mathrm{mf}}}$ is shown in the bottom right panel of Fig. 5 of the main text. While the piston is in motion, the distribution is skewed in such a way that faster moving particles are found closest to the piston. Indeed, the mean momentum of fast particles $\langle \eta \rangle$ as a function of their position $\xi$ is set by the piston momentum as,

$$\langle \eta \rangle = \frac{mp}{M+\kappa}\frac{\xi}{x} = \langle \mathbf{q} \rangle \frac{m}{M}\frac{\xi}{x}, \qquad [128]$$

through application of Eq. (124). Note that the quantity $p/(M+\kappa) = \langle \mathbf{q} \rangle / M$ corresponds to the time-averaged velocity of the piston at a given instant of time, as shown in Fig. 6 of the main text.

As the piston oscillates, the gas settles into a synchronized state with little dissipation. The synchronized equilibrium can be understood from conservation of thermodynamic (Boltzmann) entropy, not in the lab frame, but *in the moving frame.* Indeed, compare the lab and moving frame entropies,

$$S(x, q, E) = \ln \Omega(x, q, E), \quad S^{\mathrm{M}}(x, p, E^{\mathrm{M}}) = \ln \Omega^{\mathrm{M}}(x, p, E^{\mathrm{M}}), \qquad [129]$$

where,

$$\Omega(x, q, E) = \int d\vec{\xi} d\vec{\eta}\, \delta(E - H(\vec{\xi}, \vec{\eta}, x, q)) \overset{N\to\infty}{=} \frac{(2\pi m x^2)^{\frac{N}{2}}}{(N/2)!} \left( E - \frac{q^2}{2M} - \frac{1}{2}kx^2 \right)^{\frac{N}{2}}, \qquad [130]$$

and,

$$\Omega^{\mathrm{M}}(x, p, E^{\mathrm{M}}) = \int d\vec{\xi} d\vec{\eta}\, \delta(E^{\mathrm{M}} - H^{\mathrm{M}}(\vec{\xi}, \vec{\eta}; x, p)) \overset{N\to\infty}{=} \frac{(2\pi m x^2)^{\frac{N}{2}}}{(N/2)! \sqrt{1 + \frac{\kappa}{M}}} \left( E^{\mathrm{M}} - \frac{p^2}{2(M+\kappa)} - \frac{1}{2}kx^2 \right)^{\frac{N}{2}}. \qquad [131]$$

Here, $E$ denotes the total energy of the system, which is the same in the lab and moving frames, $E = E^{\mathrm{M}}$ and is conserved in time. The lower panel of Fig. S7 shows the exact time evolution of these two quantities for the setup shown in the upper panel. While the lab frame entropy $S$ exhibits reversible oscillations, the moving entropy $S^{\mathrm{M}}$ is monotonically increasing. This example illustrates how the non-trivial equilibrium state between fast particles and piston can be understood from entropy conservation in a moving frame. The upwards drift of the MBO entropy is a slow relaxation towards equilibrium which lies beyond the MBOA scheme.

## 10. Single Particle Piston

The single particle $N = 1$ case is especially instructive to highlight connections between the classical and quantum formalism. The Hamiltonian of the system is given by the main text Eq. 22 and Eq. 23 with a single fast degree of freedom,

$$H(\xi, \eta, x, q) = \frac{q^2}{2M} + \frac{\eta^2}{2m} + \frac{1}{2}kx^2 + U_0(\Theta(-\xi) + \Theta(\xi - x)). \qquad [132]$$

The fast particle's Hamiltonian is that of a particle in a box. If the system is treated quantum mechanically, the BOA assumes that the integer label of the fast particle eigenstate is conserved. The analogous conserved quantity in classical systems is the adiabatic invariant,

$$\mathcal{S}(x, p, E) = \int d\xi d\eta\, \Theta(E - \mathbf{H}(x, p)). \qquad [133]$$

The connection between these two statements follows from Born-Sommerfeld quantization, which states that the quantum number $n$ is proportional to $\mathcal{S}$. For many-particle systems, the logarithm of the adiabatic invariant is equivalent to the entropy up to sub-extensive corrections. We briefly discuss the solution of the problem within the BOA and then within the MBOA.

**BOA.** Within the BOA, the adiabatic invariant is given by,

$$\mathcal{S}(x,p,E)=\int \mathrm{d}\xi\mathrm{d}\eta\,\Theta(E-\mathbf{H}(x,p))=2x\sqrt{2mE^{\text{fast}}}, \tag{134}$$

where we have defined $E^{\text{fast}}=E-\frac{p^2}{2M}-\frac{1}{2}kx^2$. Conservation of $\mathcal{S}$ then implies the invariant $x^2E^{\text{fast}}(x)=\text{const}$. In the quantum language, the instantaneous BO energy for the $n$-th eigenstate is,

$$E_n^{\text{fast}}(x)=\frac{\pi^2n^2}{2mx^2}. \tag{135}$$

Thus, the condition $x^2E^{\text{fast}}(x)=\text{const}$, as we highlighted, corresponds to conservation of the level number $n$ or, for mixed states, the conservation of eigenstate occupation probabilities. Averaging $\mathbf{H}$ over $\xi$ and $\eta$, we recover the same effective Hamiltonian as in Eq. (103).

**MBOA.** Within the MBOA, the conserved adiabatic invariant is instead,

$$\mathcal{S}^{\text{M}}(x,p,E)=\int \mathrm{d}\xi\mathrm{d}\eta\,\Theta(E-\mathbf{H}^{\text{M}}(x,p)), \tag{136}$$

where $\mathbf{H}^{\text{M}}$ is given by Eq. (108) with $N=1$. To evaluate $\mathcal{S}^{\text{M}}$, we must calculate the phase space area enclosed by the energy contours of $\mathbf{H}^{\text{M}}$. Since $\mathbf{H}^{\text{M}}$ is a quadratic form in the fast momenta, the exact phase space trajectories $\eta(\xi)$ can be found as,

$$\eta(\xi)=\frac{p\frac{m\xi}{Mx}}{1+\frac{m\xi^2}{Mx^2}}\pm\sqrt{\frac{2m}{1+\frac{m\xi^2}{Mx^2}}\left(E-\frac{p^2}{2M(1+\frac{m\xi^2}{Mx^2})}-\frac{1}{2}kx^2\right)}\approx\frac{p\frac{m\xi}{Mx}}{1+\frac{m\xi^2}{Mx^2}}\pm\sqrt{\frac{2mE}{1+\frac{m\xi^2}{Mx^2}}}, \tag{137}$$

where we assume for simplicity that the total energy $E$ is mostly dominated by the contribution from the fast particle, and not by the piston. Then, expanding Eq. (137) to leading order in the mass ratio,

$$\eta(\xi)\approx p\frac{m\xi}{Mx}\pm\sqrt{2mE}\left(1-\frac{m\xi^2}{2Mx^2}\right)+\mathcal{O}\left(\frac{m^2}{M^2}\right). \tag{138}$$

Within this approximation, the adiabatic invariant is found as,

$$\mathcal{S}^{\text{M}}(x,p,E)\approx 2\sqrt{2mE}\int_0^x\left(1-\frac{m}{2M}\frac{\xi^2}{x^2}\right)\mathrm{d}\xi=2x\sqrt{2Em}\left(1-\frac{m}{6M}\right). \tag{139}$$

Conservation of $\mathcal{S}^{\text{M}}$ requires $Ex^2=\text{const}$, exactly as in the multi-particle case.

The next step is to average the moving Hamiltonian along the constant $\mathcal{S}^{\text{M}}$ contour. Neglecting order $\mathcal{O}\big(m^2/M^2\big)$ contributions, each term gives,

$$\left\langle\frac{p\xi\eta(\xi)}{Mx}\right\rangle\approx\frac{p^2}{M}\int_0^x\mathrm{d}\xi\frac{m\xi^2}{Mx^2}=\frac{p^2}{M}\frac{m}{3M}, \tag{140}$$

$$\left\langle\frac{\xi^2\eta^2(\xi)}{2Mx^2}\right\rangle\approx\frac{m}{M}E\int_0^x\mathrm{d}\xi\frac{\xi^2}{x^2}=E\frac{m}{3M}, \tag{141}$$

$$\left\langle\frac{\eta^2(\xi)}{2m}\right\rangle\approx\frac{p^2}{2M}\frac{m}{M}\int_0^x\mathrm{d}\xi\frac{\xi^2}{x^2}+E\left(1-\frac{m}{M}\int_0^x\mathrm{d}\xi\frac{\xi^2}{x^2}\right)=\frac{p^2}{2M}\frac{m}{3M}+E\left(1-\frac{m}{3M}\right). \tag{142}$$

Finally,

$$\mathcal{H}^{\text{MBO}}=\frac{p^2}{2M}\left(1-\frac{m}{3M}\right)+\frac{1}{2}kx^2+E_0\left(\frac{x_0}{x}\right)^2\approx\frac{p^2}{2(M+\frac{m}{3})}+\frac{1}{2}kx^2+E_0\left(\frac{x_0}{x}\right)^2, \tag{143}$$

which again agrees with the multi-particle expression.

## 11. Mean-Field MBOA

In this section we discuss a mean-field approximation, valid in the large $N$ limit of many fast DOFs, to compute the MBO Hamiltonian and related quantities. Like in section 9, we are often interested in evaluating multi-dimensional phase-space integrals weighted by a Gibbs distribution $\rho(x,p)\sim e^{-\beta^{\text{M}}\mathbf{H}^{\text{M}}(x,p)}$. However, the $\mathbf{A}_x^2$ term in $\mathbf{H}^{\text{M}}$ can be problematic, since it induces all-to-all interactions between the fast DOFs that result in non-separable integrals. The following mean-field approximation allows us to turn the interacting problem back into a single particle non-interacting one.

Like in standard mean-field calculations, the idea is to expand in the fluctuations of the AGP,

$$\mathbf{A}_{\mathbf{x}}=\langle\mathbf{A}_x\rangle+\delta\mathbf{A}_{\mathbf{x}}, \tag{144}$$

where $\langle \mathbf{A}_x \rangle$ denotes the AGP mean in the MBO equilibrium state. The moving Hamiltonian can be written as,

$$\mathbf{H}^{\mathrm{M}}(x,p) = \frac{p^2}{2M} + V(x) + \mathbf{H}_{\mathrm{int}}(x) - \frac{p}{M}(\langle \mathbf{A}_x \rangle + \delta \mathbf{A}_x) + \frac{\langle \mathbf{A}_x \rangle^2}{2M} + \frac{\langle \mathbf{A}_x \rangle \delta \mathbf{A}_x}{M} + \mathcal{O}(\delta \mathbf{A}_x^2). \qquad [145]$$

The mean-field Hamiltonian is obtained by neglecting $\mathcal{O}(\delta \mathbf{A}_x^2)$ terms in Eq. (145) and re-expressing in terms of $\langle \mathbf{A}_x \rangle$ and $\mathbf{A}_x$,

$$\mathbf{H}^{\mathrm{M}}_{\mathrm{mf}}(x,p) \equiv \frac{p^2}{2M} + V(x) - \frac{\langle \mathbf{A}_x \rangle^2}{2M} + \mathbf{H}_{\mathrm{int}}(x) - \left( \frac{p - \langle \mathbf{A}_x \rangle}{M} \right) \mathbf{A}_x. \qquad [146]$$

Note that the quantity $(p - \langle \mathbf{A}_x \rangle)/M = \langle \hat{\mathbf{q}} \rangle / M$ exactly corresponds to the mean velocity of the slow DOF at a given moment in time (Fig. 6(a) of the main text). Therefore, up to a constant that is independent of $\vec{\xi}$ and $\vec{\eta}$, Eq. (146) coincides with the Hamiltonian $\mathbf{H}_{\mathrm{int}}(x) - \dot{x}\mathbf{A}_x$ that one would obtain by transforming to *time-dependent* dilated coordinates that are co-moving with the piston (2) where $x(t)$ is treated as an external (not dynamical) parameter. We stress that in Eq. (146) the velocity that appears coupled to the AGP is an average over a probability distribution of the fast particles (over the density matrix in the quantum case). The instantaneous velocity of the piston is, on the contrary, a fluctuating (noisy) quantity. The mean-field approximation Eq. (146) can be justified at large $N$ (if the AGP is extensive) using standard methods like saddle point analysis (20). Its validity is intuitively clear in the many-particle limit as fluctuations of the AGP are directly related to velocity fluctuations of the piston, which are expected to be relatively small if there are many fast particles and hence many collisions.

The AGP mean $\langle \mathbf{A}_x \rangle$ is still an undetermined parameter in Eq. (146). As usual, it must be determined self-consistently. We require,

$$\langle \mathbf{A}_x \rangle = \int \mathrm{d}\vec{\xi}\mathrm{d}\vec{\eta}\, \frac{e^{-\beta^{\mathrm{M}} \mathbf{H}^{\mathrm{M}}_{\mathrm{mf}}}}{Z^{\mathrm{M}}_{\mathrm{mf}}} \mathbf{A}_x, \qquad Z^{\mathrm{M}}_{\mathrm{mf}} = \int \mathrm{d}\vec{\xi}\mathrm{d}\vec{\eta}\, e^{-\beta^{\mathrm{M}} \mathbf{H}^{\mathrm{M}}_{\mathrm{mf}}}. \qquad [147]$$

The evaluation of the self-consistency integral Eq. (147) now depends on the specific form of the AGP. In the case where the AGP is quadratic in the fast phase-space variables, as is the case for translations ($\mathbf{A}_x = \eta$), dilations ($\mathbf{A}_x = \xi\eta/x$), and rotations ($\mathbf{A}_x = L_\phi$), the momentum integrals are Gaussian and can be done exactly.

For concreteness, let us discuss the result for the piston and gas example. The phase-space integrals can be evaluated using the same techniques as detailed in Sec. 9, and Eq. (147) yields,

$$\langle \mathbf{A}_x \rangle = \frac{\kappa}{M} \Big( p - \langle \mathbf{A}_x \rangle \Big), \qquad [148]$$

with $\kappa = mN/3$ the mass renormalization. Solving for $\langle \mathbf{A}_x \rangle$ in Eq. (148), we find the self-consistent condition,

$$\langle \mathbf{A}_x \rangle = \frac{\kappa p}{\kappa + M}, \qquad [149]$$

and plugging back into Eq. (146),

$$\mathbf{H}^{\mathrm{M}}_{\mathrm{mf}}(x,p) = \frac{p^2}{2M} \left( 1 - \frac{\kappa^2}{(\kappa + M)^2} \right) + V(x) + \mathbf{H}_{\mathrm{int}}(x) - \frac{p\mathbf{A}_x}{M + \kappa}. \qquad [150]$$

Crucially, as opposed to $\mathbf{H}^{\mathrm{M}}$, the mean-field Hamiltonian $\mathbf{H}^{\mathrm{M}}_{\mathrm{mf}}$ describes $N$ non-interacting particles for which the evaluation of thermodynamic quantities reduces to single-particle integrals.

## 12. Rotating Bucket

Consider a bucket of mass $M$ swinging on a rigid rod of length $R$ and carrying $N$ non-interacting "water" particles of mass $m$, as shown in Fig. 1(a) of the main text. Like in the previous piston example we will treat the water particles classically. The exact same analysis applies to quantum water molecules, where we simply have to replace functions of their positions and momenta by the corresponding quantum mechanical operators. The bucket is labelled by an angular coordinate $\theta$ and angular momentum $L_\theta$, while the water particles are labelled by polar coordinates $(r_i, \phi_i)$ and their conjugate radial and angular momenta $(p_{r_i}, L_{\phi_i})$. The interaction between the bucket and the $i$-th water particle is modelled by a rotationally invariant potential $V_b(r_i, \phi_i - \theta)$. There is also a gravitational force acting on both bucket and water particles. The Hamiltonian of the system is given by,

$$H = \frac{L_\theta^2}{2MR^2} + V_g(\vec{r}, \vec{\phi}, \theta) + H_{\mathrm{int}}(\vec{r}, \vec{\phi}, \vec{p}_r, \vec{L}_\phi, \theta), \qquad [151]$$

with,

$$H_{\mathrm{int}}(\vec{r}, \vec{\phi}, \vec{p}_r, \vec{L}_\phi, \theta) = \sum_{i=1}^{N} \left( \frac{p_{r_i}^2}{2m} + \frac{L_{\phi_i}^2}{2mr_i^2} + V_b(r_i, \phi_i - \theta) \right), \qquad V_g(\vec{r}, \vec{\phi}, \theta) = MgR\sin\theta + \sum_{i=1}^{N} mgr_i \sin\phi_i. \qquad [152]$$

In order to apply the MBOA fully, one would need to find the AGP $\mathbf{A}_\theta(\theta)$ associated with adiabatic deformations of the full Hamiltonian $\mathbf{V}_g(\theta) + \mathbf{H}_{\mathrm{int}}(\theta)$, including both gravitational and bucket-water interactions. In the absence of the gravitational potential, the AGP is simply the total angular momentum of the water molecules because the adiabatic transformations associated with the motion of the bucket are rotations. The gravitational interaction breaks the rotational invariance of the water plus bucket and the AGP becomes a non-trivial model-dependent operator. In the end of this section we will show how one can find the exact AGP for the special case where the bucket potential is represented by an anisotropic harmonic trap. It is intuitively clear that, for fast rotations, the difference between the exact AGP and the angular momentum becomes subdominant and can be neglected. This happens because water remains pinned to the floor of the bucket (as in the right panel of Fig. 1(a) of the main text) due to the centrifugal force and the adiabatic transformations are almost equivalent to rotations.

The transformation to the rotating frame is defined by the new water canonical variables $\delta\phi_i(\theta) = \phi_i - \theta$ and $L_{\phi_i}$, and slow variables $\theta$ and $\mathcal{L}$, which play a role analogous to $x$ and $p$ for the piston,

$$\mathcal{L} = L_\theta + \mathbf{A}_\theta = L_\theta + \sum_{i=1}^{N} L_{\phi_i}. \qquad [153]$$

Clearly the canonical momentum $\mathcal{L}$ corresponds to the total angular momentum of the bucket+water system. As before, the redefinition of $L_\theta$ is necessary to preserve the Poisson bracket $\{\delta\phi_i(\theta), \mathcal{L}\} = 0$.

To apply the MBOA, we assume that the water particles follow a Gibbs state $\rho^{\rm M}_\beta(\theta,\mathcal{L}) \sim e^{-\beta^{\rm M}\mathbf{H}^{\rm M}(\theta,\mathcal{L})}$ at fixed $\theta$ and $\mathcal{L}$, with,

$$\mathbf{H}^{\rm M}(\theta,\mathcal{L}) = \mathbf{H}(\theta,\mathcal{L}) - \frac{\mathcal{L}\mathbf{A}_\theta}{MR^2} + \frac{\mathbf{A}_\theta^2}{2MR^2}. \qquad [154]$$

Note the crucial difference between first transforming to the rotating frame and only after assuming equilibrium at fixed values of the slow parameters $(\theta, \mathcal{L})$. The moving Hamiltonian contains a new term $-\mathcal{L}\mathbf{A}_\theta/MR^2$, which is proportional to the angular momentum of water within our approximations, that can counteract the gravitational force. Indeed, notice that by completing the square,

$$\frac{L_{\phi_i}^2}{2mr_i^2} - \frac{\mathcal{L}L_{\phi_i}}{MR^2} = \frac{\left(L_{\phi_i} - \mathcal{L}\frac{mr_i^2}{MR^2}\right)^2}{2mr_i^2} - \frac{\mathcal{L}^2 mr_i^2}{2M^2R^4} \qquad [155]$$

we obtain a centrifugal potential $\sim -r_i^2$ that favors large values of the radius. In the limit when this centrifugal term dominates over the gravitational potential, i.e. when $\mathcal{L}$ is sufficiently large, the water equilibrates in a state that is pinned to the floor of the bucket and has a non-zero angular momentum $\langle \mathbf{A}_\theta \rangle$.

Let us now demonstrate this last statement by explicitly calculating the expectation value $\langle \mathbf{A}_\theta \rangle$ in the moving Gibbs state. In general one has to integrate over both angular and radial water degrees of freedom. The radial integral cannot be done without specifying the microscopic interactions between the bucket and the water. We can, however, simplify the analysis, by assuming that the radial integral is sharply peaked around the value $r_i = R$, which is true in the regime where the centrifugal force is much greater than the gravitational one or when the radial confinement is tight. Then,

$$\langle \mathbf{A}_\theta \rangle \approx \frac{1}{Z^{\rm M}(\theta,\mathcal{L},\beta)} \int_{-\infty}^{\infty} \mathrm{d}\vec{L}_\phi \left(\sum_k L_{\phi_k}\right) \exp\left\{-\beta^{\rm M} \sum_i \frac{\left(L_{\phi_i} - \frac{m}{M}\mathcal{L}\right)^2}{2mR^2} - \beta^{\rm M} \sum_{ij} \frac{L_{\phi_i}L_{\phi_j}}{2MR^2}\right\}, \qquad [156]$$

where we have absorbed constant terms independent of $\vec{L}_\phi$ into the definition of the partition function. This expression is a Gaussian integral that can be evaluated exactly to give,

$$\langle \mathbf{A}_\theta \rangle \approx \sum_k \langle L_{\phi_k} \rangle \approx \frac{mN\mathcal{L}}{M + mN} \equiv mN\omega R^2, \qquad [157]$$

where $\omega \equiv \mathcal{L}/(M + mN)R^2$ corresponds to the mean (time-averaged) angular frequency of the bucket. Note that $\omega$ is different from the instantaneous angular frequency $\dot{\theta} = L_\theta/MR^2$, which evolves in time in a noisy fashion due to collisions between the particles and the bucket. In contrast, $\omega$ evolves smoothly in time. Indeed, notice that,

$$\left\langle \dot{\theta} \right\rangle = \frac{\langle \mathbf{L}_\theta \rangle}{MR^2} = \frac{\mathcal{L} - \langle \mathbf{A}_\theta \rangle}{MR^2} = \frac{\mathcal{L} - \frac{mN}{M+mN}\mathcal{L}}{MR^2} = \frac{\mathcal{L}}{(M+mN)R^2} \equiv \omega\,. \qquad [158]$$

Similarly, one can find the MBO Hamiltonian for the slow DOF, $\mathcal{H}^{\rm MBO}(\theta,\mathcal{L})$, by averaging $\mathbf{H}^{\rm M}$ in the moving Gibbs state. We will do it in the large $N$-limit, where the fluctuations of the AGP are negligible, as described in Sec. 11. The mean-field moving Hamiltonian is given by,

$$\mathbf{H}^{\rm M}_{\rm mf}(\theta,\mathcal{L}) = \frac{\mathcal{L}^2}{2MR^2} - \frac{\langle \mathbf{A}_\theta \rangle^2}{2MR^2} + \mathbf{V}_g(\theta) + \mathbf{H}_{\rm int}(\theta) - \left(\frac{\mathcal{L} - \langle \mathbf{A}_\theta \rangle}{MR^2}\right)\mathbf{A}_\theta. \qquad [159]$$

We now average Eq. (159) over the mean-field equilibrium state $\rho^{\rm M}_{\rm mf} \propto e^{-\beta^{\rm M}\mathbf{H}^{\rm M}_{\rm mf}}$ with $\langle \mathbf{A}_\theta \rangle$ given by Eq. (157). As before, we assume that both radial and angular integrals are sharply peaked around the bucket coordinates so that we can simply replace $\phi_i \approx \theta$ and $r_i \approx R$. We find,

$$\mathcal{H}^{\rm MBO}(\theta,\mathcal{L}) = \langle \mathbf{H}^{\rm M}_{\rm mf} \rangle \approx \frac{\mathcal{L}^2}{2(M+mN)R^2} + (M+mN)gR\sin\theta + \text{const.} \qquad [160]$$

In Eq. (160), we have used the fact that $\mathbf{H}^{\mathrm{M}}_{\mathrm{mf}}$ is separable and the mean-field expectation value for the angular kinetic energy gives $\langle L^2_{\phi_k}/2mr_k^2\rangle \approx 1/(2\beta^{\mathrm{M}}) + m\omega^2 R^2/2$ independently for each of the water particles. Here, $\omega$ is given by Eq. (158) and corresponds to the mean (time-averaged) angular velocity of the bucket. The fact that water particles acquire extra kinetic energy due to the motion of the bucket results in a renormalization $M \to M + mN$ of the inertial mass appearing in the equations of motion for $\mathcal{L}$. The Hamiltonian Eq. (160) can be used to describe the motion of the bucket and find $\theta(t)$ and $\mathcal{L}(t)$ in the regime where the centrifugal force is large or the radial confinement is tight.

**Calculation of Exact AGP for Harmonic Bucket.** The exact AGP in the presence of gravity can be found for the special case where the water-bucket interaction is assumed to be harmonic. Without loss of generality, we show all calculations for a single fast particle. We set the bucket potential to have different spring constants $k_r$ and $k_\theta$ associated with displacements along the radial and angular directions, respectively, so that,

$$V_b(\vec{r}) = \frac{1}{2}k_r\left[(\vec{r}-\vec{R})\cdot\hat{R}\right]^2 + \frac{1}{2}k_\theta\left[(\vec{r}-\vec{R})\cdot\hat{\theta}\right]^2, \tag{161}$$

$$= \frac{1}{2}k_r(r\cos(\theta-\phi)-R)^2 + \frac{1}{2}k_\theta r^2\sin^2(\theta-\phi). \tag{162}$$

In addition, each particle has a gravitational potential energy $V_g(\vec{r}) = mg\vec{r}\cdot\hat{e}_y = mgr\sin\phi$.

We notice that even in the presence of gravity, the total potential $V_b + V_g$ is still harmonic with the same spring constants, but with a shifted minimum at position $\vec{r}_\star(\theta)$. The new minimum can be compactly expressed in terms of the mean and difference between the spring constants $k \equiv (k_r + k_\theta)/2$ and $\delta k \equiv k_r - k_\theta$,

$$\vec{r}_\star(\theta) = \vec{R}(\theta) + \frac{mgk}{k^2-\delta k^2/4}\left(\frac{\delta k\sin 2\theta}{2k}\hat{e}_x - \left(1+\frac{\delta k\cos 2\theta}{2k}\right)\hat{e}_y\right). \tag{163}$$

Clearly, if $g=0$ then the minimum is simply found at $\vec{r}_\star = \vec{R}(\theta) = R\cos\theta\,\hat{e}_x + R\sin\theta\,\hat{e}_y$, the position of the bucket.

The canonical transformation that preserves time-averaged trajectories is found by redefining the origin to be at position $\vec{r}_\star(\theta)$ and aligning the coordinate axes with the radial/angular directions. It is given by,

$$\vec{r}\,'(\theta) = \mathcal{R}(\theta)\left(\vec{r}-\vec{r}_\star(\theta)\right), \qquad \vec{p}\,'(\theta) = \mathcal{R}(\theta)\vec{p}, \qquad \mathcal{R}(\theta) = \begin{pmatrix}\cos\theta & \sin\theta\\ -\sin\theta & \cos\theta\end{pmatrix}. \tag{164}$$

Eq. (164) preserves the Poisson brackets $\{r'_i, p'_j\} = \delta_{ij}$ and, importantly, one can verify that rewriting the total potential $V_b + V_g$ in the primed variables yields,

$$V_b(\vec{r}\,') + V_g(\vec{r}\,') = \frac{1}{2}k_r x'^2 + \frac{1}{2}k_\theta y'^2 + V_b(\vec{r}_\star(\theta)) + V_g(\vec{r}_\star(\theta)), \tag{165}$$

where the only $\theta$-dependence appears as an overall energy shift $V_b(\vec{r}_\star(\theta)) + V_g(\vec{r}_\star(\theta))$. Hence, time-averaged trajectories in the primed frame are preserved under adiabatic changes of $\theta$.

The AGP associated with this transformation is given by,

$$A_\theta = L_\theta + \big(\partial_\theta(\mathcal{R}\vec{r}_\star)\big)\cdot\big(\mathcal{R}\vec{p}\big) = L_\theta + \frac{mgk}{k^2-\delta k^2/4}\left(\left(-1+\frac{\delta k\cos 2\theta}{2k}\right)p_x + \frac{\delta k\sin 2\theta}{2k}p_y\right), \tag{166}$$

where $L_\theta = xp_y - yp_x$ is the angular momentum in the original (unprimed) coordinates. Since $k^2 - \delta k^2/4 = k_r k_\theta$, the denominator in the second term is strictly positive. One can easily see that by setting $g=0$ the AGP reduces to simply the angular momentum operator. However, in the general case it is a linear combination of generators of both rotation and translation. The gravitational corrections to the AGP are expected to be negligible when $mgR/(kR^2) \ll 1$ (assuming small $\delta k$). This physically means that the spring constants are stiff such that the bucket potential energy is dominant over the gravitational one.

Let us now prove Eq. (166). To do so, it is first convenient to express the AGP in the primed coordinates. Define the skew-symmetric matrix,

$$\mathcal{J} \equiv \left[\partial_\theta\mathcal{R}(\theta)\right]\mathcal{R}^T(\theta) = \begin{pmatrix}0 & 1\\ -1 & 0\end{pmatrix}, \tag{167}$$

such that the angular momentum can be written as $L_\theta = -(\mathcal{J}\vec{r})\cdot\vec{p}$. By inverting Eq. (164), we find,

$$\begin{aligned} A_\theta &= -\Big(\mathcal{J}(\mathcal{R}^T\vec{r}\,' + \vec{r}_\star)\Big)\cdot(\mathcal{R}^T\vec{p}\,') \;+\; \partial_\theta\big(\mathcal{R}\vec{r}_\star\big)\cdot\vec{p}\,'\\ &= -(\mathcal{J}\vec{r}\,')\cdot\vec{p}\,' - (\mathcal{R}\mathcal{J}\vec{r}_\star)\cdot(\vec{p}\,') + \partial_\theta\big(\mathcal{R}\vec{r}_\star\big)\cdot\vec{p}\,'\\ &= L'_\theta + \Big[(\partial_\theta\mathcal{R})\vec{r}_\star + \mathcal{R}\partial_\theta\vec{r}_\star - \mathcal{R}\mathcal{J}\vec{r}_\star\Big]\cdot\vec{p}\,' \;=\; L'_\theta + \big(\mathcal{R}\,\partial_\theta\vec{r}_\star\big)\cdot\vec{p}\,', \end{aligned} \tag{168}$$

where we have used the properties (i) $[\mathcal{R},\mathcal{J}]=0$ and (ii) $\partial_\theta\mathcal{R} = \mathcal{J}\mathcal{R} = \mathcal{R}\mathcal{J}$. Finally, on the one hand from Eq. (164) we find,

$$\partial_\theta\vec{r}\,' = (\partial_\theta\mathcal{R})(\vec{r}-\vec{r}_\star) - \mathcal{R}\partial_\theta\vec{r}_\star = \left[(\partial_\theta\mathcal{R})\mathcal{R}^T\right]\vec{r}\,' - \mathcal{R}\partial_\theta\vec{r}_\star \equiv \mathcal{J}\vec{r}\,' - \mathcal{R}\partial_\theta\vec{r}_\star, \tag{169}$$

On the other hand,

$$\{A_\theta, \vec{r}\,'\} = \{L'_\theta + \mathcal{R}(\partial_\theta \vec{r}_\star)\cdot \vec{p}\,', \vec{r}\,'\} = -\{(\mathcal{J}\vec{r}\,')\cdot \vec{p}\,', \vec{r}\,'\} + \{\mathcal{R}(\partial_\theta \vec{r}_\star)\cdot \vec{p}\,', \vec{r}\,'\} = \mathcal{J}\vec{r}\,' - \mathcal{R}\partial_\theta \vec{r}_\star. \quad [170]$$

By comparing Eq. (169) and Eq. (170), we obtain the expected result,

$$\frac{\partial \vec{r}\,'(\theta)}{\partial \theta} = \{A_\theta, \vec{r}\,'\}. \quad [171]$$